\documentclass[a4paper]{article}

\usepackage{geometry}
\usepackage[T1]{fontenc}
\usepackage{graphicx} 
\usepackage[fleqn]{amsmath}
\usepackage{amssymb}
\usepackage{mathtools}
\usepackage{enumitem}
\usepackage{todonotes}
\usepackage{xspace}
\usepackage[ruled]{algorithm2e}
\usepackage{blkarray}
\usepackage{subcaption}
\usepackage{pgfplots}
\usepackage{bbm}
\usepackage{amsthm}
\usepackage{color}
\usepackage{tikz} 
\usepackage{xparse}

\pgfplotsset{compat=1.18}

\usepackage{siunitx}
\usepackage{authblk} 
\usepackage{booktabs}
\usepackage[colorlinks=true, allcolors=blue]{hyperref}
\usepackage[capitalize]{cleveref}
\usepackage{orcidlink} 

\newtheorem{definition}{Definition}

\newtheorem{theorem}[definition]{Theorem}
\newtheorem{lemma}[definition]{Lemma}
\newtheorem{proposition}[definition]{Proposition}
\newtheorem{corollary}[definition]{Corollary}
\newtheorem{conjecture}[definition]{Conjecture}
\newtheorem{example}[definition]{Example}
\numberwithin{definition}{section}

\newcommand{\conv}{\mathrm{conv}}
\newcommand{\lineal}{\mathrm{lineal}}
\newcommand{\proj}{\mathrm{proj}}
\newcommand{\cone}{\mathrm{cone}}
\newcommand{\rec}{\mathrm{rec}}
\newcommand{\spn}{\mathrm{span}}
\DeclareDocumentCommand\rank{}{\mathop{rk}}
\DeclareDocumentCommand\localrank{}{\mathop{lrk}}

\newcommand{\cleaninst}{all}

\newcommand{\bracket}[2]{[#1,#2]}

\DeclareDocumentCommand\transpose{m}{#1^{\intercal}}
\DeclareDocumentCommand\Z{}{\mathbb{Z}}
\DeclareDocumentCommand\Znonneg{}{\Z_{\geq 0}}
\DeclareDocumentCommand\Q{}{\mathbb{Q}}
\DeclareDocumentCommand\R{}{\mathbb{R}}
\DeclareDocumentCommand\Rnonneg{}{\R_{\geq 0}}
\DeclareDocumentCommand\M{}{\mathbb{MI}}

\DeclareDocumentCommand\zerovec{}{\mathbbm{O}}
\DeclareDocumentCommand\onevec{}{\mathbbm{1}}
\DeclareDocumentCommand\idmat{}{\mathbb{I}}

\DeclareDocumentCommand\unitvec{m}{\mathbbm{e}_{#1}}

\DeclareDocumentCommand\cplxNP{}{\mathsf{NP}}
\DeclareDocumentCommand\cplxcoNP{}{\mathsf{coNP}}
\DeclareDocumentCommand\orderO{o}{\mathcal{O}\IfValueTF{#1}{\left(#1\right)}{}}

\title{Implied Integrality in Mixed-Integer Optimization}
\author{
Rolf van der Hulst \orcidlink{0000-0002-5941-3016},
Matthias Walter \orcidlink{0000-0002-6615-5983}
}
\affil{
University of Twente, Enschede, The Netherlands\\
\sffamily
r.p.vanderhulst@utwente.nl, m.walter@utwente.nl
\rmfamily
}

\providecommand{\keywords}[1]
{
  \small	
  \textbf{{Keywords:}} #1
}
\date{\today}

\begin{document}

\maketitle

\begin{abstract}
    Implied-integer detection is a well-known presolving technique that is used by many Mixed-Integer Linear Programming solvers.
    Informally, a variable is said to be \emph{implied integer} if its integrality is enforced implicitly by integrality of other variables and the constraints of a problem.
   In this work we formalize the definition of implied integrality by taking a polyhedral perspective.
    Our main result characterizes implied integrality as occurring when a subset of integer variables is fixed to integer values and the polyhedron on the remaining variables is integral.
    While integral polyhedra are well-understood theoretically, existing detection methods infer implied integrality only for one variable at a time.
    We introduce new detection methods based on the detection of integral polyhedra, extending existing techniques to multiple variables.
    Additionally, we discuss the computational complexity of recognizing implied integers.
    We conduct experiments using a new detection method that uses totally unimodular submatrices to identify implied integrality.
    For the MIPLIB 2017 collection dataset our results indicate that, on average, \SI{18.8}{\percent} of the variables are classified as implied integer after presolving, compared to just \SI{3.3}{\percent} identified by state-of-the-art techniques. Moreover, we are able to reduce the average percentage of variables whose integrality needs to be enforced after presolving from 70.2\% to 59.0\%.
\end{abstract}

\section{Introduction}

In order to solve practical Mixed-Integer Linear Programming (MILP) problems, many solvers apply a presolve procedure to the given model with the aim of reducing its the size, to strengthen the linear programming (LP) relaxation and to infer additional information that can be exploited further during execution of a branch-and-bound algorithm~\cite{Land1960}.
This work is motivated by the technique of \emph{implied integers}, which falls under the latter category of presolving methods. 

In previous works, most notably by Achterberg, Bixby, Gu, Rothberg and Weninger~\cite{Achterberg2016, AchterbergBGRW20}, there exist multiple definitions of implied integrality.
In~\cite{Achterberg2016}, a continuous variable was considered to be \emph{implied integer} if it can only take integer values in any feasible solution.
In the journal version~\cite{AchterbergBGRW20} of the same article, the authors consider a set of continuous variables to be implied integer if the existence of an optimal solution for a MIP implies an optimal solution for which these variables take integral values.

Both definition lack an indication of which integer variables cause this implication.
Before providing a more rigorous definition based on a polyhedral point of view we introduce useful notation.

We consider an MILP problem given by a polyhedron $P = \{ x \in \R^N \mid A x \leq b \}$, where $A \in \R^{M \times N}$ and $b \in \R^M$ for index sets $M$ and $N$, and a set $I \subseteq N$ of variables $x_I$ that are required to be integral. 
For any $S \subseteq N$, we use $\M^S \coloneqq  \{ x \in \R^N \mid x_i \in \Z ~\forall i \in S \}$ to denote the corresponding mixed-integer set.
For $J \subseteq M$ and $S \subseteq N$, we use $A_{J,S}$ to denote the $|J| \times |S|$ submatrix of $A$ induced by $I$ and $J$. If $J$ or $S$ is $\star$ then this shall indicate all rows or columns, respectively.
Additionally, we shall use $\zerovec$ and $\onevec$ to represent the vectors whose entries are all zeros and ones, respectively.
Finally, by $\Rnonneg$ and $\Znonneg$ we denote the nonnegative reals and integers, respectively.

Now, let us define implied integrality. 

\begin{definition}[implied integrality]
  \label{def_implied_integer}
  Let $P \subseteq \R^N$ be a rational polyhedron and let $S,T \subseteq N$.
  We say that \emph{$x_T$ is implied integer by $x_S$ for $P$} if $\conv(P \cap \M^S) = \conv(P \cap \M^{S \cup T})$ holds. 
  If $S = \emptyset$, we also say that \emph{$x_T$ is implied integer} and that \emph{$P$ is $T$-integral}. Furthermore, we say that $S$ are the \emph{implying variables} and $T$ are the \emph{implied variables}.
\end{definition}

\Cref{def_implied_integer} differs from the previously proposed definitions~\cite{AchterbergBGRW20}.
It is more general since the implied set $T$ is not necessarily restricted to continuous variables, and since $T$ can be of arbitrary size, rather than just a single-variable set.
However, \cref{def_implied_integer} is independent of the objective function.
In \cref{sec_local_implied_integrality} we consider \emph{local implied integrality}, which takes the objective into account.

Note that \cref{def_implied_integer} for $S = \emptyset$ and $T = N$ is equivalent to the usual definition of integral polyhedra, such as in \cite[Chapter 16]{Schrijver86}.
Next, we briefly review a few concepts related to integral polyhedra; for more detailed definitions and explanations, we refer to \cite[Chapter 16]{Schrijver86}.
For a set $X\subseteq \R^N$, the \emph{recession cone} $\rec(X)\coloneqq \{ y\in \R^N \mid x+\lambda y \in X  \forall x \in X~ \forall \lambda \geq 0\}$ indicates the directions in which $X$ is unbounded.
The \emph{lineality space} $\lineal(X) \coloneqq \{y \in \rec(X) \mid - y \in \rec(X) \}$ is the largest linear subspace contained in $X$.
Moreover, for two sets $X,Y\subseteq \R^N$, we denote the Minkowski sum by $X + Y \coloneqq \{ x + y \mid x \in X,~ y \in Y \}$.
A polyhedron $P$ is \emph{rational} if there exist rational $A \in \Q^{m\times n}$ and $b \in \Q^{m}$ such that $P = \{x\in \R^n \mid A x \leq b\}$ holds. A polyhedron $P$ is said to be \emph{pointed} if $\lineal(P) = \{ \zerovec \}$ holds, which is equivalent to $P$ having ($0$-dimensional) vertices as its minimal faces.
A rational polyhedron $P \subseteq \R^N$ is said to be \emph{integral} if $P = \conv(P \cap \Z^N)$ holds.
Note that we require rationality of $P$, as otherwise $\conv(P\cap\Z^N)$ may not be a polyhedron. A fundamental theorem by Meyer~\cite{Meyer74} shows that if $P$ is a rational polyhedron then $\conv(P \cap \M^I)$ is a rational polyhedron with the same recession cone unless it is empty.

\begin{proposition}[Meyer~\cite{Meyer74}; see also Theorem~16.1 in \cite{Schrijver86}]
  \label{thm_meyer}
  Let $P \subseteq \R^N$ be a rational polyhedron.
  For any $I \subseteq N$, $\conv(P \cap \M^I)$ is a rational polyhedron and, if non-empty, has the same recession cone as $P$.
\end{proposition}

From a practical point of view, rationality is not a limitation and hence we focus on rational polyhedra.
There are many known classes of integral polyhedra, such as those generated by totally unimodular matrices~\cite{HoffmanK56}, k-regular matrices~\cite{Appa2004}, balanced matrices~\cite{Berge72} or independence polytopes of intersections of matroids~\cite{Edmonds70}.

\subsection{Related work}

Implied integrality can also be interpreted as a method to reduce the number of necessary integrality constraints, which has been explored in prior research. In~\cite{Paat2022}, Paat, Schl{\"o}ter and Weismantel show lower bounds on the number of necessary integrality constraints for integer programs after reformulation as a mixed-integer program for integer programs with bounded subdeterminants.
Bader, Hildebrand, Weismantel and Zenklusen~\cite{BaderHWZ18} introduce the notion of an \emph{affine TU decomposition}, which decomposes a constraint matrix $A \in \Z^{m \times n}$ as $A = \tilde{A} + U W$, such that $\begin{bmatrix}
  \transpose{\tilde{A}} & \transpose{W} 
\end{bmatrix}$ is totally unimodular and $U \in \Z^{m\times k}$ holds.
If an integer program admits such an affine TU decomposition then it can be reformulated as a mixed-integer program with only $k$ integrality constraints.
Together with Hupp, the same authors also describe specific affine TU decompositions for several families of integer programs, including the resource-constrained bipartite matching problem and the master knapsack problem~\cite{Hupp2017}.
For the two-stage stochastic bipartite $b$-matching problem, Hupp~\cite{Hupp2017} evaluates several reformulations based on affine TU decompositions, and shows that, in general, reformulation using affine TU decompositions is helpful and leads to smaller root gaps and reduced solution times.
In~\cite[Chapter 3]{DeKruijff2019}, de Kruijff considers lot-sizing and production planning models and investigates whether the inclusion of integrality constraints for inventory variables is helpful.
De Kruijff proposes a dynamic method that decides, based on the root LP solution, whether or not to include integrality constraints.
In simplified terms, their algorithm decides to include integrality constraints for variables if their value in the root LP solution is relatively small.
De Kruijff also recommends to investigate the phenomenon of implied integrality more generally.

Implied integers are used in state-of-the-art MILP solvers such as SCIP~\cite{bolusani2024scipoptimizationsuite90}, HiGHS~\cite{HuangfuH17} and Gurobi~\cite{AchterbergBGRW20}.
In contrast to the theoretical results for integral polyhedra, the detection of implied integers for general mixed-integer programming solvers received little attention in the literature.
To the best of our knowledge, it has only been investigated in \cite{AchterbergBGRW20}, and received a quick mention in Achterberg's thesis~\cite[Section 14.5.1]{Achterberg2008}.
In~\cite{AchterbergBGRW20} a primal and a dual detection method are presented, and the computational impact of enabling implied integer detection for the Gurobi solver is highlighted. We illustrate the primal and dual detection method from~\cite{AchterbergBGRW20} in \cref{primal_example} and \cref{dual_example}.

\begin{example}[Primal detection]
  \label{primal_example}
  We consider a mixed-integer program that has the constraint $a x + b y + z = c$ for $a,b,c \in \Z$ with integer variables $x, y \in \Z$.
  After fixing $x = \bar{x}$ and $y = \bar{y}$ for arbitrary integer values $\bar{x},\bar{y}$, we can infer that $z = c - a \bar{x} - b \bar{y}$ must be integral.
  Since this holds for any choice of integral $\bar{x},\bar{y}$, we can freely choose to omit or enforce the constraint $z \in \Z$. 
\end{example}

\begin{example}[Dual detection]
  \label{dual_example}
  Consider the following mixed-integer program:
  \begin{subequations}
    \label{dual_detection_example}
    \begin{alignat}{7}
      &\text{maximize }
        & 10 x + 10 y + z \\
      &\text{s.t. }
        & 3 x + 2 y + z &\leq 4 \label{dual_example_ineq1} \\
      & &   x + 3 y - z &\leq 3 \label{dual_example_ineq2} \\
      & &             z &\geq 0 \label{dual_example_ineq_3} \\
      & &          x, y &\in \Znonneg
    \end{alignat}
  \end{subequations}
  Assume that $x$ and $y$ are fixed to integral values $\bar{x}$ and $\bar{y}$, respectively.
  Then the remaining linear program that maximizes $z$ only has integral bounds for the single variable $z$, since $4 - 3 \bar{x} - 2 \bar{y}$ and $3 - 3\bar{y} - \bar{x}$ and $0$ are integral.
  This implies that an optimal solution with integral $z$ exists.
  Thus, one can freely choose to omit or enforce the constraint $z \in \Z$.
  Alternatively, one could use complementary slackness to argue that one of the inequalities  \eqref{dual_example_ineq1}, \eqref{dual_example_ineq2} or \eqref{dual_example_ineq_3} must hold with equality in an optimal solution, and use the same argument as for the primal detection method to infer that $z$ is implied integer.

  By using $z \in \Z$, stronger cutting planes can be derived.
  The initial solution of the LP relaxation of \eqref{dual_detection_example} is $(x^\star,y^\star,z^\star) = (\frac{6}{7},\frac{5}{7},0)$ with objective $\frac{110}{7}\approx 15.71$.
  The Gomory-Mixed-Integer cut~\cite{Gomory60} from the Simplex tableau for variable $y$ is given by $21 x + 49 y \leq 43 + 33 z$.
  Adding this cut to \eqref{dual_detection_example} and optimizing the LP relaxation yields $(x_1^\star, y_1^\star, z_1^\star) = (\frac{1}{2},1,\frac{1}{2})$, which has objective value $15.5$.
  The Gomory-Mixed-Integer cut that additionally exploits the integrality of $z$ is given by $21 x + 49 y \leq 43 + 19 z$ and has a stronger coefficient for $z$.
  In particular, one can easily verify that $(x_1^\star, y_1^\star, z_1^\star)$ is not feasible for this stronger cut.
  Resolving the LP relaxation in \eqref{dual_detection_example} augmented with this stronger cut yields the solution $(x_2^\star, y_2^\star, z_2^\star) = (0, \frac{119}{87}, \frac{110}{87})$, which has an objective value of approximately $14.94$.
\end{example}

\noindent
The main benefit of implied integers is that the integrality of a variable can be used during the solution process without having to explicitly enforce its integrality. 
Implied integer variables that are continuous variables in the original model may be used in branching and can use integrality to derive stronger coefficients in cutting planes, which is highlighted in \cref{dual_example}.
Moreover, new classes of cutting planes that require rows with integer variables only may become usable.
Furthermore, propagation techniques and conflict analysis can infer stronger bounds by using the detected integrality.
For implied integer variables that are integer variables in the original model it is not necessary to enforce their integrality during branch-and-bound, which can reduce the size of branch-and-bound tree.
This can be particularly useful in primal heuristics that successively solve linear programs and fix variables, such as diving methods or local-neighborhood search methods, to limit the number of branching candidates that are to be fixed or explored. 
Additionally, dual presolving techniques that work only on the continuous variables, such as fixing by complementary slackness~\cite[section 7.5]{Achterberg2016}, may become available for the implied integer variables.
One result from~\cite{AchterbergBGRW20} is that enabling implied integers reduces Gurobi's total solve time by \SI{13}{\percent} and the total node count by \SI{18}{\percent} on average over all models in their testset that take longer than \SI{10}{\second} to solve.
Compared to the other presolving methods tested in \cite{AchterbergBGRW20}, this is a significant reduction and highlights that implied integer detection is one of the most impactful presolving methods for mixed-integer optimization.

\subsection{Contribution and outline}
As the focus in~\cite{AchterbergBGRW20} is clearly on computational impact, a formal treatment of the topic is missing.
In this work, we close this gap by studying implied integrality from a polyhedral perspective and apply it in various ways:
\begin{itemize}
\item
  We formally define implied integers as the equivalence of two mixed-integer hulls with two different sets of integer variables.
  As our main tool, we apply the concept of a \emph{fiber}~\cite{BilleraS92}, which arises from the LP relaxation by fixing a subset of the variables to integers.
  We characterize implied integrality for a polyhedron in terms of (partial) integrality of all fibers.
\item 
  We present conditions under which integrality of all fibers can be established by well-known concepts for establishing perfect formulations, such as (total) unimodularity.
  We show that this generalizes the dual detection method.
\item
  We show that partial integrality of the fibers may arise from fixed variables in the fibers, which generalizes the primal detection method. Furthermore, we investigate the structure of fixed integer linear expressions in the fibers and highlight their connection with lattice theory.
\item 
  We highlight that by using the objective, one can derive additional implied integrality relationships.
\item
  Our new methods yield questions related to the computational complexity of recognizing implied integers, some of which we answer.
\item 
  We complement our theoretical results with a computational study by detecting implied integers using totally unimodular submatrices.
  For the MIPLIB~2017 collection dataset our results indicate that, on average, \SI{18.8}{\percent} of the variables are classified as implied integer after presolving, compared to just \SI{3.3}{\percent} identified by the primal and the dual detection method. Moreover, we can significantly reduce the average percentage of variables whose integrality needs to be enforced after presolving from \SI{70.2}{\percent} to \SI{59.0}{\percent}.
\end{itemize}

\medskip
\noindent \textbf{Outline.}
In \cref{sec_polyhedral} we present a polyhedral view on implied integrality. In \cref{sec_block_structure,sec_implied_fixed}, we use this view to derive new methods to detect implied integrality that generalize the two methods proposed in literature.
In \cref{sec_local_implied_integrality}, we expand the notion of implied integrality to include information from the objective.
In \cref{sec_complexity} we present complexity-theoretic consequences of our more general concept.
Finally, in \cref{sec_computations} we complement the theoretical results by a computational study that highlights the practical impact of the presented ideas.



\section{A polyhedral view on implied integrality}
\label{sec_polyhedral}

\noindent
The primal and dual detection methods in \cref{primal_example,dual_example} use integrality of some variable set $x_S$ to conclude that another single variable $x_t$ can be assumed to be integral. In \cref{def_implied_integer} we generalized this idea to multiple variables $x_T$ and phrased it in terms of polyhedra. An example is depicted in \Cref{fig_implied_integrality_example}.

\subsection{Characterizations using fibers}

As noted in the introduction, \cref{def_implied_integer} contains the definition of integral polyhedra with $S = \emptyset$ and $T = N$.
For the latter there exist several equivalent statements, such as~\cite[Theorem~4.1]{ConfortiCZ14} and~\cite[Section~16.3]{Schrijver86}. 
Note that for $S = \emptyset$ we typically have a description of $\conv(P \cap \M^S) = P$ in terms of linear inequalities.
We will add two more equivalent statements, both of which reduce implied integrality to this case using the concept of a fiber.

\begin{definition}[fiber;  see~\cite{BilleraS92}]
  Let $P \subseteq \R^N$ be a polyhedron and let $S \subseteq N$.
  The \emph{$S$-integral fibers} are the sets $Q_S(\bar{x}) \coloneqq \{ x \in P \mid x_S = \bar{x} \}$, parameterized by $\bar{x} \in \Z^S$.
\end{definition}

The fibers for the example depicted in \Cref{fig_implied_integrality_example} are shown in \Cref{fig_implied_integrality_fractional_fiber}.
Note that a fiber $Q_S(\bar{x})$ is non-empty if and only if $\bar{x}$ lies in the projection of $P$ onto the $x_S$-variables. Hence, to deal with the odd case of $S = \emptyset$, it is convenient to define $Z^\emptyset$ as a set consisting of a single element $\bar{x}^\emptyset$ that satisfies $Q_\emptyset(\bar{x}^\emptyset) = P$ in this case.
Moreover, if the constraints defining $P$ are $Ax_S + Bx_{N \setminus S} \leq c$ then
$
  Q_S(\bar{x}) = \{ x \in \R^N \mid x_S = \bar{x},~ B x_{N \setminus S} \leq c - A \bar{x} \}
$ is an inequality description of the fiber.

We now turn to the characterization of implied integrality.
We formulate our result similarly to \cite[Section 16.3]{Schrijver86}.

\begin{theorem}
  \label{thm_implied_integrality}
  Let $P \subseteq \R^N$ be a rational polyhedron and let $S, T\subseteq N$.
  Then the following are equivalent:
  \begin{enumerate}[label=(\roman*)]
  \item \label{thm_implied_integrality_definition}
    $x_T$ is implied integer by $x_S$ for $P$, i.e., $\conv(P \cap \M^S) = \conv(P \cap \M^{S\cup T})$.
  \item \label{thm_implied_integrality_minimal_faces}
    Every minimal face of $\conv(P \cap \M^S)$ contains a point $x \in \M^{S \cup T}$.
  \item \label{thm_implied_integrality_faces}
    Every non-empty face of $\conv(P \cap \M^S)$ contains a point $x \in \M^{S \cup T}$.
  \item \label{thm_implied_integrality_maxima}
    $\max \{c^\intercal x \mid x \in \conv(P \cap \M^S)\}$ has some optimal solution $x^\star \in \M^{S \cup T}$ for each $c$ for which the maximum is finite.
  \item
    \label{thm_implied_integrality_fiber_minimal_faces}
    For each $\bar{x} \in \Z^S$ and every minimal face $F$ of $Q_S(\bar{x})$,  $F \subseteq \conv(P \cap \M^{S \cup T})$ holds.
  \item
    \label{thm_implied_integrality_fiber_contained}
    $Q_S(\bar{x}) \subseteq \conv(P \cap \M^{S \cup T})$ holds for each $\bar{x} \in \Z^S$.
  \end{enumerate}
\end{theorem}

\begin{proof}
  \ref{thm_implied_integrality_definition} $\Rightarrow$ \ref{thm_implied_integrality_minimal_faces}:
  Let $F$ be a minimal face of $\conv(P \cap \M^S)$.
  By assumption, $F$ is thus a non-empty face of $\conv(P \cap \M^{S \cup T})$, and hence $F \cap (P \cap \M^{S \cup T}) \neq \emptyset$ holds.
  
\noindent
  \ref{thm_implied_integrality_minimal_faces} $\Rightarrow$ \ref{thm_implied_integrality_faces}:
  Let $F$ be a non-empty face of $\conv(P \cap \M^S)$.
  Then $F$ must contain some minimal face $F'\subseteq F$, with $x \in F'$ such that $x_{S \cup T}$ is integral.
  Clearly, $x \in F$ holds too.

\noindent
  \ref{thm_implied_integrality_faces} $\Rightarrow$ \ref{thm_implied_integrality_maxima}:
  For any $c$ with a finite maximum, consider $\max \{ \transpose{c} x \mid x \in \conv(P \cap \M^S) \} = \delta < \infty$.
  By definition, $F = \{x \in \conv(P \cap \M^S) \mid \transpose{c} x = \delta \}$ is a non-empty face.
  By assumption, $F$ contains a point $x^\star$ such that $x^\star_{S \cup T}$ is integral.
  Clearly, $x^\star$ attains $\max \{ \transpose{c} x \mid x \in \conv(P \cap \M^S) \}$ and has $x^\star_{S \cup T}$ integral.

\noindent
  \ref{thm_implied_integrality_maxima} $\Rightarrow$ \ref{thm_implied_integrality_definition}:
   The inclusion $\conv(P \cap \M^{S\cup T}) \subseteq \conv(P \cap \M^S)$ follows from $P \cap \M^{S \cup T} \subseteq P \cap \M^S$ and monotonicity of the convex hull.
   Assume that the reverse inclusion does not hold, i.e., $\conv(P \cap \M^S ) \not\subseteq \conv(P \cap \M^{S \cup T})$.
   Thus, there exists an objective vector $c$ such that $\delta^S \coloneqq \max\{ \transpose{c}x \mid x \in \conv(P \cap \M^S) \} > \delta^{S \cup T} \coloneqq \max\{ \transpose{c}x \mid x \in \conv(P\cap\M^{S \cup T}) \}$ holds.
   Note that by applying \ref{thm_implied_integrality_maxima} with $c = \zerovec$, $P$ contains a point in $\M^{S \cup T}$.
   Hence, by \cref{thm_meyer}, the recession cones of $\conv(P \cap \M^S)$ and of $\conv(P \cap \M^{S \cup T})$ are both identical to that of $P$, which implies that both maxima are attained, i.e., $\delta < \infty$.
   By \ref{thm_implied_integrality_maxima}, the face of $\conv(P \cap \M^S)$ induced by the valid inequality $\transpose{c}x \leq \delta$ (which is set to equality) contains a minimal face which in turn contains a point $x \in \M^{S \cup T}$.
   Clearly, $x \in \conv(P \cap \M^{S\cup T})$, which contradicts $\delta^S > \delta^{S \cup T}$.
   We conclude that $\conv(P\cap\M^S) \subseteq \conv(P\cap\M^{S\cup T})$ (and hence equality) holds.

\noindent
   \ref{thm_implied_integrality_definition} $\Rightarrow$ \ref{thm_implied_integrality_fiber_minimal_faces}:
   Follows from $F \subseteq Q_S(\bar{x}) \subseteq \conv(P \cap \M^S) = \conv(P \cap \M^{S\cup T})$.

\medskip
\noindent
  \ref{thm_implied_integrality_fiber_minimal_faces} $\Rightarrow$ \ref{thm_implied_integrality_fiber_contained}:
  Follows since $Q_S(\bar{x})$ is the convex hull of all its minimal faces.

\medskip
\noindent
  \ref{thm_implied_integrality_fiber_contained} $\Rightarrow$ \ref{thm_implied_integrality_definition}:
  As above, $\conv(P \cap \M^{S \cup T}) \subseteq \conv(P \cap \M^{S})$ trivially holds.
  For the reverse inclusion, observe that 
  \[
    P \cap \M^S
    = \bigcup_{\bar{x} \in \Z^S} Q_S(\bar{x})
    \subseteq \bigcup_{\bar{x} \in \Z^S} \conv(P \cap \M^{S \cup T})
    = \conv(P \cap \M^{S \cup T})
  \]
  holds.
  Since the right-hand side is convex, we can conclude with
  $\conv(P \cap \M^S) \subseteq \conv(P \cap \M^{S \cup T})$.
\end{proof}

The following corollary follows from \cref{thm_implied_integrality}\ref{thm_implied_integrality_fiber_contained} since $x \in Q_S(\bar{x}) \cap \M^T$ (for $\bar{x} \in \Z^S$) implies $x \in P \cap \M^{S \cup T}$, and is useful for establishing implied integrality.

\begin{corollary}
  \label{thm_implied_integer_sufficiency}
  Let $S, T \subseteq N$ and let $P \subseteq \R^N$ be a rational polyhedron such that $Q_S(\bar{x})$ is $T$-integral for each $\bar{x} \in \Z^S$.
  Then $x_T$ is implied integer by $x_S$ for $P$.
\end{corollary}

Note that $T$-integrality of all fibers is, in general, not a necessary condition, as shown in \Cref{fig_implied_integrality_fractional_fiber}.
However, in case of binary $x_S$-variables, $T$-integrality of all fibers turns out to also be necessary:

\begin{corollary}
  \label{thm_implied_integer_binary}
  Let $S, T \subseteq N$ and let $P \subseteq [0,1]^S \times \R^{N \setminus S}$ be a rational polyhedron.
  Then the following are equivalent:
  \begin{enumerate}[label=(\alph*)]
  \item
    \label{eq_implied_integer_binary_def}
    $x_T$ is implied integer by $x_S$ for $P$.
  \item
    \label{eq_implied_integer_binary_minface}
    For each $\bar{x} \in \{0,1\}^S$, every minimal face of $Q_S(\bar{x})$ contains a point in $\M^T$.
  \item
    \label{eq_implied_integer_binary_emptyset}
    For each $\bar{x} \in \{0,1\}^S$,  $Q_S(\bar{x})$ is $T$-integral.
  \end{enumerate}
\end{corollary}

\begin{proof}
  \ref{eq_implied_integer_binary_def} $\Rightarrow$ \ref{eq_implied_integer_binary_minface}:
  Consider a vector $\bar{x} \in \{0,1\}^S$ and a minimal face $F$ of $Q_S(\bar{x})$.
  By definition, we have $\conv(P \cap \M^S) = \conv(P \cap \M^{S \cup T})$.
  Since $0 \leq x_j \leq 1$ is valid for all $j \in S$, $Q_S(\bar{x})$ is a face of $\conv(P \cap \M^S) = \conv(P \cap \M^{S \cup T})$, where equality follows by definition of implied integrality.
  Hence, $F$ is a face of $\conv(P \cap \M^{S \cup T})$ and thus contains a point $x \in \M^{S \cup T} \subseteq \M^T$.

\medskip
\noindent
  \ref{eq_implied_integer_binary_minface} $\Rightarrow$ \ref{eq_implied_integer_binary_def}:
  Let $\bar{x} \in \Z^S$. 
  If $\bar{x} \notin \{0,1\}^S$ then $Q_S(\bar{x}) = \emptyset$, which has no minimal face.
  Otherwise, there is a point $x \in \M^T$ in a minimal face $F$ of $Q_S(\bar{x})$.
  From $x_S = \bar{x} \in \Z^S$ we obtain $x \in \M^{S \cup T}$ and hence $x \in \conv(P \cap \M^{S \cup T})$.
  Moreover, the lineality space $L\coloneqq \lineal(Q_S(\bar{x}))$ satisfies $L\subseteq \lineal(P) = \lineal(\conv(P \cap \M^{S \cup T}))$, where containment follows from monotonicity and equality follows from \cref{thm_meyer}. 
  We conclude that even $F = x + L \subseteq \conv(P \cap \M^{S \cup T})$ holds.
  Thus, \ref{thm_implied_integrality_fiber_minimal_faces} in \cref{thm_implied_integrality} is satisfied, and the claim follows from the theorem.

\medskip
\noindent
  \ref{eq_implied_integer_binary_minface} $\Leftrightarrow$ \ref{eq_implied_integer_binary_emptyset}:
  Follows by definition of implied integrality.
\end{proof}

The core argument used in \cref{thm_implied_integer_binary} is that, due to $\zerovec \leq x_S \leq \onevec$, the $S$-integral fibers are actually faces of $\conv(P \cap \M^S)$.

\medskip

Unfortunately, $T$-integrality (of the $S$-integral fibers) may be difficult to establish.
However, in case $Q_S(\bar{x}) = Q_1 \times Q_2$ with $Q_1 \subseteq \R^T$ and $Q_2 \subseteq \R^{N \setminus (S\cup T)}$ holds, it reduces to integrality of $Q_1$, for which well-known techniques can be utilized.
The following lemma makes this formal and turns out to be useful when investigating structured MILPs in the next section.

\begin{lemma}
  \label{thm_block_integrality}
  The Cartesian product $Q_1 \times Q_2$ of two polyhedra $Q_1 \subseteq \R^{N_1}$ and $Q_2 \subseteq \R^{N_2}$ is $T$-integral if and only if $Q_i$ is $(T \cap N_i)$-integral for $i=1,2$.
\end{lemma}
\begin{proof}
  The statement follows from the well-known fact that every face of the product is the Cartesian product of a face of $Q_1$ and a face of $Q_2$.
  Although some textbooks refer to this fact \cite{Ziegler01}, we could not find a proof of it, so we provide one for the sake of completeness.

  Let $F$ be any face of $Q_1 \times Q_2$.
  By definition of a face, there exists some $(c_1,c_2) \in \R^{N_1} \times \R^{N_2}$ and $\delta = \max \{ \transpose{c}_1 x + \transpose{c}_2 y \mid (x,y) \in Q_1\times Q_2 \}$ such that $F = (Q_1 \times Q_2) \cap \{ (x,y) \in \R^{N_1} \times \R^{N_2} \mid \transpose{c}_1 x + \transpose{c}_2 y = \delta\}$.
  First, notice that since $x$ and $y$ are independent in $Q_1\times Q_2$, we can optimize over them separately.
  Thus, $\delta = \max\{\transpose{c}_1 x + \transpose{c}_2 y \mid (x,y) \in Q_1\times Q_2\} = \max \{\transpose{c}_1 x \mid x \in Q_1 \} + \max\{ \transpose{c}_2 y \mid y \in Q_2\} = \delta_1 + \delta_2$ holds, and $\transpose{c}_1 x = \delta_1$ and $\transpose{c}_2 y = \delta_2$ must both hold for any point $(x,y) \in F$.
  By rewriting the definition of $F$, we obtain \begin{equation*}
    F =  \{x \in Q_1 \mid \transpose{c}_1 x = \delta_1 \} \times \{y \in Q_2  \mid \transpose{c}_2 y = \delta_2 \}.
  \end{equation*}
  Note that $F_1 = \{x \in Q_1 \mid \transpose{c}_1 x = \delta_1 \}$ is a face of $Q_1$ and, similarly, $F_2 = \{y \in Q_2 \mid \transpose{c}_2 y = \delta_2\}$ is a face of $Q_2$.
  Thus, every face $F$ of $Q_1 \times Q_2$ is the Cartesian product of a face of $Q_1$ and $Q_2$.
\end{proof}

\subsection{Combined implied integrality relations}

In our definition of implied integrality in \cref{def_implied_integer} we consider precisely the two sets $S$ and $T$.
One downside of this characterization is that we cannot additionally require integrality of $x_j$ for any $j \in N \setminus (S \cup T)$ and add $j$ to $S$ since the derived implied integrality of $x_T$ may no longer be valid (see \Cref{fig_implied_integrality_not_totally_implied}).
This motivates the definition of a stronger implied integrality concept.

\begin{definition}[total implied integrality]
  \label{def_total_implied_integer}
  Let $P \subseteq \R^N$ be a rational polyhedron.
  For $S, T \subseteq N$, we say that \emph{$x_T$ is totally implied integer by $x_S$ for $P$} if $x_T$ is implied integer by $x_{S'}$ for every $S' \supseteq S$ (for $P$).
\end{definition}

\begin{figure}[htpb]
  \begin{subfigure}[t]{0.30\textwidth}
    \centering%
    \begin{tikzpicture}
      \draw[blue, fill=blue!20, dash pattern=on 2pt off 4pt, dash phase=-1.0pt, line width=2pt] (-0.2, 0) -- (2.25,0) -- (2.25,2.125) -- (-0.2,0.9) -- cycle;
      \draw[green!60!black, dash pattern=on 2pt off 4pt, dash phase=0.7pt, line width=2pt] (0, 0) -- (2,0) -- (2,2) -- (0,1) -- cycle;
      \draw[red, dash pattern=on 2pt off 4pt, dash phase=2.7pt, line width=2pt] (0, 0) -- (2,0) -- (2,2) -- (0,1) -- cycle;

      \foreach \y in {0,1,2}
      {
        \draw (-0.4,\y) -- (2.4,\y);
      }
  
      \foreach \x in {0,1,2}
      {
        \draw (\x,-0.4) -- (\x,2.4);
      }

    \end{tikzpicture}
    \subcaption{%
      $\textcolor{red}{P_I} = \textcolor{green!60!black}{P_1} \subsetneqq \textcolor{blue}{P}$
      \label{fig_implied_integrality_example}
    }
  \end{subfigure}
\hfill
  \begin{subfigure}[t]{0.30\textwidth}
    \centering%
    \begin{tikzpicture}
      \draw[blue, fill=blue!20, dash pattern=on 2pt off 4pt, dash phase=-1.0pt, line width=2pt] (-0.2, 0) -- (2.25,0) -- (2.25,2.125) -- (-0.2,0.9) -- cycle;

      \foreach \y in {0,1,2}
      {
        \draw (-0.4,\y) -- (2.4,\y);
      }
  
      \foreach \x in {0,1,2}
      {
        \draw (\x,-0.4) -- (\x,2.4);
      }

      \draw[purple, line width=2.5pt] (0,0) -- (0,1);
      \draw[purple, line width=2.5pt] (1,0) -- (1,1.5);
      \draw[purple, line width=2.5pt] (2,0) -- (2,2);

    \end{tikzpicture}
    \subcaption{%
      \textcolor{purple}{$\{1\}$-fibers} of \eqref{fig_implied_integrality_example}; one has a fractional vertex.
      \label{fig_implied_integrality_fractional_fiber}
    }
  \end{subfigure}
\hfill
  \begin{subfigure}[t]{0.30\textwidth}
    \centering%
    \begin{tikzpicture}
      \draw[blue, fill=blue!20, dash pattern=on 2pt off 2pt, dash phase=0pt, line width=2pt] (-1, -0.5) -- (0,0.5) -- (1,0.5) -- (0,-0.5) -- cycle;

      \draw[green!60!black, dash pattern=on 2pt off 2pt, dash phase=2pt, line width=2pt] (-1, -0.5) -- (0,0.5) -- (1,0.5) -- (0,-0.5) -- cycle;

      \foreach \y in {-1,0,1}
      {
        \draw (-1.4,\y) -- (1.4,\y);
      }
  
      \foreach \x in {-1,0,1}
      {
        \draw (\x,-1.4) -- (\x,1.4);
      }

      \draw[orange, line width=1.5pt] (0.5,0) -- (-0.5,0);

      \node[red, fill=red, circle, inner sep=1.5pt] at (0,0) {};
    \end{tikzpicture}
    \subcaption{%
      $\textcolor{red}{P_I} \subsetneqq \textcolor{orange}{P_2} \subsetneqq \textcolor{blue}{P} = \textcolor{green!60!black}{P_1}$
      \label{fig_implied_integrality_not_totally_implied}
    }
  \end{subfigure}
  \caption{%
    Examples of a polyhedron $\textcolor{blue}{P} \subseteq \R^2$ together with $\textcolor{green!60!black}{P_1 \coloneqq \conv(P \cap (\Z \times \R))}$, $\textcolor{orange}{P_2 \coloneqq \conv(P \cap (\R \times \Z))}$ and $\textcolor{red}{P_I \coloneqq \conv(P \cap \Z^2)}$.
    In \eqref{fig_implied_integrality_example}, $x_2$ is implied integer by $x_1$.
    The corresponding fibers are depicted in \eqref{fig_implied_integrality_fractional_fiber}.
    The second fiber has a fractional vertex, which is, however not a vertex of $\textcolor{green!60!black}{P_1}$ because it is a convex combination of points from other fibers.
    In \eqref{fig_implied_integrality_not_totally_implied}, $x_1$ is implied integer (by $x_\emptyset$) since $\textcolor{green!60!black}{P_1} = \textcolor{blue}{P}$, but not totally implied integer due to $\textcolor{red}{P_I} \neq \textcolor{orange}{P_2}$.
  }
  \label{fig_implied_integrality}
\end{figure}

The main use case of total implied integrality is the combination of several implied integrality relations.
Suppose for example, that for disjoint sets $S, T, U \subseteq N$ that $x_U$ is totally implied integer by $x_T$, and $x_T$ is totally implied integer by $x_S$.
It follows that $x_{T \cup U}$ is implied integer by $x_S$, more specifically that $\conv(P \cap \M^S) = \conv(P\cap \M^{S\cup T}) = \conv(P\cap \M^{S\cup T \cup U})$ holds.

In \cref{thm_totally_implied_integer_some_binary} we show that, to establish total implied integrality, it suffices to consider sets obtained from $S$ by adding elements not in $T$ whose variables are not binary.

\begin{theorem}
  \label{thm_totally_implied_integer_some_binary}
  Let $S,T,U \subseteq N$ and consider a rational polyhedron $P \subseteq \R^{N \setminus U} \times [0,1]^U$.
  If $x_T$ is implied integer by $x_S$ then $x_T$ is implied integer by every $x_{S'}$ with $S \subseteq S' \subseteq S \cup T \cup U$.
\end{theorem}

\begin{proof}
  Let $S$ and $S'$ be as in the theorem, let $U' \subseteq U$ and $T' \subseteq U$ be such that $S' = S \cup U' \cup T'$ holds, where the three sets are pairwise disjoint, e.g., $U' \coloneqq (S' \cap U) \setminus S$ and $T' \coloneqq S' \setminus (S \cup U')$.

  Consider $\bar{x} \in \Z^{S \cup U'}$ and a minimal face $F$ of $Q_{S \cup U'}(\bar{x})$ and let $x' \coloneqq \bar{x}_S \in \Z^S$.
  Let $\mathcal{A} \coloneqq \{ x \in \R^N \mid x_j = \bar{x}_j ~\forall j \in U' \}$.
  For any $X \subseteq [0,1]^U \times \R^{N \setminus U}$ we thus have $\conv(X) \cap \mathcal{A} = \conv( X \cap \mathcal{A} )$ since the bound constraints define a face.
  Hence, $Q_{S \cup U'}(\bar{x})$ is a face of $Q_S(x')$, so $F$ is also a minimal face of $Q_S(x')$, which implies $F \subseteq \conv(P \cap \M^{S \cup T})$ by the equivalence of~\ref{thm_implied_integrality_definition} and~\ref{thm_implied_integrality_fiber_minimal_faces} in \cref{thm_implied_integrality}.
  Intersecting with $\mathcal{A}$ yields
  \begin{multline*}
    F 
    = F \cap \mathcal{A}
    \subseteq \conv(P \cap \M^{S \cup T}) \cap \mathcal{A}
    = \conv( P \cap \M^{S \cup T} \cap \mathcal{A} ) 
    \subseteq \conv( P \cap \M^{S' \cup T}).
  \end{multline*}
  Again, equivalence of~\ref{thm_implied_integrality_fiber_minimal_faces} and~\ref{thm_implied_integrality_definition} in \cref{thm_implied_integrality} implies that $x_T$ is implied integer by $x_{S \cup U'}$ for $P$.

  We now consider the addition of $T'$ to $S \cup U'$.
  From the obvious inclusion $\M^{S \cup U'} \subseteq \M^{S \cup U' \cup T'} \subseteq \M^{S \cup U' \cup T}$ we obtain
  \[
    \conv( P \cap \M^{S \cup U'}) \subseteq \conv( P \cap \M^{S \cup U' \cup T'} ) \subseteq \conv( P \cap \M^{S \cup U' \cup T} ).
  \]
  The last and the first set are equal since $x_T$ is implied integer by $x_{S \cup U'}$.
  Hence, $\conv(P \cap \M^{S'}) = \conv(P \cap \M^{S \cup U' \cup T'}) = \conv(P \cap \M^{S' \cup T})$ holds, which concludes the proof.
\end{proof}

The following corollaries highlight the consequences of \cref{thm_totally_implied_integer_some_binary}.

\begin{corollary}
  \label{thm_totally_implied_integer_relevant}
  Let $S,T,U \subseteq N$ and consider a rational polyhedron $P \subseteq \R^{N \setminus U} \times [0,1]^U$.
  Then $x_T$ is totally implied integer by $x_S$ for $P$ if and only if $x_T$ is implied integer by every $x_{S'}$ (for $P$) that satisfies $S \subseteq S' \subseteq S \cup (N \setminus (T \cup U))$.
\end{corollary}

\begin{corollary}
  \label{thm_totally_implied_all_binary}
  Let $S,T \subseteq N$ and consider a rational polyhedron $P \subseteq \R^{S \cup T} \times [0,1]^{N \setminus (S \cup T)}$.
  Then $x_T$ is totally implied integer by $x_S$ for $P$ if and only if $x_T$ is implied integer by $x_S$ for $P$.
\end{corollary}

Total implied integrality keeps $T$ fixed and varies the set $S$.
In our next result, we consider the opposite case where $S$ is fixed but where $T$ may vary.
We show that for pointed polyhedra the set $T$ can be decomposed into implied integer relations for its elements.

\begin{theorem}
    \label{thm_pointed_implied_decomposition}
    Let $S,T\subseteq N$ and let $P$ be a rational and pointed polyhedron. Then $x_T$ is implied integer by $x_S$ for $P$ if and only if for all $t\in T$, $x_t$ is implied integer by $x_S$ for $P$.
\end{theorem}
\begin{proof}

  First assume that $x_T$ is implied integer by $x_S$ for $P$ and consider a $t \in T$.
  From $\M^{S \cup T} \subseteq \M^{S \cup \{t\}} \subseteq \M^S$ we have
  \begin{equation*}
    \conv(P \cap \M^{S \cup T} )
    \subseteq \conv(P \cap \M^{S \cup \{t\}} )
    \subseteq \conv(P \cap \M^S ).
  \end{equation*}
  Due to our assumption, the first and the last convex hulls are equal.
  Hence, we have equality throughout, which implies that $x_t$ is implied integer by $x_S$ for $P$.

  For the reverse direction, assume that for each $t\in T$, $x_t$ is implied integer by $x_S$. Since $\conv(P\cap \M^S)$ has the same recession cone as $P$ by \cref{thm_meyer}, both polyhedra also have identical lineality spaces, and it follows that $\conv(P\cap \M^S)$ is pointed because $P$ is pointed. Thus, each minimal face $F$ of $\conv(P\cap \M^S)$ consists of a unique vertex $x^F$. By \cref{thm_implied_integrality}\ref{thm_implied_integrality_minimal_faces}, $x^F_t$ must then be integral. Since $x^F$ is the unique point in the face and implied integrality holds for all $t\in T$, it then follows that $x^F_T$ is integral. Thus, it follows that $x^F\in \M^{S\cup T}$, since $x^F$ must also be $S$-integral by its definition. Since this holds for every minimal face $F$ of $\conv(P\cap \M^S)$, it follows from \cref{thm_implied_integrality}\ref{thm_implied_integrality_definition} that $x_T$ is implied integer by $x_S$ for $P$.
    
\end{proof}

\Cref{thm_pointed_implied_decomposition} has two interesting implications for pointed polyhedra. First of all, it shows that for pointed polyhedra, implied integrality decomposes into implied integrality for single-variable target sets. This shows that for the purpose of fully characterizing implied integrality, it suffices to characterize for each set $S$ those $t\in N$ for which $x_t$ is implied integer by $x_S$.
Secondly, we show in \cref{thm_pointed_unique_maximizer} that for each set $S$, there exists a maximal set $T$ such that $x_T$ is implied integer by $x_S$.

\begin{corollary}
  \label{thm_pointed_unique_maximizer}
  Let $P \subseteq \R^N$ be a rational and pointed polyhedron and let $S \subseteq N$.
  Then there exists a unique inclusion-wise maximal set $T \subseteq N$ such that $x_T$ is implied integer by $x_S$ for $P$.
\end{corollary}

\begin{proof}
  Suppose that there exists some $T'\subseteq N$ for which $x_{T'}$ is implied integer by $x_S$ for $P$ and such that $T'$ is not a subset of $T$. 
  By applying \cref{thm_pointed_implied_decomposition}, it follows that for all $t \in T \cup T'$, $x_{t}$ is implied integer by $x_S$.
  However, this implies that $x_{T \cup T'}$ is implied integer by $x_S$, using \cref{thm_pointed_implied_decomposition} again.
  Since $T \subset T \cup T'$ holds, this contradicts maximality of $T$.
\end{proof}

Given \cref{thm_pointed_implied_decomposition}, one may wonder if a similar property holds for non-pointed polyhedra. In \Cref{example_nonpointed_no_decomposition}, we show an example of a non-pointed polyhedron where implied integrality does not decompose over its elements.

\begin{example}
  \label{example_nonpointed_no_decomposition}
  Consider the rational polyhedron $P=\{(x,y)\in \R^2 \mid x+y = \frac{1}{2}\}$, that is also an affine space.
  Then $P = \conv(P\cap \M^x) = \conv(P\cap \M^y)$ and $\conv(P\cap \Z^2) = \emptyset$ hold, and thus $P \neq \conv(P\cap \Z^2)$.
  In particular, this shows that $x$ and $y$ are both implied integer by the the empty variable set, but $\{x,y\}$ is not due to $P \neq \conv(P \cap \Z^2)$.
\end{example}

\section{Implied integrality for block-structured problems}
\label{sec_block_structure}
In \cref{thm_implied_integer_sufficiency}, we observed that, in order to derive that $x_T$ is implied integer by $x_S$, it suffices to show that the fiber $Q_S(\bar{x})$ is $T$-integral for every $\bar{x}\in \Z^S$.
To show $T$-integrality of $Q_S(\bar{x})$, we observed in \cref{thm_block_integrality} that, if $Q_S(\bar{x})$ is the Cartesian product of multiple polyhedra, then it suffices to show that the ones involving any $x_T$-variable are $T$-integral polyhedra.
This allows us to apply known results on integral polyhedra to our setting.

\subsection{Exploiting totally unimodular submatrices}

Recall that a matrix $A$ is \emph{totally unimodular} if and only if the determinant of every square submatrix is $-1$, $0$ or $+1$.
A matrix $A \in \Z^{M \times N}$ of rank $r$ is said to be \emph{unimodular} if for every submatrix $B$ consisting of $r$ linearly independent columns of $A$, the greatest common divisor of the subdeterminants of $B$ of order $r$ equals $1$.
Hoffman and Kruskal showed that (totally) unimodular matrices are closely related to integral polyhedra~\cite{HoffmanK56}:

\begin{proposition}[Hoffman \& Kruskal \cite{HoffmanK56}]
  \label{thm_totally_unimodular_integral}
  For $A \in \Z^{m \times n}$ the following hold:
  \begin{enumerate}[label=(\roman*)]
  \item
    \label{thm_totally_unimodular_integral_eqns}
    The polyhedron $\{ x \in \R^n \mid Ax = b,~ \ell \leq x \leq u \}$ is integral for all $b \in \Z^m$ and all $\ell,u$ with $\ell_j,u_j \in \Z \cup \{\pm \infty\}$ for $j = 1,2,\dotsc,n$ if and only if $A$ is unimodular.
  \item
    \label{thm_totally_unimodular_integral_ineqs}
    The polyhedron $\{ x \in \R^n \mid Ax \leq b,~ \ell \leq x \leq u \}$ is integral for all $b \in \Z^m$ and all $\ell,u$ with $\ell_j,u_j \in \Z \cup \{\pm \infty\}$ for $j = 1,2,\dotsc,n$ if and only if $A$ is totally unimodular.
  \end{enumerate}
\end{proposition}

Assuming $P = \{ x \in \R^N \mid Ax \leq b \}$, an inequality description of each fiber $Q_S(\bar{x})$ is given by $Q_S(\bar{x}) = \{ A_{\star,N \setminus S} x_{N \setminus S} \leq b - A_{\star,S} \bar{x} \}$.
Note that $A_{\star, N \setminus S}$ is identical for each fiber, and that the fibers' descriptions thus only differ in the right-hand sides.
Unimodular and totally unimodular matrices fit this structure well since they imply that the underlying polyhedron is integral for \emph{every} integral right hand side. Thus, we can show integrality of all fibers by showing that $A_{\star,N \setminus S}$ is totally unimodular and that $b - A_{\star,S} \bar{x}$ is integral for all $\bar{x} \in \Z^S$. Our main result for showing how total unimodularity yields (total) implied integrality reads as follows.

\begin{theorem}
  \label{thm_tu_implied}
  Consider a rational polyhedron of the form
  \begin{equation}
    P = \{ (x,y,z) \in \R^S \times \R^T \times \R^U \mid
      Ax + By \leq c,~ \ell \leq y \leq u,~ Dx + Ez \leq f
    \},
    \label{eq_tu_implied}
  \end{equation}
  where $A \in \Z^{M \times S}$, $c \in \Z^M$ and $\ell_j,u_j \in \Z \cup \{\pm \infty\}$ for all $j \in T$.
  If $B$ is totally unimodular then $y$ is totally implied integer by $x$ for $P$.
\end{theorem}

\begin{proof}
  By \cref{thm_totally_implied_integer_relevant}
  we need to show that $y$ is implied integer by $(x,z_{U'})$ for arbitrary subsets $U' \subseteq U$.
  We will establish this using \cref{thm_implied_integer_sufficiency}.
  To this end, consider an arbitrary vector $(\bar{x},\bar{z}_{U'}) \in \Z^S \cup \Z^{U'}$.
  Its $(S \cup U')$-fiber reads
  \begin{multline*}
    Q_{S\cup U'}(\bar{x},\bar{z}_{U'}) = \{ (x,y,z) \in \R^{S \cup T \cup U} \mid x = \bar{x},~ z_{U'} = \bar{z}_{U'}, 
    By \leq c - Ax,~ \ell \leq y \leq u,~ Ez \leq f - Dx \}.
\end{multline*}
Note that $Q_{S\cup U'}(\bar{x},\bar{z}_{U'})$ is the Cartesian product of the three polyhedra $\{ \bar{x} \} \subseteq \R^S$, $Q^y(\bar{x}) \coloneqq \{ y \in \R^T \mid By \leq c - A\bar{x},~ \ell \leq y \leq u \} \subseteq \R^T$ and $\{ z \in \R^U \mid z_{U'} = \bar{z}_{U'},~ Ez \leq f - D\bar{x} \} \subseteq \R^U$.
Since $c - A \bar{x}$, $\ell$ and $u$ are integral (or infinite) and $B$ is totally unimodular, \cref{thm_totally_unimodular_integral}~\ref{thm_totally_unimodular_integral_ineqs} shows that $Q^y(\bar{x})$ is an integral polyhedron.
By \cref{thm_block_integrality} we conclude that $Q_{S\cup U'}(\bar{x},\bar{z}_{U'})$ is $T$-integral.
Since this holds for every $(\bar{x}, \bar{z}_{U'})$, \cref{thm_implied_integer_sufficiency} shows that $y$ is totally implied integer by $x$ for $P$, which concludes the proof.
\end{proof}

As a special case we obtain the dual detection method  from~\cite{AchterbergBGRW20} by observing that, after scaling, it treats a single-column totally unimodular submatrix.

\begin{corollary}[dual detection]
  \label{thm_dual_detection}
  Consider a rational polyhedron $P = \{x\in \R^N \mid A x \leq b\}$ with $A \in \Q^{M \times N}$ and $b \in \Q^M$.
  For some $k \in N$, let $M_k \coloneqq \{ j \in M \mid A_{j,k} \neq 0 \}$ be the rows that contain a nonzero of column $k$.
  Let $S \coloneqq \{ k' \in N \setminus \{k\} \mid \text{ there exists } j \in M_k \text{ with } A_{j,k'} \neq 0 \}$ be the other columns having nonzeros in $M_k$.
  If $\frac{A_{j,s}}{A_{j,k}} \in \Z$ and $\frac{b_j}{A_{j,k}} \in \Z$ holds for all $j \in M_k$ and all $s \in S$ then $x_k$ is totally implied integer by $x_S$ for $P$.
\end{corollary}

\begin{proof}
  First, we obtain $A'$ and $b'$ from $A$ and $b$ by scaling each row in $j \in M_k$ by $\frac{1}{|A_{j,k}|}$.
  This does not change the defined polyhedron, that is, $P = \{ x \in \R^N \mid Ax \leq b \}$.
  Moreover, $A'_{M_k,k}$ only contains entries $\pm 1$, and hence is totally unimodular.
  Note that $A'_{M_k,S}$ is integral and $b'_{M_k}$ is integral, and $A'_{M \setminus M_k, k} = 0$ by definition.
  Thus, the conditions of \cref{thm_tu_implied} are satisfied, and $x_k$ is totally implied integer by $x_S$ for $P$. 
\end{proof}

\Cref{thm_tu_implied} is strictly more general than \cref{thm_dual_detection} because we can apply \cref{thm_tu_implied} to rows of the MILP that involve multiple continuous variables.
In contrast to this, \cref{thm_dual_detection} requires a column that may only appear in rows where all other variables are integral.

It might seem unlikely that the conditions \cref{thm_tu_implied} hold, as it imposes strong integrality requirements and requires a totally unimodular submatrix.
However, many MILP formulations of common problems actually satisfy these requirements.
For example, in the canonical formulation of the \emph{fixed-charge network flow problem}~\cite{Hirsch68} with integer capacities, the flow variables are implied integer by the decision variables that activate an arc.
Many models for production scheduling and network design problems satisfy these conditions, such as \emph{lot sizing problems}~\cite{Karimi2003} or the canonical formulation of the \emph{uncapacitated facility location problem}.
Other examples of models that satisfy \cref{thm_tu_implied} are the \emph{natural-dates formulation} of the \emph{single machine scheduling problem}~\cite{Balas1985} and several formulations of the \emph{periodic event scheduling problem}~\cite{Liebchen2006}.

Note that in the proof of \cref{thm_tu_implied}, we only use total unimodularity to show that $Q_y(\bar{x})$ is an integral polyhedron.
Using almost exactly the same proof but relying on \cref{thm_totally_unimodular_integral}~\ref{thm_totally_unimodular_integral_eqns}, we can derive a similar result for (the more general) unimodular matrices involved in equations.

\begin{theorem}
  \label{thm_unimodular_implied}
  Let $P$ be a rational polyhedron
  \begin{equation*}
    P = \{ (x,y,z) \in \R^S \times \R^T \times \R^U \mid
      Ax + By = c, ~ \ell \leq y \leq u,~ Dx + Ez \leq f
    \},
  \end{equation*}
  where $A \in \Z^{M \times S}$, $c \in \Z^M$ and $\ell_j,u_j \in \Z \cup \{\pm \infty\}$ for all $j \in T$.
  If $B$ is unimodular then $y$ is totally implied integer by $x$ for $P$.
\end{theorem}


\subsection{\texorpdfstring{Exploiting $k$-regularity and balancedness}{Exploiting $k$-regularity and balancedness}}

In \cref{thm_tu_implied} and \cref{thm_unimodular_implied}, we highlighted how totally unimodular submatrices together with block-structure can cause implied integrality. In this section, we highlight how similar ideas can be used to detect implied integers using $k$-regular and balanced submatrices. 

Appa and Kotnyek studies $k$-regular matrices~\cite{Appa2004}, which are a natural generalization of total unimodularity.
A matrix is said to be \emph{$k$-regular} if for each non-singular square submatrix $R$, $k R^{-1}$ is integral.
They show the following:

\begin{proposition}[Appa \& Kotnyek~\cite{Appa2004}]
  \label{thm_regular_integral}
  For $A \in \R^{M \times N}$ and $k\in \Z_{\geq 1}$, $P=\{x\geq \zerovec, A x\leq k b\}$ is integral for all $b\in \Z^M$ if and only if $A$ is $k$-regular.
\end{proposition}

This allows us to show a more general result than \cref{thm_tu_implied}:
  
\begin{theorem}
  \label{thm_regular_implied}
  Consider a rational polyhedron of the form
  \begin{equation*}
    P = \{ (x,y,z) \in \R^S \times \R^T \times \R^U \mid
      Ax + By \leq c, ~ y\geq \zerovec, ~ Dx + Ez \leq f
    \},
  \end{equation*}
  If for $k\in \Z_{\geq 1}$, matrix $B$ is $k$-regular and $c - A \bar{x} \equiv \zerovec \mod k$ holds for all $\bar{x} \in \proj_S(P) \cap \Z^S$ then $y$ is totally implied integer by $x$ for $P$.
\end{theorem}
\begin{proof}
      By \cref{thm_totally_implied_integer_relevant}
  we need to show that $y$ is implied integer by $(x,z_{U'})$ for arbitrary subsets $U' \subseteq U$.
  We will establish this using \cref{thm_implied_integer_sufficiency}.
  To this end, consider an arbitrary vector $(\bar{x},\bar{z}_{U'}) \in \Z^S \cup \Z^{U'}$.
  Its $(S \cup U')$-fiber reads
  \begin{multline*}
    Q_{S\cup U'}(\bar{x},\bar{z}_{U'}) = \{ (x,y,z) \in \R^{S \cup T \cup U} \mid x = \bar{x},~ z_{U'} = \bar{z}_{U'}, By \leq c - Ax,~ y \geq \zerovec,~ Ez \leq f - Dx \}.
\end{multline*}
Note that $Q_{S\cup U'}(\bar{x},\bar{z}_{U'})$ is the Cartesian product of the three polyhedra $\{ \bar{x} \} \subseteq \R^S$, $Q^y(\bar{x}) \coloneqq \{ y \in \R^T \mid By \leq c - A\bar{x},~ y \geq \zerovec \} \subseteq \R^T$ and $\{ z \in \R^U \mid z_{U'} = \bar{z}_{U'},~ Ez \leq f - D\bar{x} \} \subseteq \R^U$.
Since $c - A \bar{x} = \zerovec \mod k$ holds for all $\bar{x} \in \proj_S(P)\cap \Z^S$ and $B$ is $k$-regular, $Q^y(\bar{x})$ is integral by \cref{thm_regular_integral}. For all other $\bar{x}$, we must have $Q_{S}(\bar{x})=\emptyset$ by definition, which implies that $Q_{S\cup U'}(\bar{x},\bar{z}_{U'}) = \emptyset$ holds, which is also integral.
By \cref{thm_block_integrality} we conclude that $Q_{S\cup U'}(\bar{x})$ is $T$-integral.
Since this holds for every $(\bar{x}, \bar{z}_{U'})$, \cref{thm_implied_integer_sufficiency} shows that $y$ is implied integer by $x$ for $P$, which concludes the proof.
\end{proof}

Note that $k$-regular matrices include many important classes of matrices.
\emph{Node-edge incidence matrices} of undirected graphs are 2-regular and occur in MILP formulations of important combinatorial optimization problems such as the \emph{matching problem}.
The \emph{vertex cover} and the \emph{independent set} problems on graphs contain \emph{edge-node incidence matrices} and also admit MILP formulations with 2-regular constraint matrices.
\emph{Binet matrices} generalize node-edge incidence matrices and are also 2-regular~\cite{Appa2004}.

\DeclareDocumentCommand\relationBalanced{}{\ensuremath{{\lessgtr}}}

Another relevant class of perfect formulations are those generated by balanced matrices. A 
A $\{-1,0,1\}$-matrix $A$ is \emph{balanced} if $A$ if for every submatrix $H$ with two nonzeros in every row and column, the sum of the entries of $H$ is divisible by 4.
Balanced matrices yield important classes of perfect formulations for set packing, set partitioning and set covering polytopes. Let $n(A)$ and $p(A)$ denote the column vector whose $i$'th component indicates the number of $-1$ (respectively $+1$ ) entries in the $i$'th row of $A$.
For a $\{-1,0,1\}$-matrix $A$ and a relation $\relationBalanced \in \{\leq, =, \geq\}$, we use $R^\relationBalanced(A) \coloneqq \{ x \in [0,1]^N,~ A x ~\relationBalanced~ \onevec - n(A)\} $ to denote the set packing, set partitioning and set covering polytopes, respectively.
\begin{proposition}[Conforti \& Cornuejols \cite{ConfortiC95}]
  \label{thm_balanced_integral}
  For $A\in \{-1,0,1\}^{M\times N}$,  the following are equivalent:
  \begin{enumerate}[label=(\roman*)]
  \item
    $A$ is balanced.
  \item
    For each submatrix $B$ of $A$, the set packing polytope $R^\leq(B)$ is integral.
  \item
    For each submatrix $B$ of $A$, the set partitioning polytope $R^=(B)$ is integral.
  \item
    For each submatrix $B$ of $A$, the set covering polytope $R^\geq(B)$ is integral.
  \end{enumerate}
\end{proposition}

\begin{theorem}
  Consider, for $\relationBalanced \in \{\leq, =, \geq \}$ and matrices $A \in \{-1,0,1\}^{M \times S}$ and $B \in \{-1,0,1\}^{M \times T}$, a rational polyhedron of the form
  \begin{equation*}
    P^\relationBalanced = \{ (x,y,z) \in [0,1]^S \times [0,1]^T \times \R^U \mid Ax + By ~\relationBalanced~ \onevec - n(A) - n(B),~ Dx + Ez \leq f \}.
  \end{equation*}
  If $B$ is balanced, then $y$ is totally implied integer by $x$ for $P^\relationBalanced$.
\end{theorem}

\begin{proof}
  By \cref{thm_totally_implied_integer_relevant}
  we need to show that $y$ is implied integer by $(x,z_{U'})$ for arbitrary subsets $U' \subseteq U$.
  We will establish this using \cref{thm_implied_integer_sufficiency}.
  To this end, consider an arbitrary vector $(\bar{x},\bar{z}_{U'}) \in \Z^S \cup \Z^{U'}$.
  Its $(S \cup U')$-fiber reads
  \begin{multline*}
    Q^\relationBalanced_{S \cup U'}(\bar{x}, \bar{z}_{U'}) = \{ (x,y,z) \in [0,1]^S \times [0,1]^T \times \R^U \mid \\ x = \bar{x},~ z_{U'} = \bar{z}_{U'},~ By ~\relationBalanced~ \onevec - Ax - n(A) -n(B),~ Ez \leq f - Dx \}.
  \end{multline*}
  Note that $Q_{S \cup U'}(\bar{x}, \bar{z}_{U'})$ is the Cartesian product of the three polyhedra $\{x \in [0,1]^S \mid \bar{x} = x \} \subseteq \R^S$,  $Q^{\relationBalanced, y}(\bar{x}) \coloneqq \{ y \in [0,1]^T \mid By ~\relationBalanced~ \onevec - A \bar{x} - n(A) -n(B) \} \subseteq \R^T$ and $\{ z \in \R^U \mid z_{U'} = \bar{z}_{U'},~ Ez \leq f - D\bar{x} \} \subseteq \R^U$.
  Since for any $\bar{x} \in \Z^S \setminus \{0,1\}^S$ the bound constraints on $x$ imply that $Q^{\relationBalanced}_{S\cup U'}(\bar{x},\bar{z}_{U'})$ is empty, it suffices to show for $\bar{x}\in\{0,1\}^S$ that $Q^\relationBalanced_{S \cup U'}(\bar{x}, \bar{z}_{U'})$ is $T$-integral.

  We first show $T$-integrality of the fibers for $P^\geq$, that is, for the case in which $\relationBalanced$ is equal to $\geq$.
  Note that for $Q^{\geq,y}(\bar{x}) = \{ y \in [0,1]^T \mid By \geq \onevec - A\bar{x} - n(A) - n(B) \}$ the rows $M_1$ of $By \geq \onevec - A\bar{x} - n(A) - n(B)$ that satisfy $A_{M_1,*}\bar{x} \geq \onevec_{M_1} -  n(A)_{M_1}$ are redundant since they are implied by the bound constraints of $y$.
  By considering the signs of $A$ and the fact that $\bar{x}$ is binary, the only other possibility for the other rows $M_2 = M\setminus M_1$ is that $A_{M_2,\star} \bar{x} = n(A_{M_2,\star})$ holds.
  Thus, we have that $Q^{\geq,y}(\bar{x}) = \{ y \in [0,1]^T \mid B_{M_2,\star} y \geq \onevec_{M_2} - n(B_{M_2,\star}) \} $.
  Since $B_{M_2,*}$ is a submatrix of $B$, \cref{thm_balanced_integral} shows that $Q^{\geq,y}(\bar{x})$ is integral.
  By \cref{thm_block_integrality} we conclude that $Q_{S \cup U'}(\bar{x},\bar{z}_{U'})$ is $T$-integral.
  Since this holds for every $(\bar{x}, \bar{z}_{U'})$, \cref{thm_implied_integer_sufficiency} shows that $y$ is totally implied integer by $x$ for $P^\geq$.

  We now show $T$-integrality of the fibers for $\relationBalanced ~\in \{ \leq, = \}$.
  We consider $Q^{\relationBalanced,y}(\bar{x}) = \{ y \in [0,1]^T \mid By ~\relationBalanced~ \onevec - A\bar{x} - n(A) - n(B) \}$.
  Note that $Q^{\relationBalanced,y} \neq \emptyset$ implies that $\onevec - A \bar{x} - n(A) \in \{0,1\}^M$ must be binary. Let $M_1 = \{m\in M \mid (\onevec - A\bar{x} - n(A))_m = 0\}$ indicate those rows for which this expression is zero, and let $M_2 = M\setminus M_1$. We use $T_1$ to denote all columns that have nonzeros in some row of $M_1$, and $T_2 = T\setminus T_1$ otherwise. Note that in all rows $M_1$, the remaining constraints take the form $B_{M_1, T_1} y_{T_1}  \relationBalanced n(B_{M_1})$, which only has a single feasible solution $y_{T_1} = \bar{y}_{T_1}$ based on the sign pattern of $B$. Then, for the remaining problem we obtain that
  $Q^{\relationBalanced,y}(\bar{x}) = \{ (\bar{y}_{T_1}, y_{T_2}) \in [0,1]^{T_1} \times [0,1]^{T_2} \mid B_{M_2, T_2} y_{T_2} ~\relationBalanced~ \onevec - n(B_{M_2,T_2}) \}$.

  Hence, we again have a set packing or partitioning problem at hand, and \cref{thm_balanced_integral} shows that $Q^{\relationBalanced, y}(\bar{x})$ is integral.
  By \cref{thm_block_integrality} we conclude that $Q^{\relationBalanced}_{S \cup U'}(\bar{x},\bar{z}_{U'})$ is $T$-integral.
  Since this holds for every $(\bar{x}, \bar{z}_{U'})$, \cref{thm_implied_integer_sufficiency} shows that $y$ is totally implied integer by $x$ for $P^\leq$ and $P^=$.
\end{proof}

One important aspect of the results in this section is that they provide a practical motivation to formulate and implement efficient recognition methods for classes of matrices that generate perfect formulations.
As integral polyhedra occur rarely in practice, recognition of them is from a computational point of view less relevant because if one wants to optimize over the integer hull, one typically optimizes over the linear relaxation as a first step.
In contrast to this, implied integrality can be exploited algorithmically to speed up the solution algorithms. In particular, we have shown that if matrices that generate classes of integral polyhedra occur as submatrices, that they may induce implied integrality in the problem.
There are known polynomial time recognition algorithms for totally unimodular matrices~\cite{Truemper1990}, binet matrices~\cite{Musitelli2007, Musitelli2010} and balanced matrices~\cite{ConfortiCR99,Zambelli05}. 

\section{Implied fixed integrality}
\label{sec_implied_fixed}

The primal detection method from~\cite{AchterbergBGRW20}, as illustrated in \cref{primal_example}, works on any equation of a MILP problem with suitable coefficients. Importantly, it does not require a diagonal block structure as in~\eqref{eq_tu_implied}, which is the key idea that we used in the previous section to detect implied integrality.
Moreover, observe that in \cref{primal_example}, $z$ has a \emph{unique} solution that is integral after $x$ and $y$ are fixed.
This motivates us to define \emph{implied fixed integers}.
\begin{definition}[implied fixed integrality]
  \label{def_implied_fixed}
  Let $P \subseteq \R^N$ be a rational polyhedron and let $S,T \subseteq N$.
  We say that \emph{$x_T$ is implied fixed integer by $x_S$ for $P$} if for each $\bar{x} \in \Z^S$, the set $\proj_{x_T}(Q_S(\bar{x}))$ is either empty or consists of a unique vector that is integer.
\end{definition}

First, we show that implied fixed integrality is a form of implied integrality. In fact, we even observe in \cref{thm_implied_fixed_integer_closed,thm_implied_fixed_totally} that it is a form of total implied integrality.

\begin{lemma}
  \label{thm_implied_fixed_integer_closed}
  Let $P \subseteq \R^N$ be a rational polyhedron and let $S,T\subseteq N$.
  If $x_T$ is implied fixed integer by $x_S$ for $P$ then, for any $S'\subseteq N$ with $S'\supseteq S$, $x_T$ is implied fixed integer by $x_{S'}$.
\end{lemma}
\begin{proof}
    Consider any $S'\subseteq N$ with $S'\supseteq S$. For any $\bar{x}\in \proj_{S'}(P)\cap Z^{S'}$, we consider $Q_{S'}(\bar{x})$; note that $Q_{S'}(\bar{x})$ is non-empty by definition. Because $S\subseteq S'$ holds, $Q_{S'}(\bar{x}) \subseteq Q_S(\bar{x}_S)$ holds. Additionally, note that $\bar{x}_S$ must lie in $\proj_S(P)$, as otherwise this contradicts that $\bar{x}$ lies in $\proj_{S'}(P)$. In particular, this implies that any point $x\in Q_{S'}(\bar{x})$ must have $x_T = \hat{x}$ where $\hat{x}$ is integer, because $\proj_T(Q_S(\bar{x}_S))$ consists of a unique vector that is integer. Since this holds for any $x\in Q_{S'}(\bar{x})$, we must have $\proj_T(Q_{S'}(\bar{x})) = \{\hat{x}\}$, which is a unique vector that is integral. Since this holds for any $\bar{x}\in \proj_{S'}(P)\cap \Z^{S'}$, $T$ is implied fixed integer by $S'$.
\end{proof}

\begin{corollary}
    \label{thm_implied_fixed_totally}
  Let $P \subseteq \R^N$ be a rational polyhedron and let $S, T \subseteq N$.
  If $x_T$ is implied fixed integer by $x_S$ for $P$ then $x_T$ is totally implied integer by $x_S$ for $P$.
\end{corollary}

\begin{proof}
  By \cref{thm_implied_fixed_integer_closed} it suffices to show implied integrality instead of total implied integrality.
  To this end, assume that $x_T$ is implied fixed integer by $x_S$ for $P$.
  Consider any $\bar{x} \in \Z^S$.
  If $Q_S(\bar{x}) = \emptyset$ then it is trivially $T$-integral.
  Otherwise, since $T$ is implied fixed integer by $S$, the projection $\proj_T(Q_S(\bar{x}))$ consists of a unique vector that is integer, and hence any $x \in Q_S(\bar{x})$ must have $x_T \in \Z^T$.
  Thus, any face of $Q_S(\bar{x})$ contains a $T$-integral point, and by \cref{thm_implied_integrality}, $Q_S(\bar{x})$ is $T$-integral.
  The result follows from \cref{thm_implied_integer_sufficiency}.
\end{proof}

\subsection{Exploiting unimodularity}
As uniqueness often arises as a result of solving a full-rank system of linear equations, \cref{thm_unimodular_implied} stands out.
In particular, if, for $B y = c - A\bar{x}$, $B$ is unimodular and has full column rank then the solution $y$ is not just integral, but also unique for the given $\bar{x}$. 

\begin{theorem}
  \label{thm_unimodular_implied_fixed}
  Consider a rational polyhedron of the form
  \begin{equation*}
    P = \{ (x,y,z) \in \R^S \times \R^T \times \R^U \mid
      Ax + By = c,~ Dx + Ey + F z \leq g
    \},
  \end{equation*}
  where $A \in \Z^{M \times S}$ and $c \in \Z^M$.
  If $B$ is unimodular and has full column rank then $y$ is implied fixed integer by $x$ for $P$.
\end{theorem}

\begin{proof}
  Let $\bar{x} \in \Z^S$.
  Since $B$ has full column rank, there is at most one feasible solution $\bar{y} \in \R^T$ such that $B \bar{y} = c - A \bar{x}$.
  If $Q_S(\bar{x})$ is non-empty, then there also exists such a solution $\bar{y}$, together with some vector $\bar{z} \in \R^U$ such that $(\bar{x},\bar{y},\bar{z}) \in P$.
  Unimodularity of $B$ implies integrality of $\bar{y}$ by \cref{thm_totally_unimodular_integral} \ref{thm_totally_unimodular_integral_eqns}.
  We conclude that $y$ is implied fixed integer by $x$ for $P$.
\end{proof}

We observe that \cref{thm_unimodular_implied_fixed} generalizes the primal detection method from~\cite{AchterbergBGRW20} by realizing that the 1-by-1 matrix with entry $\pm 1$ is unimodular and has full column rank.

\begin{corollary}[primal detection]
  \label{thm_primal_detection}
  Let $S \subseteq N$ and $k \in N \setminus S$.
  Consider the polyhedron $P = \{x \in \R^N \mid a_S x_S + a_k x_k = b,~ Gx \leq d \}$ with $a_S \in \Q^S$, $a_k \in \Q \setminus \{0\}, b \in \Q$, $G \in \Q^{M \times N}$ and $d \in \Q^M$.
  If $\frac{ a_{S} }{ a_k } \in \Z^S$ and $\frac{ b }{ a_k } \in \Z$, then $x_k$ is implied fixed integer by $x_S$ for $P$.
\end{corollary}

\begin{proof}
  After dividing the equation $a_S x_S + a_k x_k = b$ by $a_k$, it is of the form $a'_S x_S + x_k = b'$ for integral $a'_S \in \Z^S$ and $b' \in \Z$.
  This does not affect the feasible region $P$.
  Since $a'_S$ and $b'$ are integral and the $1$-by-$1$ matrix with entry $1$ is unimodular with full column rank, \cref{thm_unimodular_implied_fixed} yields the desired result.
\end{proof}
One advantage of \cref{thm_primal_detection} is that it is rather easy to detect and use in practice.
However, easy examples where it may fail to detect implied fixed integrality exist.
For example, if $3x + 2y = 3$ and $5x + 3 y = 4$ are part of a MILP model, the associated 2-by-2 matrix is unimodular and has full column rank, so $x$ and $y$ are implied fixed integers.
However, \cref{thm_primal_detection} fails to detect implied fixed integrality of both $x$ and $y$ in this case.

Although this example may make \cref{thm_primal_detection}
might seem like quite a weak result, it is in a sense as powerful as \cref{thm_unimodular_implied_fixed} if we additionally consider row transformations of the system of equations. Consider the setting of \cref{thm_unimodular_implied_fixed}, where $Ax + B y = c$ is included in the constraint matrix of $P$ with $A,c$ integral and $B$ is a square unimodular matrix. Since $B$ is a square unimodular matrix, its inverse $B^{-1}$ is also a square unimodular matrix. Multiplying the equation system from the left by $B^{-1}$ corresponds to a row transformation and does not change the feasible region of $P$. Thus, we can obtain the equivalent equation system $B^{-1} A x + I y = B^{-1} c$. Since $B^{-1}$ is unimodular, $B^{-1} A$ and $B^{-1} c$ are both integral. Now, note that \cref{thm_primal_detection} can be applied independently to every row of the equation system to argue that $y_i$ is implied fixed integer by (a subset of) the $x$ variables. Combining these relations, this shows that $y$ is implied fixed integer by $x$.

\subsection{Nonlinear relations}

In \cref{thm_unimodular_implied_fixed}, we used given equations and observed that implied fixings for $y$ have a direct linear relationship with the $x$-variables.
The following example shows that implied fixed integers may also arise from inequalities, and that the unique fixings for $y$ may depend nonlinearly on $x$.

\begin{example}
    \label{thm_fortet_implied}
  Consider $P \coloneqq \{ x \in \R^n \mid A x \leq b \}$ and suppose that Fortet's inequalities~\cite{Fortet1960} for linearizing $x_3 = x_1 x_2$ for binary variables $x_1,x_2,x_3$ are valid for $P$, i.e., $A x \leq b$ contains a subsystem of the form
  \begin{align*}
    x_3 &\leq x_1, &
    x_3 &\leq x_2, &
    x_3 &\geq x_1 + x_2 - 1, &
    0 &\leq x_i \leq 1 ~\forall i \in \{1,2,3\}.
  \end{align*}
  Then $x_3$ is implied fixed integer by $(x_1,x_2)$.
\end{example}

\begin{proof}
    In order to check the definition of implied integrality, we need to check all the values of $(x_1,x_2) \in \Z^2$. It suffices to check $(x_1,x_2)\in \{0,1\}^2$, as all non-binary vectors are clearly infeasible by the constraints, and thus must have an empty projection onto $x_3$. If $x_1=0$ or $x_2= 0$ holds, the given constraints fix $x_3 = 0$. Otherwise, if $x_1=x_2 = 1$, then the given constraints fix $x_3 = 1$. Thus, the projection onto $x_3$ always consists of a unique vector that is integer, and $x_3$ is implied fixed integer by $(x_1,x_2)$. 
\end{proof}

Using \cref{thm_fortet_implied}, one can show that linearizations of binary quadratic mixed-integer programming problems with $n$ original variables can have $\orderO(n^2)$ linearization variables for quadratic terms that are implied integer by only $n$ original variables.
This also highlights that implied integer relations may exist in practice where a large number of variables is implied integer by a much smaller number of variables.

\subsection{Linear expressions inducing fixed integrality}

In \cref{thm_unimodular_implied_fixed} we observed that implied fixed integrality follows from a square unimodular submatrix. Square and unimodular submatrices are also relevant for lattices, as they represent lattice bases of the lattice given by $\Z^N$.
To establish this connection more clearly, we define the implied fixed integrality of linear combinations $\transpose{v} x$ for arbitrary vectors $v \in \R^N$.

\begin{definition}
  \label{def_expression_implied}
  Let $P\subseteq \R^N$ be a rational polyhedron, let $S \subseteq N$, and let $v \in \R^N$.
      We say that \emph{$v$ induces implied fixed integrality by $x_S$ for $P$} if there exists a function $\alpha : \Z^S \to \R$ such that for any $\bar{x} \in \Z^S$, $\alpha(\bar{x})$ is integral and the equation $\transpose{v} x = \alpha(\bar{x})$ is valid for $Q_S(\bar{x})$.
  We denote the set of such vectors as
  \begin{equation*}
    \Psi_S(P) \coloneqq \{ v \in \R^N \mid v \text{ induces implied fixed integrality by $x_S$ for $P$}\}.
  \end{equation*}
\end{definition}

Note that every equation is valid for the empty polyhedron, so in the definition above we only need to consider $\bar{x} \in \Z^S$ for which $Q_S(\bar{x})$ is non-empty.
Implied fixed integrality of linear combinations is naturally related to implied fixed integrality as we defined it in \cref{def_implied_fixed}.
We use $\unitvec{i} \in \R^N$ to denote the $i$'th unit vector. 

\begin{proposition}
  \label{thm_strong_equivalence}
  Let $P\subseteq \R^N$ be a rational polyhedron and let $S,T \subseteq N$.
  Then $x_T$ is implied fixed integer by $x_S$ for $P$ if and only if $\unitvec{i} \in \R^N$ induces implied fixed integrality by $x_S$ for $P$ for all $i \in T$. 
\end{proposition}

\begin{proof}
  First, we assume that $x_T$ is implied fixed integer by $x_S$ for $P$, and show that $\unitvec{i}$ induces implied fixed itnegrality by $x_S$ for $P$ for all $i\in T$. 
  By definition, for each $\bar{x} \in \Z^S$, $\proj_{x_T}(Q_S(\bar{x}))$ is either empty (which implies $Q_S(\bar{x}) = \emptyset$ for which any equation is valid) or it consists of a single vector $\hat{x} \in \R^T$, which may depend on $\bar{x}$.
  Now, for each $i \in T$, $\unitvec{i}$ induces implied fixed integrality by $x_S$ for $P$ since the equation $\transpose{\unitvec{i}}x = \alpha(\bar{x})$ is valid if we choose $\alpha(\bar{x}) \coloneqq \hat{x}_i$. 

  For the reverse direction, assume that $\unitvec{i}$ induces fixed integrality for each $i \in T$ and let $\alpha_i : \Z^S \to \R$ be corresponding functions such that the equation $\transpose{\unitvec{i}}x = \alpha_i(\bar{x})$ is valid for $Q_S(\bar{x})$ and that $\alpha_i(\bar{x}) \in \Z$ for all $\bar{x} \in \Z^S$.
  Hence, for each $\bar{x} \in \Z^S$, every $x \in Q_S(\bar{x})$ satisfies $(x_i =)~ \transpose{\unitvec{i}}x = \alpha_i(\bar{x})$ for all $i \in T$, which means that the projection $\proj_{x_T}(Q_S(\bar{x}))$ consists of at most one point.
  If the latter point exists, then its integrality follows from integrality of $\alpha_i(\bar{x})$ for all $i \in T$.
\end{proof}

It turns out that the set $\Psi_S(P)$ always constitutes a \emph{closed vector group}, which we first define.

\begin{definition}[(closed) vector group, (local) rank~\cite{Siegel1989}]
  \label{def_closed_vector_group}
  A set $G \subseteq \R^N$ is called \emph{vector group} if it is non-empty and if, for any two vectors $v^1,v^2 \in G$, also $v^1 - v^2 \in G$ holds.
  It is called \emph{closed} if, for any converging sequence of vectors in $G$, the limit vector also lies in $G$.
  For $\varepsilon > 0$, let $\rank_\varepsilon(G) \in \Znonneg$ denote the maximum number of linearly independent vectors $v^1, v^2, \dotsc, v^{r(\varepsilon)} \in G$ that satisfy $||v^i|| \leq \varepsilon$ for $i = 1,2,\dotsc,r(\varepsilon)$.
  The \emph{rank of $G$} is defined as $\rank(G) \coloneqq \max_{\varepsilon > 0} r(\varepsilon)$ and the \emph{local rank of $G$} as $\localrank(G) \coloneqq \min_{\varepsilon > 0} r(\varepsilon)$.
\end{definition}

The local rank of a vector group is the dimension of its largest contained vector space.
The next proposition establishes that $\Psi_S(P)$ has such a vector group structure.

\begin{proposition}
  \label{thm_vector_group_closed}
  Let $P \subseteq \R^N$ be a rational polyhedron and let $S \subseteq N$.
  Then $\Psi_S(P)$ is a closed vector group.
  Moreover, for any converging sequence $v^i \in \Psi_S(P)$ with limit $v^\star \coloneqq \lim_{i \to \infty} v^i \in \Psi_S(P)$ and every $\bar{x} \in \Z^S$ with $Q_S(\bar{x}) \neq \emptyset$, the values $\alpha_i(\bar{x})$ of the fixing functions $\alpha_i$ for $v^i$ also converge such that $\transpose{(v^\star)} x = \lim_{i \to \infty} \alpha_i(\bar{x}) = \alpha_k(\bar{x})$ holds for sufficiently large $k$.
\end{proposition}

\begin{proof}
  First, observe that $\Psi_S(P) \neq \emptyset$ since clearly $\zerovec \in \Psi_S(P)$.

  Second, let $v^1, v^2 \in \Psi_S(P)$ such that $\transpose{(v^i)} x = \alpha_i(\bar{x}) \in \Z$ for $i=1,2$.
  Then $\transpose{(v^1 - v^2)} x = \transpose{(v^1)} x - \transpose{(v^2)} x = \alpha_1(\bar{x}) - \alpha_2(\bar{x})$, which is also integral.
  We conclude that $(v^1 - v^2) \in \Psi_S(P)$.

  Third, consider a converging sequence $v^i \in \Psi_S(P)$ with limit $v^\star \coloneqq \lim_{i \to \infty} v^i$.
  For any $\bar{x} \in \Z^S$ with $Q_S(P) = \emptyset$ we define $\alpha^\star(\bar{x}) \in \Z$ arbitrarily.
  For any $\bar{x} \in \Z^S$ with $Q_S(P) \neq \emptyset$, consider the function $f: \Psi_S(P) \to \Z$ that maps each $v \in \Psi_S(P)$ to the right-hand side $\alpha(\bar{x})$ of the equation $\transpose{v}x = \alpha(\bar{x})$ that is valid for $Q_S(\bar{x})$.
  Clearly, $f$ is continuous and hence we can define $\alpha^\star(\bar{x}) \coloneqq \lim_{i \to \infty} f(v^i)$.
  By construction, $\alpha^\star : \Z^S \to \R$ shows $v^\star \in \Psi_S(P)$.
  Moreover, for any $\tau > 0$, there exists some positive integer $k$ such that, for all $i \geq k$, $||f(v^{i+1}) - f(v^i)|| < \tau$ hold.
  For $\tau \leq 1$ this implies $f(v^{i+1}) = f(v^i)$ since both values are integer due to $\alpha_{i+1},\alpha_i \in \Z$.
  This shows the last statement and concludes the proof.
\end{proof}

The structure of such closed vector groups is well known.
The next proposition shows that it is the Minkowski sum of a lattice and of a vector spaces whose dimension equals its local rank.

\begin{proposition}[Siegel~\cite{Siegel1989}, Theorems 21 and 22]
  \label{thm_vector_group_decomposition}
  Let $G$ be a closed vector group.
  Then there exist linearly independent vectors $g_1, g_2, \dotsc, g_{\rank(G)} \in G$ such that
  \[
    G = \left\{ \sum_{i=1}^{\rank(G)} \lambda_i g_i \mid \lambda_i \in \R ~\forall i \in \{1,2,\dotsc,\localrank(G)\} \text{ and } \lambda_i \in \Z ~\forall i \in \{\localrank(G)+1,\localrank(G)+2,\dotsc,\rank(G)\} \right\}.
  \]
\end{proposition}

Our next theorem characterizes, for $\Psi_S(P)$, the associated vector space.
It essentially states that the (largest) vector space contained in $\Psi_S(P)$ consists of precisely those vector $v \in \R^N$ for which $\transpose{v}x = 0$ is a valid for all $x \in P \cap \M^S$.
Clearly, the dimension of this subspace is the local rank.

\begin{theorem}
  \label{thm_fixed_group_decomposition}
  Let $P \subseteq \R^N$ be a rational polyhedron and let $S \subseteq N$.
  Then there exist integers $0 \leq \ell \leq k$ and linearly independent vectors $v^1, v^2, \dotsc, v^k \in \Psi_S(P)$ such that
  \begin{equation}
    \Psi_S(P) = \left\{ \sum_{i=1}^k \lambda_i v^i \mid \lambda_i \in \R ~\forall i \in \{1,2,\dotsc,\ell\} \text{ and } \lambda_i \in \Z ~\forall i \in \{\ell+1,\ell+2,\dotsc,k\} \right\},
    \label{eq_fixed_group_decomposition}
  \end{equation}
  where the fixing functions $\alpha_i : \Z^S \to \Z$ for $v^i$ satisfy $\alpha_i(\bar{x}) = 0$ for all $\bar{x} \in \Z^S$ with $Q_S(\bar{x}) \neq \emptyset$ if and only if $i \leq \ell$ holds.
  For any choice of such vectors, every $v \in \Psi_S(P)$ whose fixing function $\alpha : \Z^S \to \Z$ satisfies $\alpha(\bar{x}) = 0$ for all $\bar{x} \in \Z^S$ with $Q_S(\bar{x}) \neq \emptyset$ satisfies $v \in \spn(v^1,v^2,\dotsc,v^\ell)$.
Moreover, the fixing functions $\alpha_{\ell+1},\alpha_{\ell+2},\dotsc,\alpha_k$ are linearly independent.

\end{theorem}

\begin{proof}
  Let $\varepsilon > 0$ be such that $\rank_{\varepsilon}(\Psi_S(P)) = \localrank(\Psi_S(P))$.
  The existence of $v^1, v^2, \dotsc, v^k$ (for $k \coloneqq \rank(\Psi_S(P))$ and $\ell \coloneqq \localrank(\Psi_S(P)))$ follows from \cref{thm_vector_group_decomposition}.
  Let $\mathcal{V} \coloneqq \spn\{ v^1,v^2, \dotsc, v^{\ell} \}$ denote the largest vector space contained in $\Psi_S(P)$.
  It remains to show that the vectors in $\mathcal{V}$ are exactly those for which the corresponding equations have constant right-hand side $0$.

  To this end, consider a vector $v \in \Psi_S(P)$ with $\transpose{v}x = 0$ for all $x \in P \cap \M^S$.
  The scaled vector $v' \coloneqq \frac{\varepsilon}{||v||} \cdot v$ clearly lies in $\mathcal{V}$.

  For the reverse direction, consider a vector $v \in \mathcal{V}$.
  Suppose the corresponding fixing function $\alpha$ satisfies $\alpha(\bar{x}) \neq 0$ for some $\bar{x} \in \Z^S$ with $Q_S(\bar{x}) \neq \emptyset$.
  The scaled vector $v' \coloneqq \frac{2 \alpha(\bar{x})}{||v||} \cdot v$ clearly induces the equation $\transpose{(v')} x = 1/2$ valid for $Q_S(\bar{x})$.
  This, however, establishes $v' \notin \Psi_S(P) \supseteq \mathcal{V}$, a contradiction to $v \in \mathcal{V}$.\\

  Let us now consider a vector $v \in \Psi_S(P)$ all of whose equations have constant right-hand side $0$.
  From the above we have $v \in \mathcal{V}$, and thus $v \in \spn(v^1,v^2,\dotsc,v^\ell)$.\\
  It remains to show linear independence of the fixing functions for $i > \ell$.
  Consider a linear combination of fixing functions that yields the zero function, that is, a multipliers $w_{\ell+1},w_{\ell+2},\dotsc,w_k \in \R$ with $\sum_{i=\ell+1}^k w_i \alpha_i(x) \equiv 0$.
  We will show $w = \zerovec$.
  The vector $\bar{v} \coloneqq \sum_{i=\ell+1}^k w_i v^i$ satisfies
  \[
    \transpose{\bar{v}} x
    = \sum_{i=\ell+1}^k w_i \transpose{(v^i)} x
    = \sum_{i=\ell+1}^k w_i \alpha_i(x)
    = 0
  \]
  for each $x \in \Q_S(\bar{x})$ and $\bar{x} \in \Z^S$ for which $Q_S(\bar{x}) \neq \emptyset$, where the second equality follows from validity of $\transpose{(v^i)} x = \alpha_i(\bar{x})$ for $Q_S(\bar{x})$.
  Hence, $\bar{v} \in \mathcal{V}$ and thus $\bar{v} \in \spn(v^1,v^2,\dotsc,v^\ell)$.
  By construction, $\bar{v} \in \spn(v^{\ell+1},v^{\ell+2},\dotsc,v^k)$ holds, and linear independence of all $v^i$ yields $\bar{v} = \zerovec$.
  Thus, $\sum_{i=\ell+1}^k w_i v^i = \bar{v} = \zerovec$, and linear independence of $v^i$ implies $w = \zerovec$.
  We conclude that the fixing functions $\alpha_{\ell+1},\alpha_{\ell+2},\dotsc,\alpha_k$ are linearly independent.
\end{proof}

Unfortunately, determining (a suitable basis of) $\Psi_S$ is not straight-forward at all.
We provide a constructive result in case our polyhedron in question is an affine space. The hardness of the general case is considered later in \cref{thm_implied_fixed_complexity_hard}.

\begin{theorem}
  \label{thm_affine_fixed_group_decomposition}
  Given a rational matrix $A \in \Q^{M \times N}$, a rational vector $b \in \Q^M$, which define an affine space $\mathcal{A} \coloneqq \{ x \in \R^N \mid Ax = b \}$, as well as a subset $S \subseteq N$, we can determine in polynomial time the integers $0 \leq \ell \leq k$ and a basis $v^1, v^2, \dotsc, v^k$ of $\Psi_S(\mathcal{A})$ satisfying~\eqref{eq_fixed_group_decomposition} for $P = \mathcal{A}$.
\end{theorem}

\DeclareDocumentCommand\lattice{}{\ensuremath{\Lambda}}

We first prove an auxiliary result about projections of vector groups described by a linear space and integer variables, which we were unable to find in the literature.
For a finite set $X \subseteq \R^n$ of generators, we denote by $\lattice(X) \coloneqq \{ \sum_{x \in X} \lambda_x x \mid \lambda \in \Z^X \}$ the generated lattice.

\DeclareDocumentCommand\L{}{\mathcal{L}}
\DeclareDocumentCommand\X{}{\mathcal{X}}

\begin{lemma}
  \label{thm_project_affine_mixed_integer}
  There is a polynomial-time algorithm that, for a linear space $\L \subseteq \R^N$ given by means of rational linear equations and subsets $S,T \subseteq N$ of variables, computes numbers $k,\ell \in \Znonneg$ and linearly independent vectors $x^1,x^2,\dotsc,x^k,\bar{x}^1,\bar{x}^2,\dotsc,\bar{x}^\ell \in \Q^S$ such that the projection $\X$ of $\L \cap \M^T$ on the variables $x_S$ is given by
  \begin{equation}
    \label{eq_projection_affine_mixed_lattice}
    \X = \lattice( \{x^1,x^2,\dotsc,x^k\} ) + \spn( \{ \bar{x}^1,\bar{x}^2,\dotsc,\bar{x}^\ell \} ).
  \end{equation}
\end{lemma}

\begin{proof}[Proof of \cref{thm_project_affine_mixed_integer}]
  We can assume that $S$ and $T$ are disjoint since otherwise we can duplicate variables and add equations to enforce equality among these, which only causes polynomial overhead.
  Hence, our algorithm may receive as input the matrices $A \in \Q^{M \times S}$, $B \in \Q^{M \times U}$ and $C \in \Q^{M \times T}$ such that
  \[
    \L = \{ (x,y,z) \in \R^{S \cup U \cup T} \mid Ax + By + Cz = \zerovec \}
  \]
  holds.
  As a first step we carry out elementary row operations to obtain the equivalent equation system
  \begin{subequations}
    \label{eq_affine_projection}
    \begin{alignat}{7}
      \widetilde{B}^1 y &+& \widetilde{A}^1 x &+& \widetilde{C}^1 z &= \zerovec \label{eq_affine_projection_first} \\
      & & \widetilde{A}^2 x &+& \widetilde{C}^2 z &= \zerovec \label{eq_affine_projection_second} \\
      & &  & & \widetilde{C}^3 z &= \zerovec, \label{eq_affine_projection_third}
    \end{alignat}
  \end{subequations}
  where $\widetilde{B}^1$, $\widetilde{A}^2$ and $\widetilde{C}^3$ are upper-triangular matrices of full row rank, and such that $\widetilde{C}^3$ has only integer entries.
  As a second step we compute the Hermite Normal Form of $\widetilde{C}^3$, which in particular yields a matrix $\widetilde{H}$ of full column rank and a unimodular matrix $\widetilde{U}$ such that $\widetilde{C}^3 \widetilde{U} = \begin{bmatrix} \widetilde{H} & \zerovec \end{bmatrix} $ holds.
  Clearly, the set of integer vectors in the nullspace of $\begin{bmatrix} \widetilde{H} & \zerovec \end{bmatrix}$ is equal to $\{ \zerovec_{|T| - \ell} \} \times \Z^\ell$, where $\ell = |T| - \rank(\widetilde{H})$.
  Hence, a lattice basis of $\{ z \in \Z^T \mid \widetilde{C}^3 z = \zerovec \}$ is given by
  \[
    z^i \coloneqq \widetilde{U} \unitvec{|T| - i + 1} \qquad \forall i \in \{1,2,\dotsc,k\}.
  \]
  For each such vector $z^i$, $i = 1,2,\dotsc,k$, we compute a solution $x^i \in \Q^S$ to $\widetilde{A}^2 x^i = -\widetilde{C}^2 z^i$ and a solution $y^i \in \Q^U$ to $\widetilde{B^1} y^i = - \widetilde{A}^1 x^i - \widetilde{C}^1 z^i$.
  By construction, $(x^i,y^i,z^i)$ is feasible for~\eqref{eq_affine_projection}, which implies $x^i \in \X$.

  As a third step we compute a basis of the nullspace of $\widetilde{A}^2$, which we denote by $\bar{x}^j \in \Q^S$ for $j=1,2,\dotsc,\ell$.
  Again, the full row rank of $\widetilde{B}^1$ yields existence of some $\bar{y}^j \in \Q^U$ such that $(\bar{x}^j,\bar{y}^j,\zerovec)$ satisfies~\eqref{eq_affine_projection_first}, which shows $(\bar{x}^j,\bar{y}^j,\zerovec) \in \L \cap \M^T$ and thus $\bar{x}^j \in \X$.
  This shows ``$\supseteq$'' in~\eqref{eq_projection_affine_mixed_lattice}.

  We now show the reverse inclusion.
  To this end, let $x \in \X$, which means that there exist $y \in \R^U$ and $z \in \Z^T$ such that $(x,y,z) \in \L$.
  By construction of $z^i$ there exist integer multipliers $\lambda \in \Z^k$ such that $z = \sum_{i=1}^k \lambda_i z^i$.
  Let $(x',y',z') \coloneqq (x,y,z) - \sum_{i=1}^k \lambda_i (x^i,y^i,z^i)$, which clearly also lies in $\L$.
  Since $z' = \zerovec$ holds, $x'$ lies in the nullspace of $\widetilde{A}^2$ and hence there exist multipliers $\mu \in \R^\ell$ such that $y' = \sum_{j=1}^\ell \mu_j \bar{x}_j$.
  It now follows that $x - \sum_{j=1}^\ell \mu_j \bar{x}^j - \sum_{i=1}^k \lambda_i x^i = \zerovec$, which implies $x \in \lattice( \{x^1, x^2, \dotsc, x^k \} ) + \spn( \{ \bar{x}^1, \bar{x}^2, \dotsc, \bar{x}^\ell \} )$.
  We conclude that~\eqref{eq_projection_affine_mixed_lattice} holds.

  It remains to show that we can reduce, in polynomial time, the set $\{x^1,x^2,\dotsc,x^k,\bar{x}^1,\bar{x}^2,\dotsc,\bar{x}^\ell\}$ to a linearly independent set while maintaining~\eqref{eq_projection_affine_mixed_lattice}.
  By Gaussian Elimination we can detect linear dependence in polynomial time and either reduce $k$ or $\ell$ by $1$ as follows.
  Assume that we find multipliers $\lambda \in \Q^k$ and $\mu \in \Q^\ell$, not all of which are zero, such that
  \[
    \sum_{i=1}^k \lambda_i x^i + \sum_{j=1}^\ell \mu_j \bar{x}^j = \zerovec.
  \]
  If $\lambda = \zerovec$ holds then, without loss of generality, $\mu_\ell \neq 0$.
  We can remove $\bar{x}^\ell$ from our set due to
  \[
    \bar{x}^\ell = -\frac{1}{\mu_\ell} \sum_{j=1}^{\ell-1} \mu_j \bar{x}^j.
  \]
  Otherwise, we can assume by reindexing that $\lambda_k \neq 0$.
  Moreover, by scaling $\lambda$ and $\mu$ we can assume $\lambda \in \Z^k$ and that the greatest common divisor of all $\lambda_i$ is equal to $1$.
  We will now replace our lattice generators $x^1,x^2,\dotsc,x^k$ by $\widehat{x}^1, \widehat{x}^2, \dotsc, \widehat{x}^k$ such that $\lattice(\{x^1,x^2,\dotsc,x^k\}) = \lattice(\{ \widehat{x}^1,\widehat{x}^2,\dotsc,\widehat{x}^k \})$ as well as $\widehat{x}^k = \sum_{i=1}^k \lambda_i x^i$ hold.
  Then we can also discard $\widehat{x}^k$ due to $\widehat{x}^k = \sum_{j=1}^\ell -\mu_j \bar{x}^j$.
  To achieve this, we compute the Hermite Normal Form of the $1$-by-$k$ matrix $(\lambda_1,\lambda_2,\dotsc,\lambda_k)$.
  By permuting the first and the last coordinate (which also corresponds to a multiplication with a unimodular matrix), this yields a unimodular matrix $\bar{U} \in \Z^{k \times k}$ such that $\transpose{\lambda} \bar{U} = \gamma \cdot \transpose{\unitvec{k}}$, where $\gamma = 1$ is the greatest common divisor of all $\lambda_i$.
  We now construct its inverse $\widehat{U} \coloneqq \bar{U}^{-1}$.
  Note that it is also unimodular and that its last row consists of the multipliers, that is, $\widehat{U}_{k,\star} = \transpose{\unitvec{k}} \widehat{U} = \transpose{\lambda} \bar{U} \widehat{U} = \transpose{\lambda}$.
  We now construct the $n$-by-$k$ matrix
  \[
    \begin{bmatrix}
      \widehat{x}^1 & \widehat{x}^2 & \dots & \widehat{x}^k
    \end{bmatrix}
    \coloneqq
    \begin{bmatrix}
      x^1 & x^2 & \dots & x^k
    \end{bmatrix}
    \transpose{\widehat{U}}.
  \]
  Unimodularity of $\transpose{\widehat{U}}$ implies $\lattice(\{x^1,x^2,\dotsc,x^k\}) = \lattice(\{ \widehat{x}^1,\widehat{x}^2,\dotsc,\widehat{x}^k \})$.
  Moreover, the last vector satisfies $\widehat{x}^k = \begin{bmatrix} x^1 & x^2 & \dots & x^k \end{bmatrix} (\transpose{\widehat{U}})_{\star,k} = \sum_{i=1}^k \lambda_i x^i$.
\end{proof}

\begin{proof}[Proof of \cref{thm_affine_fixed_group_decomposition}]
  In order to compute a basis for $\Psi_S(\mathcal{A})$ we will construct a set of linear constraints in an extended space, and then apply \cref{thm_project_affine_mixed_integer}.
  Some of these linear constraints will be related to the defining equations $Ax = b$, while others are tied to fibers.
  Since there may be infinitely many fibers, our first step is to determine a suitable finite subset.

  To achieve this we first project $\mathcal{A}$ on the $x$-variables.
  This is done by transforming $Ax=b$ by means of elementary row operations into the equivalent system
  \begin{subequations}
    \label{eq_affine_fixed_group_decomposition_projection}
    \begin{alignat}{7}
      A^1 x_S &\;+\;& B^1 x_{N \setminus S} &= b^1 \label{eq_affine_fixed_group_decomposition_projection_first} \\
      A^2 x_S &     &                       &= b^2, \label{eq_affine_fixed_group_decomposition_projection_second}
    \end{alignat}
  \end{subequations}
  where $A^2$ and $B^1$ both have full row rank.
  The equations~\eqref{eq_affine_fixed_group_decomposition_projection_second} describes the projection of $\mathcal{A}$ on the $x_S$-variables since each solution $x_S$ can be lifted to some $x \in \R^N$ by solving~\eqref{eq_affine_fixed_group_decomposition_projection_first}, which has a solution due to the full rank of $B^1$.
  To obtain the fibers we solve the diophantine equation system~\eqref{eq_affine_fixed_group_decomposition_projection_second} for $x_S \in \Z^S$ (see \cite[Cor.~5.3]{Schrijver86}).
  If it has no integer solution then all fibers of $\mathcal{A}$ are empty, and we can return $k = \ell = |N|$ and $v^i \coloneqq \unitvec{i}$ for $i \in N$.

  Otherwise, we obtain a solution $\bar{x}^0 \in \Z^S$ and vectors $\bar{x}^i \in \Z^S$ for $i=1,2,\dotsc,k$ such that
  $\{ \bar{x} \in \Z^S \mid A^2 \bar{x} = b^2 \} = \bar{x}_0 + \lattice(\{ \bar{y}^1,\bar{y}^2, \dotsc, \bar{y}^k \})$.
  Let $M$ index the rows of $B^1$.
  We claim that $\Psi_S(\mathcal{A})$ is the projection of the set of $(v,\sigma,\zeta,z) \in \R^N \times \R^M \times \Z \times \Z^k$ satisfying
  \begin{subequations}
    \label{eq_affine_fixed_group_decomposition_system}
    \begin{alignat}{9}
      \transpose{v_{N \setminus S}} &\;-\;& \transpose{\sigma} B^1 & & & & &= \transpose{\zerovec} \label{eq_affine_fixed_group_decomposition_system_combine} \\
      \transpose{v_S} \bar{x}^0 &\;-\;& \transpose{\sigma} (A^1 \bar{x}^0 - b^1) &\;-\;& \zeta & & &= 0 \label{eq_affine_fixed_group_decomposition_system_0}  \\
      \transpose{v_S} \bar{y}^i &\;-\;& \transpose{\sigma} (A^1 \bar{y}^i) & & &\;-\;& z_i &= 0 &\quad& \forall i \in \{1,2,\dotsc,k\} \label{eq_affine_fixed_group_decomposition_system_i}
    \end{alignat}
  \end{subequations}
  on the $v$-variables.
  This will complete the proof since~\eqref{eq_affine_fixed_group_decomposition_system} can be constructed in polynomial time and since we can construct a basis of $\Psi_S(\mathcal{A})$ using \cref{thm_project_affine_mixed_integer}.
  
  To prove the claim we first show that every solution $(v,\sigma,\zeta,z) \in \R^N \times \R^M \times \Z \times \Z^k$ of~\eqref{eq_affine_fixed_group_decomposition_system} yields a vector $v \in \Psi_S(\mathcal{A})$.
  To this end, consider an arbitrary fiber $Q_S(\bar{x})$.
  If $Q_S(\bar{x}) = \emptyset$ then every equation is valid, so there is nothing to show.
  Otherwise, $\bar{x} \in \Z^S$ satisfies $A^2 \bar{x} = b^2$ and thus $\bar{x} \in \bar{x}^0 + \lattice(\{\bar{y}^1,\bar{y}^2,\dotsc,\bar{y}^k\})$.
  Hence, there exist multipliers $\lambda \in \Z^k$ for which $\bar{x} = \bar{x}^0 + \sum_{i=1}^k \lambda_i \bar{y}^i$ holds.
  The fiber is described by
  \begin{align}
    Q_S(\bar{x})
    &= \{ x \in \R^N \mid x_S = \bar{x},~ Ax = b \} \nonumber \\
    &= \{ x \in \R^N \mid x_S = \bar{x},~ A^1x_S + B^1x_{N \setminus S} = b^1,~ A^2x_S = b^2 \} \label{eq_affine_fixed_group_decomposition_fiber} \\
    &= \{ x \in \R^N \mid x_S = \bar{x},~ B^1x_{N \setminus S} = b^1 - A^1 \bar{x} \}, \nonumber
  \end{align}
  where the last equality follows since $A^2x_S = b^2$ is implied by $x_S = \bar{x}$.
  Combining these equations with multipliers $v_S \in \R^S$ and $\sigma \in \R^M$ yields
  \begin{align*}
    \transpose{v} x
    &= 
    \transpose{v_S} x_S + \transpose{\sigma} B^1 x_{N \setminus S}
    =
    \transpose{v_S} \bar{x}
    + \transpose{\sigma} \left( b^1 - A^1 \bar{x} \right) \\
    &=
    \transpose{v_S} \left( \bar{x}^0 + \sum_{i=1}^k \lambda_i \bar{y}^i \right)
    + \transpose{\sigma} \left( b^1 - A^1 \big( \bar{x}^0 + \sum_{i=1}^k \lambda_i \bar{y}^i \big) \right) \\
    &=
    \left( \transpose{v_S} \bar{x}^0 - \transpose{\sigma} \big( A^1 \bar{x}^0 - b^1 \big) \right)
    +\sum_{i=1}^k \lambda_i \left( \transpose{v_S} \bar{y}^i - \transpose{\sigma} \big( A^1 \bar{y}^i \big) \right)
    = \zeta + \sum_{i=1}^k \lambda_i z_i,
  \end{align*}
  where the first equality follows from~\eqref{eq_affine_fixed_group_decomposition_system_combine} and the last from~\eqref{eq_affine_fixed_group_decomposition_system_0} and \eqref{eq_affine_fixed_group_decomposition_system_i}.
  Hence, the equation $\transpose{v}x = \zeta + \sum_{i=1}^k \lambda_i z_i$ with integer right-hand side is valid for $Q_S(\mathcal{A})$.

  To prove the other direction we show that, for any $v \in \Psi_S(\mathcal{A})$, such $(\sigma,\zeta,z) \in \R^M \times \Z \times \Z^k$ exist.
  First note that, since $\bar{x}^0$ is feasible for~\eqref{eq_affine_fixed_group_decomposition_projection_second}, the fiber $Q_S(\bar{x}^0)$ is non-empty, which implies that there exists a unique right-hand side $\zeta \in \Z$ of the corresponding equation $\transpose{v}x = \zeta$ that is valid for $Q_S(\bar{x}^0)$, i.e., $\zeta \coloneqq \alpha(\bar{x}^0)$.
  Similarly to~\eqref{eq_affine_fixed_group_decomposition_fiber} the fiber is described by
  \begin{align*}
    Q_S(\bar{x}^0) = \{ x \in \R^N \mid x_S = \bar{x}^0,~ B^1x_{N \setminus S} = b^1 - A^1 \bar{x}^0 \}.
  \end{align*}
  Since $\transpose{v}x = \zeta$ is valid for $Q_S(\bar{x})$, there exist multipliers $\mu \in \R^S$ and $\sigma \in \R^M$ such that
  \begin{align*}
   && \transpose{\mu} \begin{bmatrix} \idmat_S & \zerovec \end{bmatrix} + \transpose{\sigma} \begin{bmatrix} \zerovec & B^1 \end{bmatrix} = \transpose{v}
   && \text{ and } &&&
   \transpose{\mu} \bar{x}^0 + \transpose{\sigma} (b^1 - A^1\bar{x}^0) = \zeta,
   &&
  \end{align*}
  where $\idmat_S$ denotes the identity matrix of order $|S|$.
  From the first equation we obtain $\mu = v_S$ and consequently that~\eqref{eq_affine_fixed_group_decomposition_system_combine} and~\eqref{eq_affine_fixed_group_decomposition_system_0} are satisfied.
  For the last equations~\eqref{eq_affine_fixed_group_decomposition_system_i} we need to consider more fibers.
  Let us fix any $i \in \{1,2,\dotsc,k\}$.
  By definition of $\bar{x}^0$ and $\bar{y}^i$ the vector $\bar{x}^i \coloneqq \bar{x}^0 + \bar{y}^i$ is feasible for~\eqref{eq_affine_fixed_group_decomposition_projection_second} and hence the fiber $Q_S(\bar{x}^i)$ is non-empty.
  Similarly to~\eqref{eq_affine_fixed_group_decomposition_fiber} the fiber is described by
  \[
    Q_S(\bar{x}^i) = \{ x \in \R^N \mid x_S = \bar{x}^i,~ B^1x_{N \setminus S} = b^1 - A^1 \bar{x}^i \}.
  \]
  Moreover, there exists a right-hand side $\alpha(\bar{x}^i)$ for the equation $\transpose{v}x = \alpha(\bar{x}^i)$ that is valid for $Q_S(\bar{x}^i)$.
  We define $z_i \coloneqq \alpha(\bar{x}^i) - \zeta$.
  This in turn gives rise to multipliers $\widetilde{\mu} \in \R^k$ and $\widetilde{\sigma} \in \R^M$ such that 
  \begin{align*}
    &&
    \transpose{\widetilde{\mu}} \begin{bmatrix} \idmat_S & \zerovec \end{bmatrix} + \transpose{\widetilde{\sigma}} \begin{bmatrix} \zerovec & B^1 \end{bmatrix} = \transpose{v}
    && \text{ and } &&
    \transpose{\widetilde{\mu}} \bar{x}^i + \transpose{\widetilde{\sigma}} (b^1 - A^1\bar{x}^i) = \zeta + z_i.
    &&
  \end{align*}
  We again derive $\widetilde{\mu} = v_S = \mu$ and $\transpose{\widetilde{\sigma}} B^1 = \transpose{v}$.
  Since $B^1$ has full row rank, also $\widetilde{\sigma} = \sigma$ holds.
  From the second equation we obtain 
  \[
    z_i
    = \transpose{v_S} (\bar{x}^0 + \bar{y}^i) + \transpose{\sigma}( b^1 - A^1(\bar{y}^i + \bar{x}^0)) - \zeta 
    = \transpose{v_S} \bar{y}^i - \transpose{\sigma} ( A^1\bar{y}^i ),
  \]
  that is, \eqref{eq_affine_fixed_group_decomposition_system_i} holds.
  This proves our claim, which concludes the proof.
\end{proof}

It is easy to see that, for $P \subseteq \mathcal{A}$, we also have $\Psi_S(P) \supseteq \Psi_S(\mathcal{A})$.
This implies that we can apply \cref{thm_affine_fixed_group_decomposition} to the affine hull of a given polyhedron $P$.

For a mixed-integer linear optimization problem given by a polyhedron $P\subseteq \R^N$ and some set of integer variables $I\subseteq N$, the structure of $\Psi_I(P)$ may be interesting to use in practice. In particular, one could use the integrality of expressions which are not already clearly integer in the original, i.e., where $\transpose{v} x$ has non-integer coefficients and/or where $v$ has nonzeros for the variables $N\setminus I$.  
There are potential applications of this set in branching or cutting planes, where solvers may be able to use the integrality of $\transpose{v} x$ to derive stronger coefficients or use it to impose branching disjunctions $\transpose{v} x \leq k$ and $\transpose{v} x \geq k+1$ for some $k\in \Z$. With respect to this application, especially the vectors that span the lattice part of $\Psi_S(P)$ are interesting, since the remaining vectors $v$ in the linear part of $\Psi_S(P)$ all satisfy $\transpose{v} x = 0$ for all $x\in \conv(P\cap\M^S)$, which is clearly a poor choice to branch on.

\section{Local implied integrality}
\label{sec_local_implied_integrality}

In our initial definition of implied integrality in \cref{def_implied_integer}, we consider the complete description of the polyhedron $P\subseteq \R^N$.
In practice, one typically only optimizes over $P$ or the mixed-integer hulls $\conv(P\cap \M^I)$
in some given direction $c \in \R^N$. More specifically, in the context of an optimization problem, one primarily cares about the $c$-maximal faces of $P$ and the integer hulls $\conv(P\cap \M^S)$ for $S\subseteq N$ during branch-and-bound.
For this purpose, we weaken our definition of implied integrality to be more local, in the hope that we may find more applications of it.

\begin{definition}
  \label{def_implied_integer_objective}
  Let $P\subseteq \R^N$ be a rational polyhedron and let $c\in \R^N$ be any direction that is maximized for which the maximum in $P$ is finite.
  For $S, T \subseteq N$, we say that \emph{$x_T$ is locally implied integer by $x_S$ for $(P,c)$} if $\arg\max\{ \transpose{c}x \mid x \in \conv(P \cap \M^S) \} \cap \M^{S \cup T} \neq \emptyset$, that is, if the $c$-maximal face of $\conv(P \cap\M^S)$ contains a point in $\M^{S \cup T}$.
  In the case of $S = \emptyset$ we simply say that $x_T$ is locally implied integer (for $(P,c)$).
  Moreover, we say that \emph{$x_T$ is locally totally implied integer by $x_S$ for $(P,c)$} if $x_T$ is locally implied integer by $x_{S'}$ for every $S'\supseteq S$ (for $(P,c)$).
\end{definition}

First, we highlight the relationship between local implied integrality and implied integrality. 

\begin{proposition}
  Let $P$ be a rational polyhedron and let $S,T \subseteq N$.
  Then $x_T$ is implied integer by $x_S$ if and only if $x_T$ is locally implied integer by $x_S$ for all $c \in \R^N$ for which the maximum over $P$ is finite.
\end{proposition}

\begin{proof}
  This follows directly from the equivalence of \ref{thm_implied_integrality_definition} and~\ref{thm_implied_integrality_maxima} in \cref{thm_implied_integrality}.
\end{proof}

\begin{theorem}
  \label{thm_local_implied_integrality}
  Let $P$ be a rational polyhedron, and let $c \subseteq \R^N$ be any objective vector whose maximum over $P$ is finite.
  For $S, T \subseteq N$, the following are equivalent:
  \begin{enumerate}[label=(\roman*)]
  \item
    \label{thm_local_implied_integrality_definition}
    $x_T$ is locally implied integer by $x_S$ for $(P,c)$, i.e., $\arg\max\{ \transpose{c}x \mid x \in \conv(P \cap \M^S) \} \cap \M^{S \cup T} \neq \emptyset$.
  \item
    \label{thm_local_implied_integrality_fiber}
    There exists some $\bar{x} \in \Z^S$ such that $\max\{\transpose{c} x \mid x \in P\cap \M^S\} = \max\{\transpose{c} x \mid x \in Q_S(\bar{x})\}$ holds and $x_T$ is locally implied integer for $(Q_S(\bar{x}),c)$.
  \item
    \label{thm_local_implied_integrality_maximization}
    $\max\{\transpose{c} x \mid x \in \conv(P \cap \M^S)\} = \max\{\transpose{c} x \mid x \in \conv(P \cap \M^{S\cup T})\}$.
  \end{enumerate}
\end{theorem}

\begin{proof}
  \ref{thm_local_implied_integrality_definition} $\Rightarrow$ \ref{thm_local_implied_integrality_fiber}:
  By \ref{thm_local_implied_integrality_definition} and strong LP duality there exists a point $x^\star \in \arg\max\{ \transpose{c}x \mid x \in \conv( P \cap \M^S ) \} \cap \M^{S \cup T}$.
  Let $\bar{x} \coloneqq x^\star_S$ and note that $\max\{\transpose{c} x \mid x \in Q_S(\bar{x}) \} \leq \max\{\transpose{c} x \mid x \in P \cap \M^S \}$ holds since $Q_S(\bar{x})$ is a subset of $P \cap \M^S$.
  Moreover, $\max\{\transpose{c} x \mid x \in P \cap \M^S \}$ is attained by $x^\star$, which also lies in $\arg\max\{ \transpose{c}x \mid x \in Q_S(\bar{x}) \}$ and is $T$-integral.
  Hence, $x_T$ is locally implied integer for $(Q_S(\bar{x}), c)$.

\medskip

  \ref{thm_local_implied_integrality_fiber} $\Rightarrow$ \ref{thm_local_implied_integrality_maximization}: 
  From $\conv(P\cap \M^S) \supseteq \conv(P\cap \M^{S\cup T})$, we immediately obtain $\max\{ \transpose{c} x \mid x \in \conv(P \cap \M^S)\geq \max\{ \transpose{c} x \mid x \in \conv(P \cap \M^{S \cup T} \}$.
  By \ref{thm_local_implied_integrality_fiber} there exists some $\bar{x} \in \Z^S$ for which $\max\{ \transpose{c} x \mid x \in Q_S(\bar{x})\} = \max\{ \transpose{c} x \mid x \in P \cap \M^S\}$ and (using strong LP duality) some optimal solution $x^\star \in \arg\max\{ \transpose{c}x \mid x \in \conv( Q_S(\bar{x}) \cap \M^S ) \} \cap \M^{S \cup T}$.
  Note that $x^\star_S = \bar{x}$ holds by construction.
  We now have
  \begin{multline*}
    \max\{\transpose{c} x \mid x \in \conv(P \cap \M^S)\}
    = \max\{\transpose{c} x \mid x \in P \cap \M^S\}
    = \max\{\transpose{c} x \mid x \in Q_S(\bar{x})\} \\
    = \transpose{c} x^\star \leq \max\{ \transpose{c} x \mid x \in \conv(P \cap \M^{S\cup T}) \},
  \end{multline*}
  where the last inequality follows from $x^\star \in \M^{S \cup T}$.

\medskip

  \ref{thm_local_implied_integrality_maximization} $\Rightarrow$ \ref{thm_local_implied_integrality_definition}:
  By \ref{thm_local_implied_integrality_maximization} and $P \cap \M^S \supseteq P \cap \M^{S \cup T}$ there exists an optimal solution $x^\star \in \arg\max \{ \transpose{c}x \mid x \in \conv(P \cap \M^S) \}$ that also lies in $\M^{S \cup T}$.
  This point establishes local implied integrality.
\end{proof}

Similar to \cref{thm_implied_integer_sufficiency}, it is sufficient to show that all fibers have local implied integrality. 
\begin{corollary}
  \label{thm_local_implied_integrality_sufficient}
  Let $P \subseteq \R^N$ be a rational polyhedron, let $c\in \R^N$ be any objective whose maximum over $P$ is finite, and let $S,T \subseteq N$.
  If, for all $\bar{x} \in \Z^S$, $x_T$ is locally implied integer for $(Q_S(\bar{x}),c)$ then $x_T$ is locally implied integer by $x_S$ for $(P,c)$.
\end{corollary}

\begin{proof}
  By strong LP duality, there exists an optimal solution 
  $x^\star$ of $\max\{ \transpose{c} x \mid x \in  P \cap \M^S\}$.
  Then $x^\star_S \in \Z^S$ and $Q_S(x^\star_S)$ satisfy the conditions of \cref{thm_local_implied_integrality} \ref{thm_local_implied_integrality_fiber}.
\end{proof}

Note that since our definition only includes the optimal face and not the complete polyhedron, the conditions in \cref{thm_local_implied_integrality} are less restrictive than those of \cref{thm_implied_integrality}.
The notion of local implied integrality is even less restrictive than implied integrality of the optimal face. If in the optimal face $x_T$ is implied integer, all minimal faces of the optimal face are $T$-integral. In contrast, if $x_T$ is locally implied integer, we are only guaranteed that a single $T$-integral point in the optimal face exists. 

Local implied integrality has two advantages compared to implied integrality.
First, by considering only the optimal face, we consider a smaller polyhedron, typically of a much lower dimension.
One may hope to derive valid equations and valid inequalities that hold only for the optimal face, but not for the complete polyhedron.
Valid equations can be particularly promising, as they can be used eliminate decision variables.
Second, since we introduce an objective, we can use duality.
Duality is arguably the most powerful tool in mixed-integer linear programming and underlies many key techniques used in practice.

\subsection{Local implied integrality from fixed variables}

Since fixed variables constitute an important special case of implied fixed integrality, it is natural to also consider them for local implied integrality.
In fact, there already exist multiple techniques that infer local implied integrality using fixed variables.
Both \emph{reduced cost fixing}~\cite{Dantzig1954} and \emph{objective propagation}~\cite[Chapter 7.6]{Achterberg2008} are well-known techniques used to fix variables to their bounds based on their contribution to the objective and known objective bounds.
One key property of these propagation techniques is that the fixed values for the bounds are always identical over all fibers.
Thus, these variables can simply be removed from the formulation, which means that their implied integrality is not relevant in terms of algorithmic exploitation.
In this section, we focus on detecting implied integrality for variables that are fixed in the optimal face of each fiber, but whose fixed integral values may depend on the specific fibers.

As before, we first focus on the case where $S=\emptyset$, which is well motivated by \cref{thm_local_implied_integrality_sufficient}.
 Our focus lies on proving that the $c$-maximal face of a polyhedron $P$ with given description contains an $x_T$-integral point.
In particular, note that it is sufficient to show that $x_T$ can be fixed to an integer in the optimal face.
We consider the well-known technique of \emph{dual fixing} \cite{Brearley1975}, which is a standard presolving procedure for linear programming and MILP solvers \cite{AchterbergBGRW20, Andersen1995}. 
Dual fixing strongly relies on a correspondence between variables that are fixed at their bounds in the optimal face of the primal linear program and redundant inequalities in the optimal face of the dual linear program.
First, let us formally define redundancy.

\begin{definition}
  For a polyhedron $P = \{ x \in \R^N \mid A x \leq b \}$ with $A \in \R^{M \times N}$ and $b \in \R^M$, a set of constraints $K \subseteq M$ is said to be \emph{redundant} if $P = P_{M \setminus K} \coloneqq \{ x \in \R^N \mid A_{M \setminus K, \star} x \leq b_{M \setminus K} \}$ holds, i.e., if the constraints $K \subseteq M$ can be removed without altering the feasible region.
  Moreover, the constraints $K \subseteq M$ are said to be \emph{strictly redundant} if for each $m \in K$ and $x \in P$, $A_{m,\star} x < b_m$ holds.
\end{definition}

Next, we consider dual fixing.
Although the method is well-known in literature, most authors prefer a purely computational view.
We complement this with a polyhedral point of view and prove its correctness.

\begin{proposition}[Dual Fixing]
 \label{thm_dual_fixing}
  Let $P = \{ x \in \R^N \mid A x \leq b, ~x \geq \zerovec \}$ be a polyhedron with $A \in \R^{M \times N}$ and $b\in \R^M$.
  For some objective $c \in \R^N$ that has a finite maximum over $P$, consider the linear program $\max\{ \transpose{c} x \mid x \in P \}$ and its dual $\min\{ \transpose{b} y  \mid y \in D \}$ where $D = \{ y \in \R^M \mid \transpose{A} y \geq c, ~ y \geq \zerovec \}$.
  Let $F_P$ be the primal $c$-maximal face of $P$ and let $F_D$ be the dual $b$-maximal face of $D$.
  Then the following hold:
  \begin{enumerate}
  \item
    If for $S\subseteq  N$ the constraints $\transpose{(A_{\star,S})} y \geq c_S$ are strictly redundant for $F_D$, then the equation $x_k = 0$ is valid for $F_P$ for all $k \in S$.
  \item
    If for some $S \subseteq N$ the constraints $\transpose{(A_{\star,S})} y \geq c_S$ are redundant for $F_D$, then there exists a solution $x' \in F_P$ with $x'_S = \zerovec$. 
  \end{enumerate}
\end{proposition}

\begin{proof}
  First, let us prove the first point. By complementary slackness, we must have, for each $k \in M$ and any two points $x^\star\in F_P$ and $y^\star \in F_D$, that $x^\star_k ( \transpose{(A_{\star,k})} y^\star - c_k) = 0$ holds.
  For $k\in S$ we have strict redundancy for $F_D$ by assumption, which implies that we must have  $\transpose{(A_{\star,k})} y^\star - c_k > 0$ for all $y^\star\in F_D$ and all $k\in S$.
  Thus, in order to satisfy complementary slackness, $x^\star_k = 0$ must hold for any $x^\star \in F_P$.

  To prove the second point, let $\delta \coloneqq \max\{ \transpose{c} x \mid x \in P\} = \min \{\transpose{b} y \mid y \in D\}$ be the optimal value of the primal-dual LP pair.
  Let $\bar{S} \coloneqq N \setminus S$ denote the variables associated to the other constraints of $F_D$.
  Redundancy of $\transpose{(A_{\star,S})} y \geq c_S$ implies that $F_D = \{ y \mid \transpose{(A_{\star,\bar{S}})} y \geq c_{\bar{S}},~ \transpose{b} y \leq \delta \}$ holds since $\transpose{b} y \leq \delta$ holds for any feasible solution $y$ in the face.
  Using the variables $(\hat{x},z) \in \R^{\bar{S}} \times \R$ for the constraints, the dual of $\min\{\transpose{b} y \mid y\in F_D\}$ is then given by $\delta = \max\{\transpose{c_{\bar{S}}} \hat{x} - \delta z \mid A_{\star,\bar{S}} \hat{x} - b z \leq b, ~x \geq \zerovec, z \geq 0 \}$.
   By strong duality there exist $(x^\star,z^\star) \in \R^{\bar{S}} \times \R$ such that $\delta = \transpose{c_{\bar{S}}} x^\star - \delta z^\star$, $A_{\star,\bar{S}} x^\star - b z^\star \leq b$ and $z^\star \geq 0$ hold.
  We define $\bar{x} \coloneqq \frac{x^\star}{1 + z^\star}$ and notice that $\transpose{c_{\bar{S}}} \bar{x} = \delta$ and $A_{\star,\bar{S}} \bar{x} \leq b$ hold.
  Then we define $x' \coloneqq (\bar{x},\zerovec) \in \R^{N \setminus S} \times \R^S$.
  Clearly, $A x' \leq b$ and $\transpose{c} x' = \delta$ hold by the above, since the terms for the $S$-variables vanish due to $x'_S = \zerovec$.
  Thus, $x'$ lies in $F_P$ and has $x'_S = \zerovec$, which completes our proof. 
\end{proof}

In \cref{thm_dual_fixing}, we show that dual redundancy can be a sufficient condition for finding integers in the primal optimal face. \Cref{example_redundancy} shows that redundancy is not a necessary condition for a coordinate to be fixed to zero.
\begin{example}
  \label{example_redundancy}
  Consider the linear program $\max\{ -x \mid -x \leq 0, ~ 0 x \leq 1,~ x \geq 0\}$ and its dual $\min\{y_2 \mid - y_1 \geq - 1,~ y_1 \geq 0,~ y_2 \geq 0 \}$.
  The primal optimal face is given by $x = 0$, and the dual optimal face is given by $\{ (y_1,y_2) \in \R^2 \mid 0 \leq y_1 \leq 1,~ y_2 = 0 \}$.
  Note that none of the dual constraints is redundant for the optimal face of the dual, even though the primal optimal face is given by $x = 0$.
\end{example}

Note that, for a polyhedron $P$ and an objective vector $c$, the existence of some $c$-maximal solution $x$ with $x_S = \zerovec$ is sufficient to prove that $x_S$ is locally implied integer for $(P,c)$.
In particular, dual fixing is a method to show local implied integrality. 
Although our proof works in the setting with $x \geq \zerovec$, it can easily be adapted to a setting with lower and upper bounds like $\ell \leq x \leq u$, by substituting either $x' = x - \ell$ or $x' = u - x$ and detecting redundancy for the resulting formulation.
\Cref{example_local} highlights a case where this may happen, and shows how \cref{thm_dual_fixing} can be used to infer local implied integrality.

Additionally, note that the set of redundant equations in the dual is not unique; removing a redundant inequality may turn other (previously redundant) inequalities non-redundant.
Thus, for some $j,k\in N$ with $j\neq k$, existence of a solution with $x_j=0$ and the existence of a solution with $x_k=0$ through dual redundancy does not imply that a solution exists that has $x_j=x_k = 0$.
In contrast to this, the set of \emph{strictly} redundant equations is unique; even after removing other redundant inequalities, the strictly redundant inequality may never touch the feasible region, so they always stay redundant.

\begin{example}
  \label{example_local}
  Consider the mixed-integer linear program $\max \{2y - x \mid x \in \{0,1\},~ y \in \Rnonneg,~ y \leq 2 x,~ 2x - 2y \leq 1 \}$ that is depicted in \Cref{fig:examplelocalimpliedintegrality}.
  For this problem, one can show that $y$ is locally implied integer by $x$ for the given objective, as follows.
  First, note that the fiber for $x = 0$ consists of the point $(x,y) = (0,0)$. 
  Moreover, the fiber for $x=1$ is equal to $\{(1,y) \mid y \geq 0,~ y \geq \frac{1}{2},~ y \leq 2 \}$.
  Although the polyhedron corresponding to this fiber is not integral, we maximize $2y$, which has positive objective coefficient, so we will always choose $y = 2$, which is integral.
  This can also be seen using \cref{thm_dual_fixing}:
  Let $y' = 2 - y$ and consider the linear program belonging to the fiber $x = 1$ in terms of $y'$.
  Then the local optimization problem can be written as $\max \{ -2 y' \mid y' \in \Rnonneg, ~ y' \leq \frac{3}{2}, y' \leq 2\}$.
  Its dual is given by $\min \{ \frac{3}{2} \pi_1 + 2 \pi_2 \mid \pi \in \Rnonneg^2,~ \pi_1 + \pi_2 \geq -2 \}$.
  Clearly, the constraint $\pi_1 + \pi_2 \geq -2$ is strictly redundant due to the nonnegativity constraints $\pi \geq \zerovec$, which implies that $y' = 0$ holds for all primal optimal solutions by \cref{thm_dual_fixing}, and shows that we can fix $y = 2$.
  Hence, $y$ is fixed to an integer for both $x$-fibers, with respect to the given objective, and thus $y$ is locally implied integer by $x$ by \cref{thm_local_implied_integrality_sufficient}.
  In fact, one can even infer $y = 2x$ in this case, which can be used to reduce the problem size.
\end{example}

\begin{figure}[htpb]
    \centering
    \begin{tikzpicture}
       \foreach \y in {0,1,2}
      {
        \draw (-0.4,\y) -- (2.4,\y);
      }
  
      \foreach \x in {0,1,2}
      {
        \draw (\x,-0.4) -- (\x,2.4);
      }
      \draw[blue, fill=blue!10, line width=1.5pt] (0, 0) -- (1,2) -- (1,0.5) -- (0.5,0) -- cycle;
       \draw[->,black,line width=1.5pt] (0,0) -- (-0.5,1);
    \end{tikzpicture}
    \caption{The feasible region (in blue) and the objective (black arrow) of \cref{example_local}.}
    \label{fig:examplelocalimpliedintegrality}
\end{figure}

In \cref{thm_local_relevant_sign_only}, we specialize \cref{thm_dual_detection} to detect local implied integrality as it appears in \cref{example_local}.

\begin{theorem}
  \label{thm_local_relevant_sign_only}
  Consider $P = \{x\in \R^N \mid A x \leq b, ~ x \geq \zerovec \}$ with $A \in \Q^{M \times N}$ and $b \in \Q^M$.
  For some $k \in N$, let $M^+_k \coloneqq \{ j \in M \mid A_{j,k} > 0 \}$ and $M^-_k \coloneqq \{ j \in M \mid A_{j,k} < 0 \}$ be the rows that contain a positive or negative nonzero entry in column $k$.
  For $\star\in \{-,+\}$, define $S^\star \coloneqq \{ k' \in N \setminus \{k\} \mid \text{ there exists } j \in M^\star_k \text{ with } A_{j,k'} \neq 0 \}$ as the other columns having non-zeros in $M^\star_k$.
  Let $c \in \R^N$ be the objective vector.
  Then, for the optimization problem $\max\{ \transpose{c} x \mid x \in P \}$, the following hold:
  \begin{enumerate}
  \item
    If $c_k \geq 0$ and $\frac{A_{j,s}}{A_{j,k}} \in \Z$ and $\frac{b_j}{A_{j,k}}\in \Z$ hold for all $j\in M^+_k$ and all $s\in S^+_k$, then $x_k$ is totally locally implied integer by $S^+_k$.
  \item
    If $c_k \leq 0$ and $\frac{A_{j,s}}{A_{j,k}} \in \Z$ and $\frac{b_j}{A_{j,k}}\in \Z$ hold for all $j\in M^-_k$ and all $s\in S^-_k$, then $x_k$ is totally locally implied integer by $S^-_k$.
  \end{enumerate}
\end{theorem}

For two polyhedra $P_1$ and $P_2$ we denote by $P_1 \cong P_2$ that they are affinely isomorphic, i.e., that there exist affine maps from each of them onto the other.

\begin{proof}
We only prove the first statement since the second one follows by negating $x_k$, $A_{M,k}$ and $c_k$.
To this end we consider a subset $S' \subseteq N$ of variables with $S' \supseteq S^+_k$.
If $k\in S'$ then the $c$-maximal face of $\conv(P\cap \M^{S'})$ contains some $k$-integral point and hence $k$ is locally implied integer by $S'$.
Otherwise, we will establish local implied integrality using \cref{thm_local_implied_integrality_sufficient}.
For any $\bar{x} \in \Z^{S'}$, consider the fiber $Q_{S'}(\bar{x})$.
Let $\bar{M} \coloneqq M \setminus M^+_k$ and $\bar{N} \coloneqq N \setminus \{k\}$.
Due to $S^+_k \subseteq S'$, the fiber is described by
\begin{multline*}
  Q_{S'}(\bar{x}) = \{ x \mid  A_{M^+_k,k} x_k \leq b_{M^+_k} - A_{M^+_k,S^+_k} x_{S^+_k},
  A_{\bar{M},\bar{N}} x_{\bar{N}} + A_{\bar{M},k} x_k \leq b_{\bar{M}},~ x_{S'} = \bar{x}, ~ x \geq \zerovec \},
\end{multline*}
where $A_{M^+_k,k} > 0$ and $A_{\bar{M},k} \leq 0$ hold by definition.
Now, let $\delta \coloneqq \min_{j \in M^+_k} \left\{ \frac{b_j - A_{j,S^+_k} x_{S^+_k}}{A_{j,k}} \right\}$.
Note that by the integrality conditions, $\delta$ is integral as the minimum is taken over integers.

Then we substitute $x'_k = \delta - x_k$ to obtain $Q'_S(\bar{x})$, which describes an isomorphic polyhedron for which we obtain
\begin{align*}
    Q_{S'}(\bar{x}) &= \{ x \mid  x_k \leq \delta, ~A_{\bar{M},\bar{N}} x_{\bar{N}} + A_{\bar{M},k} x_k \leq b_{\bar{M}}, ~x_{S'} = \bar{x}, ~x\geq \zerovec \} \\
    &\cong \{ x \mid  x'_k \geq 0, ~A_{\bar{M},\bar{N}} x_{\bar{N}} - A_{\bar{M},k} x'_k \leq b_{\bar{M}}, ~x_{S'} = \bar{x}, ~x_{N\setminus \{k\}} \geq \zerovec, x'_k \leq \delta \} \\
    &= \{ x \mid  x'_k \geq 0, ~A_{\bar{M},\bar{N}\setminus S'} x_{\bar{N}\setminus S'} - A_{\bar{M},k} x'_k \leq b_{\bar{M}} - A_{\bar{M},S'} \bar{x}, ~x_{N\setminus \{k\}} \geq \zerovec, x'_k \leq \delta \} \\
    &\eqqcolon Q'_S(\bar{x})
\end{align*}
Now we derive the dual of $\max\{\transpose{c} x \mid x\in Q'_S(\bar{x})\}$ using $\lambda$ for the upper bound $x'_k\leq \delta$ and $\pi$ for the other constraints. Then, the feasible region of the dual LP
$D'$ can be described as follows. Note that by replacing $x_k$ by $x'_k$, we negate the objective coefficient $c_k$.
\begin{equation*}
  D' = \{ (\lambda,\pi) \mid \transpose{\pi} A_{\bar{M},\bar{N} \setminus S'} \geq \transpose{c}_{\bar{N}\setminus S'},~ \lambda - \transpose{\pi} A_{\bar{M},k} \geq -c_k,~ \lambda \geq 0,~ \pi \geq \zerovec \},
\end{equation*}
where we consider the dual linear program $\min\{\delta \lambda + \transpose{(b_{\bar{M}}- A_{\bar{M},\bar{N}\setminus S'} \bar{x})} \pi \mid (\lambda,\pi)\in D'\}$.
Then, the constraint $\lambda - A_{\bar{M},k} \pi \geq -c_k$ is redundant because $\lambda \geq 0$, $\pi \geq 0$ and $A_{\bar{M},k} \leq 0$ already imply that $\lambda - A_{\bar{M},k} \pi \geq 0 $ holds, which dominates the former due to $c_k \geq 0$.
Since the dual constraint corresponding to variable $x_k$ is redundant for $D'$, it is redundant for the optimal face of $D'$ in particular.
Then there exists a solution with $x'_k = 0$ in the $c$-maximal face of $Q'_S(\bar{x})$ by \cref{thm_dual_fixing}, which shows that $x'_k = 0$ holds for $Q'_S(\bar{x})$.
By the affine isomorphism there exists a vector $x$ in the $c$-maximal face of $Q_S(\bar{x})$ with $x_k = \delta$.
Since $\delta$ is integral and this holds for any $\bar{x} \in \Z^S$, $x_k$ is locally implied integer by $x_{S'}$ for $(P,c)$ by \cref{thm_local_implied_integrality_sufficient}.
Since we have shown the result for any $S' \supseteq S^+_k$, we have shown that $x_k$ is \emph{totally} locally implied integer by $x_{S^+_k}$.
\end{proof}

\Cref{thm_local_relevant_sign_only} is already used in detection codes of MILP solvers due to its similarity with \cref{thm_dual_detection}.
However, neither SCIP~9.2.1~\cite{bolusani2024scipoptimizationsuite90} nor HIGHS~1.11~\cite{HIGHSrepo} exploit that for a variable $x_k$ with objective coefficient $c_k = 0$, one can choose between the implication sets $S^+_k$ and $S^-_k$, but instead require $S^+_k \cup S^-_k$ to have $\pm 1$ entries.
Hence, \cref{thm_local_relevant_sign_only} still represents an improvement over the currently known open source methods.

Since the structure of \cref{thm_local_relevant_sign_only} is similar to \cref{thm_dual_detection}, one may wonder if it can similarly be generalized to multiple variables using total unimodularity, as done in \cref{thm_tu_implied}.
However, this is much more challenging in this setting; the proof of \cref{thm_local_relevant_sign_only} explicitly exploits the fact that we treat only a single variable since redundancy is easy to recognize in dimension~1.
Already in dimension~2 this is more difficult, in particular since it has to hold for each (non-empty) fiber.

\section{Complexity of recognizing implied integrality }
\label{sec_complexity}

In this section we discuss questions related to the complexity of recognizing implied integers.
Since implied integers are a generalization of the concept of integral polyhedra, many of our results (implicitly) rely on the work of Papadimitriou and Yannakakis~\cite{Papadimitriou1990}.
They considered the following decision problem, and showed  $\cplxcoNP$-completeness.

\bigskip

\textsc{Integrality Recognition}:
Given a rational matrix $A \in \R^{M \times N}$ and rational vector $b \in \R^M$, determine whether $P(A,b)$ is integral.

\bigskip

Ding, Feng and Zhang~\cite{Ding2008} also show $\cplxcoNP$-hardness by reducing a graph problem to a  specific binary integer program.
The NO-certificate for showing containment in $\cplxcoNP$ is given by a non-integral (minimal) face $F$ of the polyhedron $P$.
Such a face is an affine subspace, and Schrijver \cite[Corollary 5.3b]{Schrijver86} shows that there always exists an equation system for this subspace and an objective vector defining $F$ that both have polynomial encoding length.
We will need a similar result for a certificate in the mixed-integer setting, which uses a very similar proof.
Although we could not find a proof of the following proposition in the literature, we suspect that the result is not original; the proof is similar to Schrijver's proof.

\begin{proposition}
  \label{thm_mixed_integer_system_polytime}
  Given rational matrices $A$ and $B$, and a rational vector $c$, we can find, in time polynomial in their encoding length, a vector $(x,y) \in \Z^n \times \R^d$ with $A x + B y = c$ or determine that no such vector exists.
\end{proposition}

\begin{proof}
    First, we project onto the $x$ variables by performing Gaussian elimination on the submatrix $B$.
    More specifically, we rewrite $Ax + By = c$ using elementary row operations such that it is of the form
    \begin{equation*}
        \begin{bmatrix}
            A' & B' \\
            A'' & 0
        \end{bmatrix}
        \begin{bmatrix}
            x\\
            y
        \end{bmatrix}
        = \begin{bmatrix}
            c'\\
            c''
        \end{bmatrix},
    \end{equation*}
    where $B'$ has full row rank.
    This can be done in polynomial time~\cite[Theorem 3.3]{Schrijver86}.
    We can determine if there exists an integral solution to $A'' x = c''$ in polynomial time using the Hermite Normal Form \cite[Corollary 5.3b]{Schrijver86}.
    Clearly, if $A''x = c''$ has no integral solution, then also $A x + B y= c$ does not have one.
    Otherwise, $A''x = c''$ has an integral solution $\bar{x} \in \Z^n$ that we also obtain from the Hermite Normal Form.
    Since $B'$ has full row rank, there exists a solution $\bar{y}$ to $B' y = c - A' \bar{x}$, which we can again find in polynomial time~\cite[Theorem 3.3]{Schrijver86}.
    Notice that if $B$ already has full row rank then $A''$ would have no rows, in which case we can use $\bar{x} = \zerovec$. 
\end{proof}

\subsection{Recognizing implied integers }
The first question we consider is the recognition of implied integrality, phrased as a decision problem.

\bigskip

\textsc{Implied Integer Recognition:}
Given a rational matrix $A \in \R^{M \times N}$, a rational vector $b \in \R^M$ and subsets $S,T \subseteq N$, determine whether
\begin{equation*}
    \conv(\{x\in \M^S \mid A x \leq b\}) = \conv(\{x\in \M^{S\cup T} \mid A x \leq b\}).
\end{equation*}

\bigskip

Note that the recognition of integral polyhedra is the special case of $S = \emptyset$ and $T = N$, which immediately yields $\cplxcoNP$-hardness of \textsc{Implied Integer Recognition}.
Completeness does not immediately follow, because we do not necessarily have a polynomial-size certificate due to the lack of an accessible description of $\conv(P\cap \M^S)$.
Note that if we restrict our problem to binary variables, then it reduces to one over the fibers using \cref{thm_implied_integer_binary}, which yields a certificate.
This motivates the dedicated treatment of this case:

\bigskip

\textsc{Binary Implied Integer Recognition:}
Given a rational matrix $A \in \Q^{M \times N}$, a rational vector $b \in \Q^M$ and subsets $S,T \subseteq N$, determine whether
\begin{equation*}
    \conv(\{x\in \M^S \mid \zerovec \leq x_S \leq \onevec,~ A x \leq b\}) = \conv(\{x\in \M^{S\cup T} \mid \zerovec \leq x_S \leq \onevec,~ A x \leq b\}).
\end{equation*}

\medskip

\begin{theorem}
  \label{thm_complexity_binary}
    \textsc{Binary Implied Integer Recognition} is $\cplxcoNP$-complete.
\end{theorem}

\begin{proof}
  Since the reduction by Ding, Feng and Zhang in \cite{Ding2008} uses a problem with only binary variables, $\cplxcoNP$-hardness follows directly.

  Now, let us show that there exists a NO-certificate that can be verified in polynomial time.
  \Cref{thm_implied_integer_binary} shows that it is sufficient to provide a vector $\bar{x} \in \Z^S$ such that the fiber $Q_S(\bar{x}) = \{ (\bar{x},y) \in [0,1]^S \times \R^{N \setminus S} \mid A_{M,N\setminus S} y \leq b - A_{M,S} \bar{x} \}$ is not $T$-integral.
  In particular, \cref{thm_implied_integrality} shows that it is sufficient to exhibit a minimal face $F$ of $Q_S(\bar{x})$ that is not $T$-integral.
  Since $F$ is a minimal face there exists a subset $M'\subseteq M$ of rows such that $F = \{ (\bar{x},y) \in \R^N \mid A_{M',N\setminus S} y = b_{M'} - A_{M',S} \bar{x} \}$.
  Now \cref{thm_mixed_integer_system_polytime} allows us to verify in polynomial time that $F$ does not contain a $T$-integral point.
  Thus, $\bar{x}$ and $M'$ form a valid NO-certificate, that can be checked in polynomial time.
  Clearly, $\bar{x}$ and $M'$ both have polynomial encoding length. 
\end{proof}

While we were able to treat the binary case, our certificate does not work for the general integer case. However, we establish containment in $\Pi_2^p$, and conjecture hardness. For more details on the complexity class $\Pi_2^p$, we refer to~\cite{Garey1979}.

\begin{theorem}
  \label{thm_complexity_hierarchy}
  \textsc{Implied Integer Recognition} is in $\Pi_2^p$.
\end{theorem}

\begin{conjecture}
  \label{conj_complexity}
  \textsc{Implied Integer Recognition} is $\Pi_2^p$-complete.
\end{conjecture}

\begin{proof}[Proof of \cref{thm_complexity_hierarchy}.]
  By $\left< \cdot \right>$ we denote the (binary) encoding length of numbers, vectors and matrices.
  We consider input matrix $A \in \Q^{M \times N}$ and right-hand side vector $b \in \R^M$ defining a polyhedron $P \coloneqq \{ x \in \R^N \mid Ax \leq b \}$ as well as index sets $S,T \subseteq N$.
  Clearly, the relevant polyhedra
  \begin{equation*}
    R \coloneqq \conv( P \cap \M^S )
    \qquad \text{ and } \qquad 
    Q \coloneqq \conv( P \cap \M^{S \cup T} )
  \end{equation*}
  always satisfy $R \supseteq Q$.
  Let $k \coloneqq |N|+1$.
  We prove containment of \textsc{Implied Integer Recognition} in $\Pi_2^p$ by showing that there exist  polynomials $p,q : \Znonneg \to \Znonneg$ (independent of the input) such that $R \subseteq Q$ holds if and only if
  \begin{subequations}
    \label{eq_hierarchy}
    \begin{align}
    & \forall x^\star \in P \cap \M^S \cap \Q_p^N ~\exists \lambda \in \Q_q^k, \exists x^{(1)},x^{(2)},\dotsc,x^{(k)} \in \Q_q^N : \label{eq_hierarchy_quantifiers} \\
    & x^\star = \sum_{i=1}^k \lambda_i x^{(i)},~
    \sum_{i=1}^k \lambda_i = 1,~ 
    \lambda_i \geq 0,~ x^{(i)} \in P \cap \M^{S \cup T}~ i=1,2,\dotsc,k \label{eq_hierarchy_conmbination}
    \end{align}
  \end{subequations}
  where $\Q_p$ and $\Q_q$ are the sets of all rational numbers of encoding length bounded by $p(\left<A,b\right>)$ and $q(\left<A,b\right>)$, respectively.

  By Meyer's theorem~\cite{Meyer74} there exists a polynomial $p$ such that, whenever $R \not\subseteq Q$ holds then also $P \cap \M^S \cap \Q^N_p \not\subseteq Q \cap \Q^N_p$ holds.
  Hence, if~\eqref{eq_hierarchy} holds then $P \cap \M^S \cap \Q^N_p \subseteq Q \cap \Q^N_p$ holds, and thus also $P \cap \M^S \subseteq Q$, which yields $R \subseteq Q$.

  Hence, it remains to prove that $R \subseteq Q$ implies~\eqref{eq_hierarchy}.
  To this end, consider any $x^\star \in P \cap \M^{S} \cap \Q_p^N$.
  Again by Meyer's theorem there exist finite sets $X \subseteq \M^{S \cup T} \cap \Q_{q'}^N$ and $Y \subseteq \Z^N \cap \Q_{q'}^N$ such that $Q = \conv(X) + \cone(Y)$ for some polynomial $q'$, i.e., such that each vector $v \in X \cup Y$ has encoding length at most $q'(\left<A,b\right>)$.
  From $x^\star \in R \subseteq Q$ it follows that $x^\star \in Q$.
  Let $\tau \in \Znonneg$ be the smallest integer such that $x^\star \in \conv(X) + \sum_{y \in Y} \conv(\{\zerovec, \tau \cdot y \})$.
  Clearly $\left< \tau \right>$ is bounded by a polynomial in $x^\star$ and $\left<A,b\right>$, and hence by $q'(\left<A,b\right>)$ only for a suitable polynomial $q'$ (that is independent of the input) due to $x^\star \in \Q_p^N$.
  Consider the set of points
  \[
    \bar{X} \coloneqq \big( \conv(X) + \sum_{y \in Y} [0,\tau] y \big) \cap \M^{S \cup T},
  \]
  which consists of only points whose encoding length is in turn bounded by a polynomial in $q'(\left<A,b\right>)$.
  Moreover, the choice of $\tau$ implies $x^\star \in \conv(\bar{X})$.
  It follows from Carath{\'e}odory's theorem~\cite{Caratheodory11} that there is a subset of points $x^{(1)},x^{(2)},\dotsc,x^{(k)} \in \bar{X}$ and multipliers $\lambda_1,\lambda_2, \dotsc, \lambda_k \in \Q$ satisfying~\eqref{eq_hierarchy_conmbination}.
  Standard encoding length bounds yield existence of a polynomial $q$ that is independent of $A$, $b$ and $x^\star$ (but depends on $q'$) such that $x^{(i)} \in \Q^N_q$ and $\lambda_i \in \Q_q$ for $i=1,2,\dotsc,k$, which concludes the proof.
  %
  %
  %
  %
  %
\end{proof}

\subsection{Maximizing the number of implied integers}

A natural question related to \cref{conj_complexity} is that of a minimal set of integer variables that yields integrality of all other variables.
In other words, we aim to find a large set of implied integers.

\bigskip

\textsc{Least Integer Hull:}
Given rational matrix $A \in \Q^{M \times N}$ and a rational vector $b \in \Q^M$ that define a polyhedron $P = \{ x \in \R^N \mid Ax \leq b \}$, and a nonnegative integer $k \in \Znonneg$, determine whether there exists a set $X \subseteq N$ with $|X| \leq k$ such that $\conv(P \cap \M^X) = \conv( P \cap \Z^N)$.

\bigskip

\textsc{Least Integer Hull} with $k = 0$ is equivalent to deciding whether $P$ is integral, and hence generalizes \textsc{Integrality Recognition}.
Therefore it must be at least $\cplxcoNP$-hard \cite{Papadimitriou1990}.
Paat, Schl{\"o}ter and Weismantel discuss the integrality number~\cite{Paat2022}, which is the minimum number of integer variables needed in an extended formulation.
\textsc{Least Integer Hull} is somewhat simpler, as it does not allow changing the relaxation.
\textsc{Least Integer Hull} has also been studied under the name \emph{disjunctive index} by Aguilera, Escalante and Nasini~\cite{Aguilera2002} who consider the problem for the clique relaxation of the independent set problem in graphs.
They show using antiblocking duality that the disjunctive index of a the clique relaxation for a given graph is equal to the disjunctive index of the complement graph.
We now show that \textsc{Least Integer Hull} is also $\cplxNP$-hard by reducing \textsc{Bipartite Deletion} to it.

\bigskip

\textsc{Bipartite Deletion}:
Given a graph $G = (V, E)$ and $k \in \Znonneg$, does there exist a set $S$ with $|S| \leq k$ such that $G[V \setminus S]$ is bipartite?

\bigskip

Lewis and Yannakakis~\cite{Lewis1980} have shown that \textsc{Bipartite Deletion} is $\cplxNP$-complete.
For a graph $G = (V, E)$, let $P_G \coloneqq \{x \in [0,1]^V \mid x_i + x_j \leq 1 ~ \forall \{i,j\} \in E \}$ denote the edge relaxation of the independent set problem.
For the latter, we exploit the well-known integrality characterization:

\begin{proposition}
  \label{thm_bipartite_integral}
  $P_G$ is integral if and only if $G$ is bipartite.
\end{proposition}

\begin{proof}
  If $G$ is bipartite, it can be shown using the Ghoulia-Houri criterion \cite{Ghouila-Houri1962} that $P_G$ is integral (see \cite[Chapter 19]{Schrijver86}).
  If $G$ is not bipartite, then one can consider any odd cycle of $G$ and show that it induces a fractional vertex in $P_G$. 
\end{proof}

\begin{theorem}
  \textsc{Least Integer Hull} is $\cplxNP$-hard.
\end{theorem}

\begin{proof}
  We show $\cplxNP$-hardness by reducing \textsc{Bipartite Deletion} to \textsc{Least Integer Hull}.
  Let $G = (V, E)$ be the given graph, and consider its edge relaxation $P_G$.
  We claim that, for any $S \subseteq V$, $\conv(P_G \cap \M^{V \setminus S}) = \conv(P_G \cap \Z^V)$ holds if and only if the induced subgraph $G[S]$ is bipartite.

  First, we assume that $\conv(P_G \cap \M^{V \setminus S}) = \conv(P_G \cap \Z^V)$ holds and show that $G[S]$ is bipartite.
  By \cref{thm_implied_integer_binary}, $Q_{V\setminus S}(\bar{x})$ is integral for all $\bar{x} \in \{0,1\}^{V \setminus S}$.
  For $\bar{x} = \zerovec$, consider $Q_{V\setminus S}(\zerovec) = \{x \in [0,1]^V \mid A_S x_{S} \leq \onevec, ~ x_{V \setminus S} = \zerovec\}$.
  The constraints $x_u + x_v \leq 1$ for any edges $\{u,v\} \in E$ with $u \in S$ or $v \in S$ are redundant for $Q_{V\setminus S}(\zerovec)$ since the inequalities $x_v \leq 1$ are already implied.
  Thus, $Q_{V\setminus S}(\zerovec) = \{(0,x_S) \mid x_S \in P_{G[S]}\}$ holds.
  Since $Q_{V\setminus S}(\zerovec)$ is integral, $P_{G[S]}$ is also integral.
  Then \cref{thm_bipartite_integral} shows bipartiteness of $G[S]$.

  For the reverse direction, we assume that $G[S]$ is bipartite and show that $\conv(P_G\cap \M^{V\setminus S}) = \conv(P_G\cap \Z^V)$ holds.
  By \cref{thm_implied_integer_binary}, it suffices to show that $Q_{V\setminus S}(\bar{x})$ is integral for each $\bar{x}\in \{0,1\}^{V \setminus S}$. 
  Consider any $\bar{x} \in \{0,1\}^{V \setminus S}$ and let $X \coloneqq \{v \in V \setminus S \mid \bar{x}_v = 1\}$.
  Note that if $X$ is not an independent set of $G[V \setminus S]$ then $Q_{V\setminus S}(\bar{x}) = \emptyset$ holds because some edge constraint is violated. 
  Otherwise, let $Y$ denote the set of nodes that are adjacent to some node in $X$.
  By the edge constraints of $P_G$, we must have $x_Y = \zerovec$ for any feasible $x\in Q_{V\setminus S}(\bar{x})$.
  This also implies that any edge constraint that does not have both end nodes in $S \setminus Y$ becomes redundant, as it is either already feasible by choice of feasible $\bar{x}$, or has some node in $Y$ in it, which makes it redundant due to $x_Y = \zerovec$.
  Thus, we can write
  $Q_{V\setminus S}(\bar{x}) = \{(\bar{x},\zerovec_{S\cap Y},x_{S\setminus Y}) \mid x_{S\setminus Y} \in P_{G[S\setminus Y]}\}$.
  Since $G[S \setminus Y]$ is an induced subgraph of $G[S]$, it is bipartite.
  \Cref{thm_bipartite_integral} shows that $P_{G[S\setminus Y]}$ is integral, and it follows that $Q_{V\setminus S}(\bar{x})$ is integral.
  Since this holds for any $\bar{x}\in \{0,1\}^{V\setminus S}$, \cref{thm_implied_integer_binary} shows $\conv( P \cap \M^{N \setminus S}) = \conv(P \cap \Z^V)$.

  We conclude that a subset $S \subseteq V$ of vertices induces a bipartite subgraph of $G$ if and only if $\conv(P_G \cap \M^S) = \conv(P_G\cap \Z^V)$ holds, which completes the reduction.
\end{proof}

Since we have shown that \textsc{Least Integer Hull} is both $\cplxNP$-hard and $\cplxcoNP$-hard, it is natural to conjecture that it lies higher in the polynomial hierarchy due to the common belief that $\cplxcoNP \neq \cplxNP$ holds.

\subsection{Combining multiple implied integrality relations}

Nearly all the results in this work concern totally implied integers, which can be combined to derive implied integrality of larger sets of variables.
In practice, we only have limited time to derive total implied integrality, which motivates us to consider the setting where a small number $n$ of these relations are given.
A natural question is whether a certain relation can easily be concluded from a set of such relations.
More precisely, we consider the following problem:

\bigskip

\textsc{Totally Implied Integer Combination}:
Given a ground set $N$, a finite family $(S_i,T_i)_{i\in I}$ of subset pairs ($S_i,T_i \subseteq N$) and $k \in \Znonneg$, determine whether there exists a subset $X \subseteq N$ such that $|X| \leq k$ and $\conv(P \cap \M^X) = \conv( P \cap \Z^N)$ holds for each rational polyhedron $P \subseteq \R^N$ for which $(S_i,T_i)_{i \in I}$ are totally-implied-integrality relations.

\bigskip

Note that in \textsc{Totally Implied Integer Combination}, the validation of the totally-implied-integer relations is not part of the problem because we already determined that doing so is at least $\cplxcoNP$-hard. This is well motivated by the fact that several methods that we use to detect total implied integrality relations can be verified in polynomial time, such as \cref{thm_tu_implied,thm_unimodular_implied_fixed}. Instead, we focus on the difficulty of combining the given totally-implied-integrality relations.\\

First, we will show that $\textsc{Totally Implied Integer Combination}$ is in $\cplxNP$. For a given instance, we consider a certificate given by $X\subseteq N$ and an ordering $(i_1,i_2,\dotsc,i_{|I|})$ of the elements of $I$. Then, we require that the following conditions hold.
  \begin{subequations}
    \label{eq_totally_i_i_combination_ordering}
    \begin{alignat}{7}
      S_{i_n} &\subseteq X \cup \bigcup_{j=1}^{n-1} T_{i_j}  
      &\quad& \forall  1 \leq n \leq |I| \label{eq_totally_i_i_combination_ordering_1} \\
      N & = X \cup \bigcup_{i\in I} T_i \label{eq_totally_i_i_combination_ordering_2}
    \end{alignat}
  \end{subequations}

\begin{lemma}
    \label{thm_totally_i_i_c_ordering_property}
    Given an instance of \textsc{Totally Implied Integer Combination} and a given set $X\subseteq N$,
    suppose that \eqref{eq_totally_i_i_combination_ordering_2} holds and that there does not exist an ordering of $I$ such that \eqref{eq_totally_i_i_combination_ordering_1} holds for all $1\leq n \leq |I|$. Then, there exists a specific ordering of $I$ and integer $m\in[0,|I|-1]$ such that \eqref{eq_totally_i_i_combination_ordering_1} holds for all $n\leq m$ and \eqref{eq_totally_i_i_combination_ordering_1} does not hold for all $n > m$. 
\end{lemma}
\begin{proof}
    We consider a proof by contrapositive, by assuming that there does not exist an ordering of $I$ and integer $m\in[0,|I|-1]$ such that \eqref{eq_totally_i_i_combination_ordering_1} holds for all $n\leq m$ and \eqref{eq_totally_i_i_combination_ordering_1} does not hold for all $n > m$. Then, for $p\in[1,|I|]$, we inductively proof that
    \begin{equation}
    \label{eq_induc_hypoth}
    \text{there exists an ordering of $I$ such that \eqref{eq_totally_i_i_combination_ordering_1} holds for all $1\leq n\leq p$.}
    \end{equation}

    First, consider the case $p=1$. By our assumption for $m=0$, there must exist some $i\in I$ such that $S_i \subseteq X$ holds. Then, any ordering that places $i$ first shows that \eqref{eq_induc_hypoth} holds for $p=1$. 
    Second, consider the case where \eqref{eq_induc_hypoth} holds for all $p=q$, where $q < |I|$. Let $\hat{i}_1,\dotsc,\hat{i}_n$ be any ordering of $I$ such that \eqref{eq_totally_i_i_combination_ordering_1} holds for all $1\leq n \leq q$. By our assumption, there must exist some integer $s$ such that $S_{\hat{i}_s} \subseteq X \cup \bigcup_{j=1}^q T_{\hat{i}_j}$ holds. Then, we construct a new ordering of $I$ by swapping $\hat{i}_s$ with $\hat{i}_{q+1}$. Since $S_{\hat{i}_s} \subseteq X \cup \bigcup_{j=1}^q T_{\hat{i}_j}$ holds for $q+1$ and we did not change the ordering of the first $q$ elements, this new ordering satisfies \eqref{eq_induc_hypoth} for $p=q+1$, completing the induction.

    Since \eqref{eq_induc_hypoth} holds for $p=|I|$, this shows that \eqref{eq_totally_i_i_combination_ordering} holds for all $1\leq n \leq |I|$, which completes our proof by contrapositive.
\end{proof}

In order to establish that \textsc{Totally Implied Integer Combination} is in $\cplxNP$, we will need a construction for a family of simple polyhedra with known totally-implied-integrality relations.

\begin{proposition}
    \label{thm_totally_i_i_np_example}
    For any $X\subseteq N$, define the polyhedron $P_X = \{(\zerovec_X,\frac{1}{2}_{N\setminus X})\}$ that is given by a single point with fractional coordinates only on $N\setminus X$. Then,  $(S,T)$ is a valid totally implied integer relation  for $P_X$ if and only if at least one of $S\not\subseteq X$ or $T\subseteq X$ holds.
\end{proposition}
\begin{proof}
  First, consider the case where $S\not\subseteq X$. Then, since $S$ contains some element of $N\setminus S$, any $S' \supseteq S$ satisfies $\conv(P_H\cap \M^{S'}) = \emptyset = \conv(P_H\cap \M^{S' \cup T})$, which shows that $T$ is totally implied integer by $S$.
  Second, consider the case $S \subseteq X$ and $T \subseteq X$, and let $S'\supseteq S$.
  If $S'\subseteq X$, then $\conv(P_X\cap \M^{S'}) = P_X = \conv(P_X\cap \M^X) = \conv(P_X\cap \M^{S'\cup T})$ holds since $S'\cup T\subseteq X$.
  Otherwise if $S'\not\subseteq X$, we again have that $\conv(P_X\cap \M^{S'}) = \emptyset = \conv(P_X\cap \M^{S' \cup T})$.
  Finally, if $S\subseteq X$ and $T\setminus X \neq \emptyset$, then $T$ is not (totally) implied integer by $S$ because $\conv(P_X\cap \M^S) \neq \conv(P_X\cap \M^{S\cup T})$ holds, since the former is equal to $P_X$ and the latter is the empty set.
\end{proof}

\begin{lemma}
  \label{thm_totally_i_i_combination_contained}
  \textsc{Totally Implied Integer Combination} is in $\cplxNP$.
\end{lemma}

\begin{proof}
  As a YES-certificate, we consider a given set $X\subseteq N$ and an ordering of the elements in $I$ that satisfies~\eqref{eq_totally_i_i_combination_ordering}.
  Suppose an ordering that satisfies~\eqref{eq_totally_i_i_combination_ordering} exists and consider any rational polyhedron $P \subseteq \R^N$ for which $(S_i,T_i)$ for $i=1,2,\dotsc,n$ are totally-implied-integrality relations.
  We obtain
  \begin{equation*}
    \conv(P\cap \M^X)
    \underset{j=1}{\overset{\eqref{eq_totally_i_i_combination_ordering_1}}{=}}
    \conv(P \cap \M^{X\cup T_1}) 
    \underset{j=2}{\overset{\eqref{eq_totally_i_i_combination_ordering_1}}{=}} \dots 
    \underset{j=n}{\overset{\eqref{eq_totally_i_i_combination_ordering_1}}{=}} \conv(P\cap \M^{X\cup (\bigcup_{i=1}^n T_i)}) 
    \overset{\eqref{eq_totally_i_i_combination_ordering_2}}{=}
    \conv(P\cap \Z^N).
  \end{equation*}

  Now suppose that such an ordering does not exist.
  We show that there exists a polyhedron satisfying the totally-implied-integer relations, but for which $\conv(P\cap \M^X) \neq \conv(P\cap \Z^N)$ holds.

  First, we consider the case in which an ordering does not exist because \eqref{eq_totally_i_i_combination_ordering_2} does not hold, which implies that $X'\coloneqq X\cup (\bigcup_{i\in I} T_i)\neq N$ holds. Then define $P_{X'}$ as in \cref{thm_totally_i_i_np_example}, and note that \cref{thm_totally_i_i_np_example} shows that $P_{X'}$ satisfies all the totally implied integer relations, since we have $T_i\subseteq X'$ for all $i\in I$. However, since $X\subseteq X'$ and $X'\neq N$ hold, it follows that $N\setminus X'\neq\emptyset$, which implies that $P_{X'} = \conv(P_{X'}\cap \M^X) \neq \conv(P_{X'}\cap \Z^N) = \emptyset$.
  Thus, we can assume that every certificate satisfies \eqref{eq_totally_i_i_combination_ordering_2}.
  
  Then, \cref{thm_totally_i_i_c_ordering_property} implies that there exists an ordering $\{1,2,\dots,n\}$ of $I$ such that for $X'\coloneqq $ $S_j\subseteq X \cup (\bigcup_{i=1}^{j-1} T_i)$ holds for all $j\leq k$ for some $k < n$ but that for all $j > k$, $S_j \not\subseteq X \cup (\bigcup_{i=1}^k T_i)$ holds (Note that $k = 0$ is allowed).
  Now, let $X' \coloneqq X\cup (\bigcup_{i=1}^j T_i)$, and consider $P_{X'}$ as defined in \cref{thm_totally_i_i_np_example}.
  Clearly, since there exists some $S_j$ that is not a subset of $X'$, we must have that $X'\neq N$.
  By the above, all $S_i$ that satisfy $S_i\subseteq X'$, which are precisely those with $i\leq k$, also satisfy $T_i \subseteq X'$ by definition of $X'$.
  Thus, $P_{X'}$ models all the given totally implied integer relations correctly.
  However, $P_{X'} = \conv(P_{X'}\cap \M^X) \neq \conv(P_{X'}\cap \Z^N) = \emptyset$ holds since $X'\neq N$ holds.
  Thus, if no ordering certificate exists, $\conv(P\cap \M^X) = \conv(P\cap \Z^N)$ does not hold for every polyhedron $P$ for which the given totally implied integer relations are valid. 

  Thus, $\conv(P\cap \M^X) = \conv(P\cap \Z^N)$ holds for every rational polyhedron $P\subseteq \R^N$ if and only if an ordering $I=\{1,2,\dots,n\}$ exists such that $S_j\subseteq X\cup (\bigcup_{i=1}^{j-1} T_i)$ for all $j=1,\dotsc,n$ and $X\cup (\bigcup_{i\in I} T_i) = N$ holds. Moreover, the certificate is clearly of polynomial size and can be verified in polynomial time, which concludes the proof.
\end{proof}

To show $\cplxNP$-hardness, we reduce \textsc{Hitting Set} to \textsc{Totally Implied Integer Combination}.
The former is is well-known to be $\cplxNP$-complete~\cite{Karp1972}.

\medskip
    
\textsc{Hitting Set}:
Given the universal (finite) set $U$, an integer $1 \leq k < |U|$, and subsets $S_i \subseteq U$ for $i = 1,2,\dotsc,m$, determine if there exists a set $C \subseteq U$ with $|C|\leq k$ such that $C\cap S_i \neq\emptyset$ for all $i = 1,2,\dotsc,m$.
    
\medskip

\begin{lemma}
  \label{thm_totally_i_i_combination_hard}
  \textsc{Totally Implied Integer Combination} is $\cplxNP$-hard.
\end{lemma}

\begin{proof}
  We augment a hitting set instance that is given by the sets $S_i$ with $|U|$ copies of $U$, so that $S_{j} = U$ for all $j=m+1,m+2,\dotsc m+|U|$.
  Clearly, these sets are trivially satisfied in any feasible instance and only serve to increase the number of sets, and do not change the set of hitting sets.
  We use $M = \{1,2,\dotsc,m+|U|\}$ to denote the the set of sets to be hit.
  For any hitting set instance as given above, we define a totally implied integer combination instance as follows.
  Let $U \times M$ be the ambient space, and let the totally implied integer relations be given by  
  $(u,i)$ for all $u\in S_i$,$i\in M$, such that each element $u$ totally implies each set $S_i$ that contains $u$.
  Additionally, we add the totally implied integer relation $(M,u)$ for all $u\in U$.
  We use $(V_i,W_i)$ to denote the union of these totally implied relations.
  It can be easily verified that the constructed Totally Implied Integer Combination instance has a polynomial size in terms of the original hitting set instance.

  We claim that there exists some set $X \subseteq U\times M$ with $|X| \leq k$ such that $\conv(P\cap \M^X) = \conv(P\cap (\Z^U\times \Z^M))$ holds for each rational polyhedron for which $(V_i,W_i)$ are if and only if there exists a hitting set $C \subseteq U$ with $|C| \leq k$.

  First, we assume that there exists a hitting set $C\subseteq U$ with $|C| \leq k$, and show that such a set $X$ exists.  
  Since $C$ is a hitting set, each set $S_i$ has at least one element from $C$ in it.
  Then, by the totally implied integer relations $(u,i)$ for all $u\in S_i$, $i\in M$, show that we must have $\conv(P\cap (\M^C \times \R^U)) = \conv(P\cap (\M^C \times \Z^U))$ (apply in any order).
  Subsequently applying the totally implied integer relations $(M,u)$, we find that $\conv(P\cap (\M^C \times \Z^M)) = \conv(P\cap (\Z^C\times \Z^M))$ holds.
  Then, since $|C| \leq k$, $\conv(P\cap (\M^C \times \R^U)) = \conv(P\cap (\Z^C \times \Z^U))$ holds by the totally implied integer relations and $C$ attains the conditions (for $X$) for \textsc{Totally Implied Integer Combination}.

  Second, we assume that there exists $X\subseteq U \times M$  such that $\conv(P\cap \M^X) = \conv(P\cap (\Z^U\times \Z^N))$ and $|X| \leq k$ hold, and show that there exists a hitting set $X'\subseteq U$ where $|X'|\leq k$ holds.
  Let $X_U = X\cap U$ and let $X_M = X\cap M$. Note that since any set $i\in X_M$ is implied integer by any of its elements, we can substitute it for one of its elements $u\in S_i$ and add it to $X_U$, since $i$ is directly implied integer by $u$.
  Starting with $X_U$ and performing this procedure for each $i\in M$, we obtain a set $X'$ such that $X'\subseteq U$, $|X'| \leq |X| \leq k$ and $\conv(P\cap (\M^{X'} \times \R^M))$ holds.
  Since $|U| \leq |M| = m + |U|$ holds, such a procedure always works and can never increase the size of $X'$.
  We claim that $X'$ is a hitting set.
  Suppose that $X'\cap S_i = \emptyset$ holds for some $S_i$, where $i=1,\dotsc,m+|U|$.
  In particular, this implies that none of the totally implied integer relations $(u,i)$ can be used initially to derive implied integrality of $i$.
  Since all in-arcs of $i$ are of this form, we must then show that some $u \in S_i$ is implied integer by $X'$ through an alternative route.
  However, since every $u\in U$ is implied integer only by the in-arc $(M,u)$, and $M$ is clearly not integer because $i\in M$ is not initially integer, there is no way to derive integrality of $i$ using totally implied integer relations, which contradicts that $\conv(P \cap (\M^{X'} \times \R^M)) = \conv(P \cap (\Z^U \times \Z^M))$ holds.
  Thus, $X' \cap S_i \neq \emptyset$ holds for all $i = 1,2,\dotsc,m+|U|$, which shows that $X'$ is a hitting set of the augmented and original instance with $|X'| \leq k$.
\end{proof}

The combination of \cref{thm_totally_i_i_combination_contained,thm_totally_i_i_combination_hard} yields the following result.

\begin{theorem}
  \label{thm_totally_i_i_combination}
  \textsc{Totally Implied Integer Combination} is $\cplxNP$-complete.
\end{theorem}

\subsection{Recognizing implied fixed integers}

Finally, we consider the problem of determining whether a vector is implied fixed integer (see \cref{sec_implied_fixed}).

\bigskip 

\textsc{Implied Fixed Integrality}:
Given a rational matrix $A \in \Q^{M \times N}$ and a rational vector $b \in \Q^M$ that define a polyhedron $P = \{ x \in \R^N \mid Ax \leq b \}$, a set $S \subseteq N$ and a rational vector $v \in \Q^N$ determine whether $v \in \Psi_S(P)$ holds.

\bigskip

\begin{lemma}
  \label{thm_implied_fixed_complexity_hard}
  \textsc{Implied Fixed Integrality} is $\cplxcoNP$-hard.
\end{lemma}

\begin{proof}
  We reduce the problem of determining infeasibility of an integer linear program to \textsc{Implied Fixed Integrality}.
  It is well known that the problem of determining if an integer linear program given by a rational polyhedron $P = \{ x \in \R^S \mid A x \leq b\}$ where $A \in \Q^{M \times S}$, $b \in \Q^M$ together with integrality constraints is non-empty is $\cplxNP$-complete~\cite{Cook1971}.
  Thus, the complement, which asks whether $\conv(P \cap \Z^S)$ is empty, is $\cplxcoNP$-complete.

  To describe our reduction, we first add an additional binary variable binary $y$ to $P$ to obtain $P' = \{(x,y) \in \R^S \times \R \mid A x \leq b,~ 0 \leq y \leq 1\}$.
  It is clear that $\conv(P \cap \Z^S) = \emptyset$ holds if and only if $\conv(P' \cap (\Z^S \times \R)) = \emptyset$.
  Then we consider $\Psi_S(P')$ and claim that the scaled unit vector $\frac{1}{2} \unitvec{y}$ lies in $\Psi_N(P')$ if and only if $\conv(P'\cap (\Z^S \times \R)) = \emptyset$ holds. 

  If $\conv(P' \cap (\Z^S \times \R)) = \emptyset$, then $\Psi_N(P') = \R^S \times \R$ holds, as any equation is valid for the empty polyhedron, and thus, also for its empty fibers.
  In particular, this shows that $\frac{1}{2} \unitvec{y} \in \Psi_S(P')$ holds. 
  Otherwise, if  $\conv(P'\cap (\Z^S \times \R)) \neq \emptyset$, there exists some feasible solution $\bar{x}\in \Z^S$ such that the fiber $Q'_N(\bar{x})$ of $P'$ is non-empty.
  In particular, it is easy to verify that this fiber is given by $\{\bar{x}\} \times [0,1]$.
  Clearly, such a fiber cannot satisfy an equation of the form $\frac{1}{2} \unitvec{y} = \alpha(\bar{x})$, since the $y$-variable is not fixed in the fiber.
  This shows that $\frac{1}{2} \unitvec{y} \notin \Psi_N(P')$ holds.
  From the above arguments, we can conclude that $\frac{1}{2} \unitvec{y} \in \Psi_N(P')$ holds if and only if $\conv(P\cap \Z^S)$ is empty.
  Hardness of determining the latter concludes the proof.
\end{proof}

We also conjecture containment in $\cplxcoNP$ and therefore $\cplxcoNP$-completeness, but it turns out to be not so easy to prove.

\begin{conjecture}
  \textsc{Implied Fixed Integrality} is $\cplxcoNP$-complete.
\end{conjecture}

Let us discuss a potential NO-certificate.
By definition of $\Psi_S(P)$, $v\notin \Psi_S(P)$ holds if and only if there exists some fiber $\bar{x}\in\Z^S$ such that $\transpose{v} x$ is not an integral constant for all $x\in Q_N(\bar{x})$.
The latter holds if and only if $\max(\transpose{v} x \mid x \in Q(\bar{x})\}$ and $\min(\transpose{v} x \mid x\in Q(\bar{x})\}$ are different or they are equal to the same fractional value.
However, it is not clear why, if there exists such a fiber, there must exist one for which $\bar{x}$ has polynomial encoding length.

\section{Implied integrality in practice}
\label{sec_computations}

\noindent
Although the presented concepts are of theoretical interest, they are also practically relevant, which is highlighted in this section.
In light of \cref{thm_tu_implied} it is tempting to search for totally unimodular matrices, which can be recognized in polynomial time~\cite{Truemper1990}, however, with significant computational cost~\cite{Walter2013}.
Hence, we restrict our detection to the special case of network matrices and their transposes (see~\cite[Chapter 19]{Schrijver86}).
To this end we use adjusted versions of the algorithms developed in \cite{BixbyWagner1988} and \cite{VanderHulst2024} to greedily grow network submatrices one column or one row at a time.
This way, we detect submatrices with the structure of \cref{thm_tu_implied}.

\subsection{Implemented algorithm}

The implemented algorithm is described in \cref{algo_implied_tu}.
It relies on a procedure $\mathrm{GrowNetwork}(A, T, j)$ for checking for a (transposed) network submatrix $A_{\star,T}$ if the addition of the column $j$ maintains the property, i.e., that $A_{\star, T\cup \{j\}}$ is a (transposed) network matrix.
For network matrices, we use the fast column-wise augmentation algorithm by Bixby and Wagner~\cite{BixbyWagner1988}.
For transposed network matrices, we use the row-wise network matrix augmentation algorithm by van der Hulst and Walter~\cite{VanderHulst2024}.
For each instance, \cref{algo_implied_tu} is run twice, once with network matrix detection and once with transposed network matrix detection.
From these two runs, the totally unimodular submatrix with the largest $|T|$ is used.

\begin{algorithm}[h]
    \caption{Detecting implied integers using \cref{thm_tu_implied}}
    \label{algo_implied_tu}

    \LinesNumbered
    \TitleOfAlgo{FindImpliedIntegers$(A,b,I)$}

  \KwIn{Constraint matrix $A\in \R^{M\times N}$, right hand side $b\in \R^M,$ integer variables $I\subseteq N$}
  \KwOut{Implied integer relation $(S,T)$}
    \SetKw{Continue}{continue}
    \SetKw{Break}{break}
    $T\gets \emptyset$\;
    $R\gets \{i\in M | \, b_i \notin \mathbb{Z}$ or $a_{ij}\notin\mathbb{Z}$ for some $j\in N\}$\;
    Let $\mathcal{K}$ be the set of connected components of $A_{M,N\setminus I}$\;
    \For{$K\in\mathcal{K}$}{
        Let $T_K\gets \emptyset$\;
        \If{$R\cap M_K\neq \emptyset$}{
            $R\gets R \cup M_K$\;
            \Continue
        }
        \For{$j\in N_K$}{
            \lIf{not $\mathrm{GrowNetwork}(A, T\cup T_K,j)$}{
                \Break
            }
            $T_K\gets T_K\cup\{j\}$\;
        }
        \uIf{$T_K = N_K$}{
         $T \gets T \cup T_K $\;
        }\Else{
            $R\gets R \cup M_K$
        }
    
    }
    $X \gets \{ j\in I \mid \mathrm{supp}(a_{\star,j}) \cap R = \emptyset\}$\;
    \For{$j\in X$}{    \label{algo_implied_tu_integer_start}
       \lIf{$\mathrm{GrowNetwork}(A, T,j)$}{ $T\gets T\cup \{j\}$}
    }
    \label{algo_implied_tu_integer_end}
    $S \gets \{j\in N \setminus T \mid \text{there exist } i\in M, t\in T\text{ such that } a_{ij}\neq 0, a_{it}\neq 0\}$\;
    \Return $S, T$
\end{algorithm}

\cref{algo_implied_tu} consists of three steps. First, it tracks the set of rows $R$ that are incompatible with implied integrality detection, which corresponds to rows in the submatrix $Dx + Ez$ in the statement of \cref{thm_tu_implied}. A row is put in $R$ if it has non-integral coefficients, a non-integral right hand side, or if it has continuous columns that are not implied integer. The algorithm initializes $R$ with the set of rows contain some non-integer nonzero or have a non-integer right-hand side.
By the requirements of \cref{thm_tu_implied}, any variables that have nonzeros in $R$ can not be implied integers.
Note that $R$ also contains rows that are 
non-integral variable bounds.
Second, note that by the structure of \cref{thm_tu_implied}, either all or none of the continuous columns in a row must be implied integer.
Thus, we select all connected components $\mathcal{K}$ of $A_{N \setminus I}$ that form a (transposed) network matrix and do not contain any rows from $R$, and add their columns to $T$.
For each component $K\in \mathcal{K}$, we use $N_K$ and $M_K$ to denote the component its rows and columns.
If a component is not (transposed) network matrix or intersects $R$, we add its rows to $R$.
Third, we greedily add integer columns to $T$ that do not contain rows from $R$.

\subsection{Results}

\noindent As a baseline, we use the default SCIP~9.2.1 implementation \cite{bolusani2024scipoptimizationsuite90}, which detects implied integers using \cref{thm_dual_detection,thm_primal_detection}.
This is compared to a version that, in addition uses a plugin that implements \cref{algo_implied_tu}.
The plugin and its source code will be part of the upcoming SCIP~10 release.
These computations were done on an Intel Gold 5217 CPU with \SI{64}{\giga\byte} of memory.

In \cref{tab:statistics} we report about results on the MIPLIB~2017 collection set~\cite{Gleixner2021}, where 5~models were excluded because they could not be presolved using \SI{64}{\giga\byte} of memory, and 25 further models were excluded because they were solved during presolve with both methods.
We observe that using \cref{algo_implied_tu} one finds nearly six times as many implied integer variables at the cost of only a small detection time.
\Cref{fig_distribution_sorted} shows the distribution of the implied integers over all the instances.
Furthermore, the relationship between the sizes of $S$ and $T$ is highlighted in \Cref{fig_ratios}.

\begin{table}[htpb]
  \caption{%
    Implied integer detection on 1035 instances of the MIPLIB 2017 collection set~\cite{Gleixner2021}.
  }
  \label{tab:statistics}
  \begin{subtable}[t]{1.0\textwidth}
        \begin{center}
    \begin{tabular}{l|r|r}
      Method & SCIP~9.2.1 & SCIP~9.2.1 + \Cref{algo_implied_tu} \\
      \hline
      affected instances & 203 & \textbf{712} \\
      mean of ratio $\frac{|T|}{|N|}$ & \SI{3.3}{\percent} & \textbf{18.8\,\%}\\
      mean of ratio $\frac{|T_{\mathrm{con}}|}{|N|}$& \SI{3.2}{\percent} & \SI{7.5}{\percent} \\
      mean of ratio $\frac{|T_{\mathrm{int}}|}{|N|}$& \SI{0.1}{\percent} & \SI{11.3}{\percent} \\
      mean of ratio $\frac{|N_{\mathrm{con}}\setminus T|}{|N|}$& \SI{26.5}{\percent} & \SI{22.2}{\percent} \\
      mean of ratio $\frac{|N_{\mathrm{int}}\setminus T|}{|N|}$& \SI{70.2}{\percent} & \textbf{59.0\,\%} \\ 
      number of variables & \num{6937} & \num{6944} 
    \end{tabular}
      \end{center}
      \caption{ Basic detection statistics. For a set $X\subseteq N$, $X_{\mathrm{int}}$ ($X_{\mathrm{con}}$) indicates the set of variables in $X$ that (do not) have integrality constraints in the original model. 
    The numbers and ratios reported are based on the presolved model, and averaged over all instances tested. 
    The number of variables reported is the shifted geometric mean of the number of variables with shift 10.
    All other means are arithmetic means.}
  \end{subtable}

  \begin{subtable}[t]{1.0\textwidth}
  \begin{tabular}{c|c|c|c|c|c|c}
        Detection time &  $[0, 1]$\SI{}{\milli\second} & $[1, 10]$\SI{}{\milli\second} & $[10, 100]$\SI{}{\milli\second} & $[100, 1000]$\SI{}{\milli\second} & $[1, 10]$\SI{}{\second} & $> 10 $\SI{}{\second} \\
        \hline
         Number of instances & 232 & 311 & 292 & 148 & 39 & 13 
   \end{tabular}
    \caption{Distribution of the running time of \Cref{algo_implied_tu}. For each time bracket, the corresponding number of instances that fall into it is indicated. }
  \end{subtable}

    \label{tab:time_distribution}
\end{table}

\begin{table}[htpb]
    \centering

\end{table}

\begin{figure}[htpb]
  \begin{subfigure}[t]{0.49\textwidth}
    \begin{tikzpicture}
      \begin{axis}[
        width=\textwidth,
        height=\textwidth,
        xlabel={Number of instances},
        ylabel={$|T| / |N|$},
        grid=major,
        xtick={0,250,500,750,1000},
        ytick={0,0.2,0.4,0.6,0.8,1.0},
        ylabel style={yshift=-1em},
        legend style={font=\footnotesize, legend pos=north west},
        legend cell align=left,
        ]
        \addplot[densely dotted, no markers, ultra thick] table[x=instance, y=old] {found_sorted.dat};
        \addplot[densely dashed, no markers, ultra thick] table[x=instance, y=new] {found_sorted.dat};
        \legend{SCIP, SCIP+\cref{algo_implied_tu}};
      \end{axis}
    \end{tikzpicture}
    \caption{%
      Distribution of ratio of implied \\ integers.
    }
    \label{fig_distribution_sorted}
  \end{subfigure}
\hfill
  \begin{subfigure}[t]{0.49\textwidth}
    \begin{tikzpicture}
      \begin{axis}[
        width=\textwidth,
        height=\textwidth,
        xlabel={$|S| / |N|$},
        ylabel={$|T| / |N|$},
        grid=major,
        xtick={0,0.2,0.4,0.6,0.8,1.0},
        ytick={0,0.2,0.4,0.6,0.8,1.0},
        ylabel style={yshift=-1em},
        legend style={font=\footnotesize, legend pos=north east},
        legend cell align=left,
        ]
        \addplot[only marks, mark size=0.5pt] table[x=x, y=y] {found_ratios.dat};
      \end{axis}
    \end{tikzpicture}
    \caption{%
      The ratios $\frac{|S|}{|N|}$ and $\frac{|T|}{|N|}$ per instance, represented by a dot.
    }
    \label{fig_ratios}
  \end{subfigure}
  \caption{%
    Results on implied integers for the MIPLIB 2017 collection dataset~\cite{Gleixner2021}.
  }
\end{figure}

\cref{tbl:rubberband_table} shows results of preliminary performance experiments. 
In this comparison, we ignore the detected implied integrality for originally integer variables, because preliminary tests showed a negative performance impact for using implied integrality for them. In particular, we skip the loop on lines \ref{algo_implied_tu_integer_start}--\ref{algo_implied_tu_integer_end} in \cref{algo_implied_tu}.
The performance results are inconclusive: both methods use on average the same amount of time and nodes.
Although we are \SI{5}{\percent} faster on instances that take more than \num{1000} seconds to solve by detecting more implied integer variables, this is achieved only on a small sample of 32 instances of the 240 instances tested, and cannot be considered conclusive.
The inconclusive results can be explained by the sensitivity of SCIP to the distribution of (implied) integers.
In fact, we already had to adapt many of its algorithms and parameters to account for the changed variable type distribution in order to obtain these neutral performance results.

\begin{table}[htpb]
  \caption{%
    Performance comparison on the MIPLIB 2017 benchmark set~\cite{Gleixner2021}.
    The subsets indicated by [$x$, tilim] contain all instances solved by at least one of the two runs, that took at least $x$ seconds to solve.
    The time and nodes reported are shifted geometric means with $\SI{1}{\second}$ and 100 nodes, respectively. These experiments  were run on a AMD Ryzen 2600 CPU with a time limit of \SI{3600}{\second} and a memory limit of \SI{12}{\giga\byte}.
  }
  \label{tbl:rubberband_table}
\scriptsize

\begin{tabular*}{\textwidth}{@{}l@{\;\;\extracolsep{\fill}}rrrrrrrrr@{}}
\toprule
&           & \multicolumn{3}{c}{SCIP~9.2.1} & \multicolumn{3}{c}{SCIP~9.2.1 + Alg. 1} & \multicolumn{2}{c}{relative} \\
\cmidrule{3-5} \cmidrule{6-8} \cmidrule{9-10}
Subset                & instances &                                 solved &       time &        nodes &   solved & time & nodes &      time &        nodes \\
\midrule
\cleaninst            &       240 &                                       118 &      730.6 &         3450 &   118 &  733.3 & 3425 &    1.00 &          0.99 \\
\cmidrule{1-10}
\bracket{0}{tilim}    &       120 &                                      118 &      147.6 &         2292 &   118 & 148.7 & 2307 &     1.01 &          1.01 \\
\bracket{1}{tilim}    &       118 &                                      116 &      159.5 &         2388 &   116 & 160.7 & 2405 &     1.01 &          1.01 \\
\bracket{10}{tilim}   &       105 &                                      103 &      246.0 &         3459 &   103 & 248.4 & 3525 &     1.01 &          1.01 \\
\bracket{100}{tilim}  &        73 &                                       71 &      617.5 &         9460 &   71 & 617.4 & 9559 &      1.00 &          1.01 \\
\bracket{1000}{tilim} &        32 &                                       30 &     1768.8 &        33695 &   30 & 1678.7 & 33664 &     0.95 &          1.00 \\
\bottomrule
\end{tabular*}
\end{table}

\newpage
\section{Conclusion and discussion}
\label{sec_conclusion_discussion}

In line with the structure of our work we start with a discussion of our theoretical insights.
One of our main contributions is the reduction of implied integrality to partial integrality of fibers via~\cref{thm_implied_integer_sufficiency,thm_implied_integer_binary}.
This property, however, does not seem to be well-understood, even in contexts where integrality characterizations exist.
We are only aware of one line of research in that direction, namely on strengths of Dantzig-Wolfe reformulations for stable-set polytopes~\cite{LuebbeckeW18}. 

Besides showing (partial) integrality, our insights relate to many other topics in MILP since important techniques like cutting planes, branch-and-bound and domain propagation typically only use the integrality of a subset of the integral variables.
Implied integrality could be used in these contexts to strengthen results that only use partial integrality to larger sets of integer variables.
For example, Balas, Ceria and Cornuejols' lift-and-project procedure~\cite{Balas1993} provides a way to strengthen $P \subseteq [0,1]^N$  to $\conv(P\cap \M^S)$ for some $S \subseteq N$ by adding valid inequalities in a procedure that iterates over the variables $S$.
If $x_T$ is implied integer by $x_S$ then it suffices to apply the lift-and-project procedure only on the $x_S$-variables to obtain $\conv(P\cap \M^{S \cup T})$, rather than on the $x_{S \cup T}$-variables.

\Cref{thm_tu_implied} can be seen as a special case of an affine TU decomposition~\cite{BaderHWZ18} in which one of the matrices is an identity matrix augmented with zero columns.
A natural question is whether more general decompositions can be detected as well.

Furthermore, the proposed framework for the geometry of fixed integers is still somewhat limited in the following sense.
\Cref{thm_fortet_implied} provides an example for which the fixing function $\alpha(\cdot)$ is nonlinear, and we were unable to derive any structural insights regarding detection of implied fixed integers in such cases.
Finally, the proposed detection method for local implied integrality through dual fixing in \cref{thm_local_relevant_sign_only} is currently limited to a single variable.
Additional results that infer integrality of multiple variables in the optimal face using duality are left as an open research question.
Several other questions remain open.
For example, given that $T$ is (totally) implied integer by $S$, it is interesting to ask for the size of the smallest subset $S' \subseteq S$ such that $T$ remains (totally) implied integer by $S'$.
Naturally, the resolution of \cref{conj_complexity} as well as extensions of implied integrality to mixed-integer nonlinear optimization constitute further research directions.

On the practical side we believe that proper usage of the additional implied integrality information in SCIP will eventually yield a performance boost.
This is justified by the impressive result in~\cite{AchterbergBGRW20} where a speed-up of \SI{13}{\percent} using the basic version of implied integrality is reported and by the small detection times that we observe in \cref{tab:statistics}.
Finally, we see a lot of unexplored potential in exploiting implied integrality in the context of decomposition approaches.
In Benders' decomposition, implied integrality may provide a major advantage since continuous variables in a subproblem are much easier to deal with than integer ones.
For Dantzig-Wolfe reformulations, however, constraints that induce integrality properties should not be reformulated as they do not contribute to strengthening of the relaxation.
Hence, we hope that presence of implied integrality can be exploited to lead to better decompositions and better automatic detection of decompositions in the future.

\medskip 
\noindent
\textbf{Acknowledgements.}
Both authors acknowledge funding support from the Dutch Research Council (NWO) on grant number OCENW.M20.151.
We thank three anonymous reviewers for their valuable suggestions and constructive feedback that led to an improved manuscript.

\bibliographystyle{plainurl}
\bibliography{implied-integers}

\begin{thebibliography}{10}

\bibitem{Achterberg2008}
Tobias Achterberg.
\newblock {\em {Constraint Integer Programming}}.
\newblock PhD thesis, TU Berlin, 2008.
\newblock \href {https://doi.org/10.14279/depositonce-1634} {\path{doi:10.14279/depositonce-1634}}.

\bibitem{Achterberg2016}
Tobias Achterberg, Robert~E. Bixby, Zonghao Gu, Edward Rothberg, and Dieter Weninger.
\newblock {Presolve Reductions in Mixed Integer Programming}.
\newblock Technical report, Zuse Institute Berlin, Berlin, 2016.

\bibitem{AchterbergBGRW20}
Tobias Achterberg, Robert~E. Bixby, Zonghao Gu, Edward Rothberg, and Dieter Weninger.
\newblock Presolve reductions in mixed integer programming.
\newblock {\em INFORMS Journal on Computing}, 32(2):473--506, 2020.
\newblock \href {https://doi.org/10.1287/ijoc.2018.0857} {\path{doi:10.1287/ijoc.2018.0857}}.

\bibitem{Aguilera2002}
N{\'{e}}stor~E. Aguilera, Mariana~S. Escalante, and Graciela~L. Nasini.
\newblock {A Generalization of the Perfect Graph Theorem Under the Disjunctive Index}.
\newblock {\em Mathematics of Operations Research}, 27(3):460--469, aug 2002.
\newblock \href {https://doi.org/10.1287/moor.27.3.460.309} {\path{doi:10.1287/moor.27.3.460.309}}.

\bibitem{Andersen1995}
Erling~D. Andersen and Knud~D. Andersen.
\newblock {Presolving in linear programming}.
\newblock {\em Mathematical Programming}, 71(2):221--245, 1995.
\newblock \href {https://doi.org/10.1007/BF01586000} {\path{doi:10.1007/BF01586000}}.

\bibitem{Appa2004}
Gautam Appa and Bal{\'{a}}zs Kotnyek.
\newblock {Rational and integral k-regular matrices}.
\newblock {\em Discrete Mathematics}, 275(1-3):1--15, 2004.
\newblock \href {https://doi.org/10.1016/S0012-365X(03)00095-5} {\path{doi:10.1016/S0012-365X(03)00095-5}}.

\bibitem{BaderHWZ18}
J{\"{o}}rg Bader, Robert Hildebrand, Robert Weismantel, and Rico Zenklusen.
\newblock {Mixed integer reformulations of integer programs and the affine TU-dimension of a matrix}.
\newblock {\em Mathematical Programming}, 169(2):565--584, 2018.
\newblock \href {https://arxiv.org/abs/1508.02940} {\path{arXiv:1508.02940}}, \href {https://doi.org/10.1007/s10107-017-1147-2} {\path{doi:10.1007/s10107-017-1147-2}}.

\bibitem{Balas1985}
Egon Balas.
\newblock {On the facial structure of scheduling polyhedra}.
\newblock {\em Mathematical Programming Study}, 24:179--218, 1985.

\bibitem{Balas1993}
Egon Balas, Sebasti{\'{a}}n Ceria, and G{\'{e}}rard Cornu{\'{e}}jols.
\newblock {A lift-and-project cutting plane algorithm for mixed 0-1 programs}.
\newblock {\em Mathematical Programming}, 58(1-3):295--324, 1993.
\newblock \href {https://doi.org/10.1007/BF01581273} {\path{doi:10.1007/BF01581273}}.

\bibitem{Berge72}
Claude Berge.
\newblock Balanced matrices.
\newblock {\em Mathematical Programming}, 2(1):19--31, 1972.
\newblock \href {https://doi.org/10.1007/BF01584535} {\path{doi:10.1007/BF01584535}}.

\bibitem{BilleraS92}
Louis~J. Billera and Bernd Sturmfels.
\newblock Fiber polytopes.
\newblock {\em Annals of Mathematics}, 135(3):527--549, 1992.
\newblock \href {https://doi.org/10.2307/2946575} {\path{doi:10.2307/2946575}}.

\bibitem{BixbyWagner1988}
Robert~E. Bixby and Donald~K. Wagner.
\newblock {An almost linear-time algorithm for graph realization}.
\newblock {\em Mathematics of Operations research}, 13(1), 1988.
\newblock \href {https://doi.org/10.1287/moor.13.1.99} {\path{doi:10.1287/moor.13.1.99}}.

\bibitem{bolusani2024scipoptimizationsuite90}
Suresh Bolusani, Mathieu Besançon, Ksenia Bestuzheva, Antonia Chmiela, João Dionísio, Tim Donkiewicz, Jasper van Doornmalen, Leon Eifler, Mohammed Ghannam, Ambros Gleixner, Christoph Graczyk, Katrin Halbig, Ivo Hedtke, Alexander Hoen, Christopher Hojny, Rolf van~der Hulst, Dominik Kamp, Thorsten Koch, Kevin Kofler, Jurgen Lentz, Julian Manns, Gioni Mexi, Erik M{\"u}hmer, Marc~E. Pfetsch, Franziska Schlösser, Felipe Serrano, Yuji Shinano, Mark Turner, Stefan Vigerske, Dieter Weninger, and Liding Xu.
\newblock {The SCIP Optimization Suite 9.0}, 2024.
\newblock \href {https://arxiv.org/abs/2402.17702} {\path{arXiv:2402.17702}}.

\bibitem{Brearley1975}
A.~L. Brearley, Gautam Mitra, and H.~Paul Williams.
\newblock {Analysis of mathematical programming problems prior to applying the simplex algorithm}.
\newblock {\em Mathematical Programming}, 8(1):54--83, dec 1975.
\newblock \href {https://doi.org/10.1007/BF01580428} {\path{doi:10.1007/BF01580428}}.

\bibitem{Caratheodory11}
Constantin Carath{\'e}odory.
\newblock {\"U}ber den variabilit{\"a}tsbereich der fourier'schen konstanten von positiven harmonischen funktionen.
\newblock {\em Rendiconti del Circolo Matematico di Palermo (1884--1940)}, 32(1):193--217, 1911.
\newblock \href {https://doi.org/10.1007/BF03014795} {\path{doi:10.1007/BF03014795}}.

\bibitem{ConfortiC95}
Michele Conforti and G{\'e}rard Cornu{\'e}jols.
\newblock Balanced 0, {\textpm}1-matrices, bicoloring and total dual integrality.
\newblock {\em Mathematical Programming}, 71(3):249--258, Dec 1995.
\newblock \href {https://doi.org/10.1007/BF01590956} {\path{doi:10.1007/BF01590956}}.

\bibitem{ConfortiCZ14}
Michele Conforti, G{\'{e}}rard Cornu{\'{e}}jols, and Giacomo Zambelli.
\newblock {\em {Integer Programming}}, volume 271 of {\em Graduate Texts in Mathematics}.
\newblock Springer International Publishing, Cham, 1 edition, 2014.
\newblock \href {https://doi.org/10.1007/978-3-319-11008-0} {\path{doi:10.1007/978-3-319-11008-0}}.

\bibitem{ConfortiCR99}
Michele Conforti, Gérard Cornuéjols, and Mendu~R. Rao.
\newblock Decomposition of balanced matrices.
\newblock {\em Journal of Combinatorial Theory, Series B}, 77(2):292--406, 1999.
\newblock \href {https://doi.org/10.1006/jctb.1999.1932} {\path{doi:10.1006/jctb.1999.1932}}.

\bibitem{Cook1971}
Stephen~A. Cook.
\newblock {The complexity of theorem-proving procedures}.
\newblock In {\em Proceedings of the third annual ACM symposium on Theory of computing - STOC '71}, pages 151--158, New York, New York, USA, 1971. ACM Press.
\newblock URL: \url{http://portal.acm.org/citation.cfm?doid=800157.805047}, \href {https://doi.org/10.1145/800157.805047} {\path{doi:10.1145/800157.805047}}.

\bibitem{Dantzig1954}
George~B. Dantzig, Delbert~Ray Fulkerson, and Selmer Johnson.
\newblock {Solution of a Large-Scale Traveling-Salesman Problem}.
\newblock {\em Journal of the Operations Research Society of America}, 2(4):393--410, nov 1954.
\newblock URL: \url{https://pubsonline.informs.org/doi/10.1287/opre.2.4.393}, \href {https://doi.org/10.1287/opre.2.4.393} {\path{doi:10.1287/opre.2.4.393}}.

\bibitem{DeKruijff2019}
Joost~T. de~Kruijff.
\newblock {\em High-tech low-volume production planning}.
\newblock PhD thesis, Technical University of Eindhoven, 2019.

\bibitem{Ding2008}
Guoli Ding, Li~Feng, and Wenan Zang.
\newblock {The complexity of recognizing linear systems with certain integrality properties}.
\newblock {\em Mathematical Programming}, 114(2):321--334, 2008.
\newblock \href {https://doi.org/10.1007/s10107-007-0103-y} {\path{doi:10.1007/s10107-007-0103-y}}.

\bibitem{Edmonds70}
Jack Edmonds.
\newblock {\em Submodular Functions, Matroids, and Certain Polyhedra}, pages 69--87.
\newblock Gordon and Breach, 1970.
\newblock \href {https://doi.org/10.1007/3-540-36478-1_2} {\path{doi:10.1007/3-540-36478-1_2}}.

\bibitem{Fortet1960}
Robert Fortet.
\newblock {L'algebre de Boole et ses applications en recherche operationnelle}.
\newblock {\em Trabajos de Estadistica}, 11(2):111--118, 1960.
\newblock \href {https://doi.org/10.1007/BF03006558} {\path{doi:10.1007/BF03006558}}.

\bibitem{Garey1979}
Michael.~R. Garey and David~S. Johnson.
\newblock {\em {Computers and Intractability A guide to the Theory of NP-completeness}}.
\newblock W.H. Freeman and Company, New York, 1979.

\bibitem{Ghouila-Houri1962}
Alain Ghouila-Houri.
\newblock {Caract{\'{e}}risation des matrices totalement unimodulaires}.
\newblock {\em Comptes Rendus Hebdomadaires des S{\'{e}}ances de l'Acad{\'{e}}mie des Science (Paris)}, 254:1192--1194, 1962.

\bibitem{Gleixner2021}
Ambros Gleixner, Gregor Hendel, Gerald Gamrath, Tobias Achterberg, Michael Bastubbe, Timo Berthold, Philipp Christophel, Kati Jarck, Thorsten Koch, Jeff Linderoth, Marco L{\"u}bbecke, Hans~D. Mittelmann, Derya Ozyurt, Ted~K. Ralphs, Domenico Salvagnin, and Yuji Shinano.
\newblock {MIPLIB 2017: data-driven compilation of the 6th mixed-integer programming library}.
\newblock {\em Mathematical Programming Computation}, 13(3):443--490, 2021.
\newblock \href {https://doi.org/10.1007/s12532-020-00194-3} {\path{doi:10.1007/s12532-020-00194-3}}.

\bibitem{Gomory60}
Ralph~E. Gomory.
\newblock An algorithm for the mixed-integer problem.
\newblock Technical Report RM-2597, RAND Corporation, 1960.

\bibitem{HIGHSrepo}
Julian~A.J. Hall, Leona. Gottwald, Ivet Galabova, and Qi~Huangfu.
\newblock Highs - linear optimization software.
\newblock \url{https://github.com/ERGO-Code/HiGHS}, 2025.

\bibitem{Hirsch68}
Warren~M. Hirsch and George~B. Dantzig.
\newblock The fixed charge problem.
\newblock {\em Naval Research Logistics Quarterly}, 15(3):413--424, 1968.
\newblock \href {https://doi.org/10.1002/nav.3800150306} {\path{doi:10.1002/nav.3800150306}}.

\bibitem{HoffmanK56}
Alan~J. Hoffman and Joseph~B. Kruskal.
\newblock Integral boundary points of convex polyhedra.
\newblock In {\em Linear Inequalities and Related Systems}, volume~38, pages 223--246. Princeton University Press, 1956.
\newblock \href {https://doi.org/10.1515/9781400881987-014} {\path{doi:10.1515/9781400881987-014}}.

\bibitem{HuangfuH17}
Qi~Huangfu and Julian~A.J. Hall.
\newblock Parallelizing the dual revised simplex method.
\newblock {\em Mathematical Programming Computation}, 1(10):119--142, 2017.
\newblock \href {https://doi.org/10.1007/s12532-017-0130-5} {\path{doi:10.1007/s12532-017-0130-5}}.

\bibitem{Hupp2017}
Lena~M. Hupp.
\newblock {\em {Integer and Mixed-Integer Reformulations of Stochastic, Resource-Constrained, and Quadratic Matching Problems}}.
\newblock PhD thesis, Friedrich-Alexander-Universit{\"{a}}t Erlangen-N{\"{u}}rnberg, 2017.
\newblock URL: \url{https://open.fau.de/items/8c897ec3-0aa5-40c9-83c4-83987cc21590/full}.

\bibitem{Karimi2003}
Behrooz Karimi, Seyyed~.M.T. {Fatemi Ghomi}, and John.~M. Wilson.
\newblock {The capacitated lot sizing problem: a review of models and algorithms}.
\newblock {\em Omega}, 31(5):365--378, oct 2003.
\newblock \href {https://doi.org/10.1016/S0305-0483(03)00059-8} {\path{doi:10.1016/S0305-0483(03)00059-8}}.

\bibitem{Karp1972}
Richard~M. Karp.
\newblock {Reducibility among Combinatorial Problems}.
\newblock In {\em Complexity of Computer Computations}, pages 85--103. Springer US, Boston, MA, 1972.
\newblock \href {https://doi.org/10.1007/978-1-4684-2001-2_9} {\path{doi:10.1007/978-1-4684-2001-2_9}}.

\bibitem{Land1960}
Alisa~H. Land and Alison~G. Doig.
\newblock {An Automatic Method of Solving Discrete Programming Problems}.
\newblock {\em Econometrica}, 28(3):497, 1960.
\newblock \href {https://doi.org/10.2307/1910129} {\path{doi:10.2307/1910129}}.

\bibitem{Lewis1980}
John~M. Lewis and Mihalis Yannakakis.
\newblock {The node-deletion problem for hereditary properties is NP-complete}.
\newblock {\em Journal of Computer and System Sciences}, 20(2):219--230, 1980.
\newblock \href {https://doi.org/10.1016/0022-0000(80)90060-4} {\path{doi:10.1016/0022-0000(80)90060-4}}.

\bibitem{Liebchen2006}
Christian Liebchen.
\newblock {\em {Periodic Timetable Optimization in Public Transport}}.
\newblock PhD thesis, TU Berlin, 2006.

\bibitem{LuebbeckeW18}
Marco Lübbecke and Jonas~T. Witt.
\newblock The strength of dantzig–wolfe reformulations for the stable set and related problems.
\newblock {\em Discrete Optimization}, 30:168--187, 2018.
\newblock \href {https://doi.org/10.1016/j.disopt.2018.07.001} {\path{doi:10.1016/j.disopt.2018.07.001}}.

\bibitem{Meyer74}
Robert~R. Meyer.
\newblock On the existence of optimal solutions to integer and mixed-integer programming problems.
\newblock {\em Mathematical Programming}, 7:223--235, 1974.
\newblock \href {https://doi.org/10.1007/BF01585518} {\path{doi:10.1007/BF01585518}}.

\bibitem{Musitelli2007}
Antoine Musitelli.
\newblock {\em {Recognition of generalized network matrices}}.
\newblock PhD thesis, EPFL Lausanne, 2007.
\newblock \href {https://arxiv.org/abs/0807.3541} {\path{arXiv:0807.3541}}, \href {https://doi.org/10.5075/epfl-thesis-3938} {\path{doi:10.5075/epfl-thesis-3938}}.

\bibitem{Musitelli2010}
Antoine Musitelli.
\newblock {Recognizing binet matrices}.
\newblock {\em Mathematical Programming}, 124(1-2):349--381, jul 2010.
\newblock \href {https://doi.org/10.1007/s10107-010-0372-8} {\path{doi:10.1007/s10107-010-0372-8}}.

\bibitem{Paat2022}
Joseph Paat, Miriam Schl{\"{o}}ter, and Robert Weismantel.
\newblock {The integrality number of an integer program}.
\newblock {\em Mathematical Programming}, 192(1-2):271--291, 2022.
\newblock \href {https://arxiv.org/abs/1904.06874} {\path{arXiv:1904.06874}}, \href {https://doi.org/10.1007/s10107-021-01651-0} {\path{doi:10.1007/s10107-021-01651-0}}.

\bibitem{Papadimitriou1990}
Christos~H. Papadimitriou and Mihalis Yannakakis.
\newblock On recognizing integer polyhedra.
\newblock {\em Combinatorica}, 10(1):107--109, 1990.
\newblock \href {https://doi.org/10.1007/BF02122701} {\path{doi:10.1007/BF02122701}}.

\bibitem{Schrijver86}
Alexander Schrijver.
\newblock {\em {Theory of Linear and Integer Programming}}.
\newblock John Wiley \& Sons, Inc., New York, NY, USA, 1986.

\bibitem{Siegel1989}
Carl~Ludwig Siegel and Komaravolu Chandrasekharan.
\newblock {\em {Lectures on the Geometry of Numbers}}.
\newblock Springer Berlin Heidelberg, Berlin, Heidelberg, 1989.
\newblock \href {https://doi.org/10.1007/978-3-662-08287-4} {\path{doi:10.1007/978-3-662-08287-4}}.

\bibitem{Truemper1990}
Klaus Truemper.
\newblock {A decomposition theory for matroids. V. Testing of matrix total unimodularity}.
\newblock {\em Journal of Combinatorial Theory, Series B}, 49(2):241--281, 1990.
\newblock \href {https://doi.org/10.1016/0095-8956(90)90030-4} {\path{doi:10.1016/0095-8956(90)90030-4}}.

\bibitem{VanderHulst2024}
Rolf van~der Hulst and Matthias Walter.
\newblock {A Row-wise Algorithm for Graph Realization}.
\newblock arXiv:2408.12869, 2024.
\newblock URL: \url{http://arxiv.org/abs/2408.12869}.

\bibitem{Walter2013}
Matthias Walter and Klaus Truemper.
\newblock {Implementation of a unimodularity test}.
\newblock {\em Mathematical Programming Computation}, 5(1):57--73, 2013.
\newblock \href {https://doi.org/10.1007/s12532-012-0048-x} {\path{doi:10.1007/s12532-012-0048-x}}.

\bibitem{Zambelli05}
Giacomo Zambelli.
\newblock A polynomial recognition algorithm for balanced matrices.
\newblock {\em Journal of Combinatorial Theory, Series B}, 95(1):49--67, 2005.
\newblock \href {https://doi.org/10.1016/j.jctb.2005.02.006} {\path{doi:10.1016/j.jctb.2005.02.006}}.

\bibitem{Ziegler01}
G{\"u}nter~M. Ziegler.
\newblock {\em {Lectures on Polytopes (Graduate Texts in Mathematics)}}.
\newblock Springer, 2001.
\newblock \href {https://doi.org/10.1007/978-1-4613-8431-1} {\path{doi:10.1007/978-1-4613-8431-1}}.

\end{thebibliography}

\end{document}